\documentclass[11pt,reqno]{amsart}

\usepackage{amssymb}
\usepackage{amsthm}
\usepackage{amsxtra}
\usepackage{amsmath}
\usepackage{wrapfig}
\usepackage[colorlinks=true, allcolors=black]{hyperref}
\usepackage{comment}
\usepackage{amsfonts}
\usepackage{fancyhdr}
\usepackage{graphicx}
\usepackage{color}
\usepackage{amsmath}
\usepackage{listings} 
\numberwithin{equation}{section}
\usepackage{float}
\usepackage{cases} 
\usepackage{threeparttable}
\usepackage{dcolumn}
\usepackage{multirow}
\usepackage{booktabs}
\usepackage[dvipsnames]{xcolor}

\usepackage{multicol}
\usepackage{caption}
\usepackage{subcaption}
\usepackage[autostyle]{csquotes}
\usepackage[english]{babel}
\usepackage{csquotes}

\newtheorem{theorem}{Theorem}[section]

\theoremstyle{definition}

\newtheorem{remark}[theorem]{Remark}

\DeclareMathOperator{\sech}{sech}
\graphicspath{ {./Images/} }

\title[Breathers in the mKdV equation]{Robustness of 
\MakeLowercase{\Large{m}}K\MakeLowercase{\Large{d}}V Breathers: a numerical perspective}

\author[C. Haight]{Chandler Haight}
\address{Department of Mathematics \& Statistics, Florida International University, Miami, FL 33199, USA}
\email{chaig004@fiu.edu}

\author[S. Roudenko]{Svetlana Roudenko}
\address{Department of Mathematics \& Statistics, Florida International University, Miami, FL 33199, USA}
\email{sroudenk@fiu.edu}

\author[D. Son]{Diana Son}
\address{Department of Mathematics \& Statistics, Florida International University, Miami, FL 33199, USA}
\email{dson003@fiu.edu}

\author[K. Yang]{Kai Yang}
\address{Key Laboratory of Nonlinear Analysis and its Applications (Ministry
of Education), and
College of Mathematics \& Statistics, Chongqing University, Chongqing, 401331, China}
\email{yangkai10@cqu.edu.cn}



\begin{document}

\begin{abstract}
We systematically investigate breather solutions in the modified Korteweg-de Vries (mKdV) equation via numerical simulations. We show that the breather solutions are stable under a variety of perturbations, including amplitude changes, modifications of internal parameters, and perturbations of the nonlinearity of the equation. 
Our results are not only consistent with the known analytical studies on the stability of breathers in integrable systems, extending them further, but also provide numerical evidence that such breather-type structures can persist and remain stable over the simulated time scales in some non-integrable settings with symmetric potentials.

\end{abstract}

\maketitle

%\bigskip
%\tableofcontents
%\bigskip

\hypersetup{linkcolor=black}
\tableofcontents
\hypersetup{linkcolor=black}

\section{Introduction}
In this paper, we examine breather solutions, their interactions, and perturbations of the modified KdV (mKdV) equation on the whole real line, as well as perturbations of the mKdV nonlinearity via combined nonlinearities. In a completely integrable system a breather solution is a {\it localized in space} entity that interacts with another breather or soliton in an elastic way, keeping the form of the envelope before and after the interaction, and the velocity. A distinct feature of a breather is its oscillatory (breathing) profile, which is {\it periodic in time}, {\it localized in space}, and is {\it modulated by an envelope} having the shape of a solitary wave (e.g., $\sech$, the hyperbolic secant). A breather can be thought of and characterized as a pocket of concentrated energy of oscillations, moving in one direction with a constant speed. 
This `breathing' pattern distinguishes a breather from a soliton, while sharing similarities such as space localization and traveling in a specific direction. This behavior is what originally attracted the attention of researchers to studying breathers in completely integrable systems. 

Originally, breathers were found by considering pairs of poles of the transmission coefficient symmetric about the imaginary axis \cite{w1973}, \cite{K1974}, \cite{KM1974}, \cite{Lamb}, while solving the integrable systems such as the 1d cubic NLS or the modified KdV (mKdV) via the inverse scattering method that was introduced in \cite{ZS}.
Investigating multi-soliton solutions in the early 1970s, Kuznetsov \cite{K1974} referred to the breather prototypes as {\it non-stationary solitary dislocations} \cite{KM1974}, and later in \cite{K1974} as {\it solitons, whose amplitude oscillated} with a certain frequency. The idea of non-purely imaginary poles as well as these breathing solitons was discussed back then with Zakharov, collaborators and visitors, \cite{K2023}, and later became termed as breathers. 
The family of breathers that we study in this paper was originally written for the mKdV equation by Wadati \cite{w1973}, who also found them via integrable-systems machinery, mentioning that when considering an $N$-soliton solution and removing the second assumption (non-pure imaginary eigenvalues for a bound state), one obtains a `new family' of solutions.

While breather solutions are less studied compared to the soliton literature, they have been recently attracting attention of researchers in terms of existence or non-existence, descriptions of their properties, stability, interactions, e.g., 
\cite{am2013}, \cite{GAV2013}, \cite{am2015}, \cite{amp2017_2}, \cite{a2018}, \cite{MP2019}, \cite{dp2020}, \cite{d2022}, \cite{HMP2023}, \cite{Sem2023}, \cite{Sem2022}, \cite{CS2023},
and their many possible applications, such as fluid dynamics \cite{MAO2023}, breather generation in oceanic models, e.g., in \cite{LPTPXK2007}, \cite{LZ2019}, or rogue waves \cite{choa2012}, \cite{DT1999}, in dispersion-managed systems \cite{TBF2012}, in nonlinear wave equations \cite{ms2021}; breathers have started to be investigated in non-integrable systems: we give some examples in this paper (see Section \ref{combined}), 
for a related non-integrable model such as Whitham equation, see \cite{KACP2022}.
\smallskip

In this paper, we systematically examine breathers and their interactions in 
the mKdV equation 
\begin{equation}\label{mKdV}
u_t + u_{xxx} + %\mathcal N(u)
u^{2}u_x = 0, \quad 
\end{equation}
where $u=u(x,t)$ is a real-valued function with $x \in \mathbb{R}$, $t \in \mathbb{R}$.  
\smallskip

We not only consider various perturbations and interactions of breathers, but we also investigate breathers in a perturbed mKdV equation by adding a small nonlinear term. 
One such version of the perturbed mKdV equation is 
\begin{equation}\label{G-general}
u_t + u_{xxx} + u^2 u_x + \varepsilon u^{q-1} u_x = 0, \qquad q>1, ~ q \ne 3, ~~~\varepsilon > 0,
\end{equation}
which is a variant of a generalized Gardner equation. Another one incorporates an absolute value (as we have considered in \cite{frrsy2022}), namely,
\begin{equation}\label{G-general-abs}
u_t + u_{xxx} + u^2 u_x + \varepsilon |u|^{q-1} u_x = 0, \qquad q>1, ~ q \ne 3, ~~~\varepsilon > 0,
\end{equation}
\smallskip

The purpose of this paper is to systematically review the breathers in the mKdV equation \eqref{mKdV}, which is an integrable system, and develop a numerical technique to identify breathers in the long-term evolution of various mKdV solutions, including perturbations of breathers, their interactions, including multiple interactions, or multi-breathers, and interactions with solitons. %as well as with solitons and other generic data. 
\smallskip

The main result of this paper is that mKdV breathers are stable under small perturbations, and under larger perturbations the solutions resolve into coherent structures such as breathers and solitons, while dispersing radiation to the left (negative direction), which provides numerical support for the {\it soliton-breather} resolution conjecture. We show the cases of negative breather interactions, and various types of breather perturbations, in some of which the breather structure remains and in some it breaks and forms solitons (or a double pole). 
We are also able to check `local' mass of breathers and the amount of radiation dispersed to the left, quantifying it as a percentage of the total mass. 
Finally, we investigate the breather-like behavior in the perturbed mKdV models, which includes a non-integrable model, where the mKdV equation is perturbed by a nonlinearity with a symmetric potential as in \eqref{G-general-abs}. 
%and it keeps exhibiting breather behaviors.  
Even in the non-integrable perturbations, we observe stable breather-like behavior, providing numerical support for the conjecture that KdV-type equations with symmetric potential have breathers (unlike the standard KdV equation). 
\medskip

The paper is organized as follows. 
In Section \ref{review}, we review the mKdV equation and its soliton and 
breather solutions. In Section \ref{interactionsmKdV}, we investigate interactions of breathers with themselves or solitons, and observe no radiation at the numerical resolution used, in agreement with the complete integrability of the mKdV model. We also show stable interactions of breathers traveling to the left. In Section \ref{perturbations}, we consider various perturbations of the mKdV breathers; to characterize their long-term behavior we develop
a numerical approach that identifies breathers by their internal parameters, providing numerical evidence for their stability or resolution into solitons and breather states. In Section \ref{combined}, we investigate perturbations of the mKdV combining it with a small lower order nonlinearity, resulting either in a Gardner-type equation or in a non-integrable mKdV-type model, and provide numerical evidence for stable breather-like dynamics in such models.

\section{Review of solitons and breathers in the mKdV equation}
We start with a review of solitons and breathers in the mKdV, as they are among the most prominent coherent structures of this completely integrable equation; their history is connected with traveling solitary waves in the KdV equation ($p=2$), going back to the observations of John Scott Russell in 1834 (see his report in \cite{JSR1839}). The Miura transform \cite{Miura1968} links the mKdV solutions with those of KdV, which produces solitary waves in the mKdV. The breathers in the mKdV were discovered and originally written a bit later in \cite{w1973}, which turned out to be a prominent feature of mKdV since the KdV equation does not possess breathers \cite{MP2019}. In this paper, we write the modified Korteweg-de Vries (mKdV) equation with the following normalization 
\begin{equation}\label{mKdV-n}
u_t+(u_{xx}+ \tfrac{1}{3} u^3)_x=0,
\end{equation}
exactly as stated in the introduction in \eqref{mKdV}.
Similar to the KdV, the mKdV is completely integrable and has traveling (to the right) solitary waves, but unlike the KdV, it also has breather solutions. 
It is well-known that the mKdV \textit{solitons} are of the form 
\begin{equation}\label{mKdVS}
u(t,x)=Q_c(x-ct), 
\end{equation}
with the parameter $c>0$ determining its height and speed, and $Q = Q_1$ being the unique positive smooth, vanishing at infinity, solution to 
\begin{equation}\label{E:S}
-Q+Q_{xx}+\frac13 Q^3=0, 
\end{equation}
which can be solved explicitly, 
\begin{equation}\label{mKdVScont}
Q(x):=\sqrt{6}\sech{(x)} \equiv 2\sqrt{6}\partial_x\left[\arctan{(e^x)}\right]. 
\end{equation}
The rescaling
\begin{equation}\label{E:mKdV-S-rescale}
Q_c(s):=\sqrt{c} \, Q(\sqrt{c}s), \quad s=x-ct, \quad c>0,
\end{equation}
solves the equation
\begin{equation}\label{mKdVssol}
-c\,Q+Q_{xx}+\frac13 \,Q^3=0, 
\end{equation}
thus, allowing one to consider a one-parameter family of solitons $\{Q_c\}_{c > 0}$ for a given $Q$. 
\smallskip

The mKdV \textit{breather} solution is a two-parameter family $B_{\alpha, \beta}$ with $\alpha,\beta>0$ and is given (e.g., \cite{w1973}, \cite{Lamb}, \cite{am2013}) by 
\begin{equation}\label{mKdV-B}
B_{\alpha, \beta}(x,t) := {2\sqrt 6} \,\partial_x \left[\text{arctan}\left(\frac{\beta\sin{(\alpha(x+\delta t))}}{\alpha\cosh{(\beta(x+\gamma t))}}\right)\right],
\end{equation} 
or differentiating and writing explicitly,
\begin{equation}\label{mKdVBS}
B_{\alpha, \beta}(x,t) := {2\sqrt 6\,} \beta \text{sech}(\beta(x+\gamma t)) \left[\frac{\text{cos}(\alpha (x+\delta t))-\frac{\beta}{\alpha}\text{sin}(\alpha(x+\delta t))\text{tanh}(\beta(x+\gamma t))}{1+(\frac{\beta}{\alpha})^2\text{sin}^2(\alpha(x+\delta t))\text{sech}^2(\beta(x+\gamma t))}\right],
\end{equation}
where
\begin{equation}\label{dg}
\delta := \alpha^2-3\beta^2,\qquad \gamma := 3\alpha^2-\beta^2.
\end{equation}
The parameter $\gamma$ determines the translation of the breather envelope: the center is located at $x=-\gamma t$, so the physical velocity is $-\gamma$. The parameter $\delta$ represents the internal oscillation rate inside the envelope. We note that $\delta\neq\gamma$, since $\alpha$ and $\beta$ are nonzero. 

Recall that breathers are periodic in time but not in space, though they are localized in space; hence, it is possible to characterize breathers as follows: there exist a period $T>0$ and $L=L(T) \in \mathbb R$ such that 
\begin{equation}\label{E:B-TL}
B_{\alpha,\beta}(x,t+T) = B_{\alpha,\beta}(x-L,t).
\end{equation}
It is also known (e.g., \cite{am2013}, \cite{Sem2022}) that the time and spatial periods are connected via the breather-envelope translation parameter $\gamma$ as $L=-\gamma T$. 
We note that it is essential that $T>0$ in \eqref{E:B-TL}, otherwise, this would indicate a soliton.

In contrast to solitons, which have constant shape for all times and only always travel to the right ($c>0$ in \eqref{mKdVS}), %\cite{frrsy2022}. 
breathers can also travel left, as seen, for example, in Figure \ref{B}. Because the physical velocity is $-\gamma$, $\gamma<0$ indicates that a breather travels \textit{to the right}, while $\gamma>0$ indicates that it travels \textit{to the left}.
One other feature that differs breathers from solitons is their zero mean, which is also conserved in time (see e.g., \cite{am2013}).

While the solitons satisfy a simple second-order ODE \eqref{E:S}, the breathers satisfy a more involved fourth-order nonlinear ODE \cite{am2013}, namely,
\begin{equation}\label{E:eqB}
B^{(4)} - 2(\beta^2-\alpha^2)\left(B^{\prime\prime} +\frac13 B^3\right) + (\alpha^2+\beta^2)^2B + \frac53 B (B^\prime)^2+\frac53 B^2B^{\prime\prime} +\frac16B^{5}=0.
\end{equation}
Here the primes denote derivatives in $x$ at fixed $t$. (In \cite[Lemma 4.1]{Sem2023} a fourth-order elliptic equation has been written for a soliton, which was used for the analysis of multi-breathers and their uniqueness.) 

The conserved quantities for the mKdV equation \eqref{mKdV} include the mass
\begin{equation}\label{mass}
M[u](t):= \int_{\mathbb{R}}^{} u^2(t,x) \,dx = M[u](0),
\end{equation}
and Hamiltonian (energy) 
\begin{equation}\label{energy}
E[u](t):=\frac{1}{2} \int_{\mathbb{R}}^{} u^2_x(t,x) \,dx - \frac{1}{12} \int_{\mathbb{R}}^{} u^4(t,x) \,dx= E[u](0).
\end{equation}
(As the completely integrable system, the mKdV equation has infinitely many conserved quantities, including the integral of the solution and higher energies, but they will not be needed for this paper, and thus, omitted.)

The mass for the mKdV soliton $Q_c$ is simple (noting that $\int_{\mathbb R} \sech^2 dx = 2$)
\begin{equation}\label{mQmass}
M[Q_{c}] = c^{1/2} M[Q] = 12 c^{1/2},
\end{equation}
while its energy is
\begin{equation}\label{mQenergy}
E[Q_c]=c^{3/2} E[Q] = -2 c^{3/2}.
\end{equation}
We note that in the mKdV (as the subcritical case of gKdV), the energy of solitons is negative, as shown in \eqref{mQenergy} (it is zero in the critical case of generalized KdV, when the nonlinearity power would be $p=5$). %, i.e., $u^3$ replaced with $u^5$ in \eqref{mKdV}.)

The mass for the mKdV breather $B_{\alpha,\beta}$ \eqref{mKdVBS} is (see \cite[Lemma 2.1]{am2013} with our normalization in \eqref{mKdVScont})
\begin{equation}\label{mBmass}
M[B_{\alpha,\beta}] = 24\beta\, ( = 2\beta M[Q]),
\end{equation}
noting that the mass of the breather is independent of $\alpha$ and only depends on $\beta$.

The energy of the breather $B_{\alpha,\beta}$ is (see \cite[Lemma 2.4]{am2013})
\begin{equation}\label{mBenergy}
E[B_{\alpha,\beta}]= 4 \beta \gamma = 2 \beta \gamma |E[Q]| \equiv 2 \beta (3\alpha^2-\beta^2) |E[Q]|.
\end{equation}

\begin{remark}
We point out that as the soliton parameter $c$ can be identified numerically from its mass or $L^\infty$ norm (for example, as it is done in \cite{KPRS}), the parameters $\alpha,\, \beta$ (and hence, $\gamma,\, \delta$) of the breather $B_{\alpha,\beta}$ can be identified numerically from its mass and energy: first, we find $\beta$ from the mass \eqref{mBmass} and then $\alpha$ from the energy \eqref{mBenergy}, thus, pinpointing exactly the breather $B_{\alpha,\beta}$, which is useful in numerical simulations when trying to understand the asymptotic profiles as well as matching the profiles after some propagation in time (see further on identification of breathers in Section \ref{S:CompTools}).
\end{remark}

\subsection{Pokhozhaev identities}
Multiplying \eqref{E:S} by $Q$ and integrating in $x$, we obtain the first Pokhozhaev identity
$$
M[Q] +\|Q_x\|^2_{L^2} = \frac13 \|Q\|^4_{L^4}.
$$
Similarly, multiplying \eqref{E:S} by $xQ_x$, integrating (via integration by parts), we also have the second Pokhozhaev identity
$$
M[Q] -\|Q_x\|^2_{L^2} = \frac16 \|Q\|^4_{L^4}.
$$
Solving for $\|Q_x\|^2_{L^2}$ and $\|Q\|^4_{L^4}$ in terms of mass $M[Q]$, we obtain
$$
\|Q_x\|^2_{L^2} = \frac13 M[Q] \quad \mbox{and} \quad \|Q\|^4_{L^4} = 4 M[Q],
$$
and hence, 
$$
E[Q] = \frac12 \|Q_x\|^2_{L^2} - \frac1{12} \|Q\|^4_{L^4} \equiv -\frac16 M[Q],
$$
which is consistent with \eqref{mQmass} and \eqref{mQenergy} when $c=1$. 
These identities and the connection between mass and energy are often useful in numerical testing and verification of solitons and accuracy of schemes.

\subsection{Concise review of mKdV breathers} %Soliton and Breather Interactions}
\label{review}

Recall that breathers are periodic in time but not in space, though well spatially localized. In several integrable systems where breathers exist, breather solutions have explicit expressions, which is not only convenient for analytical study but also can be used to identify the asymptotic behavior of breathers (e.g., their asymptotic stability). 
For example, one can extract the parameters $\alpha$ and $\beta$ from the mass and energy to identify $B_{\alpha,\beta}$ as in \eqref{mKdV-B}. We introduce this method in Section \ref{S:CompTools}, but first 
we review the dynamics of breathers in the mKdV 
and also examine the $L^\infty$ norm to further understand properties of breathers, which will later play an essential role in their identification. 

For all numerical simulations produced for this paper, the numerical schemes and approaches are detailed in \cite{Yang2022} and \cite{Yang2023}. The parameters used in this section are $N=2^{13}$, $L=50\pi$, and $dt=0.01$, unless otherwise noted. We simulate the time evolution of several breathers starting from the initial data 
$$
u_0(x)=B_{\alpha, \beta}(x,0),
$$ 
with $B_{\alpha,\beta}$ from \eqref{mKdVBS} and compare them with the exact solutions first to ensure the accuracy of our computations and also to show their dependence on parameters and properties of the time period as well as the localization; we then investigate the influence of various parameters on their properties. %\\
\smallskip

\begin{figure}[h!]
\begin{center}
\includegraphics[width=0.24\textwidth]{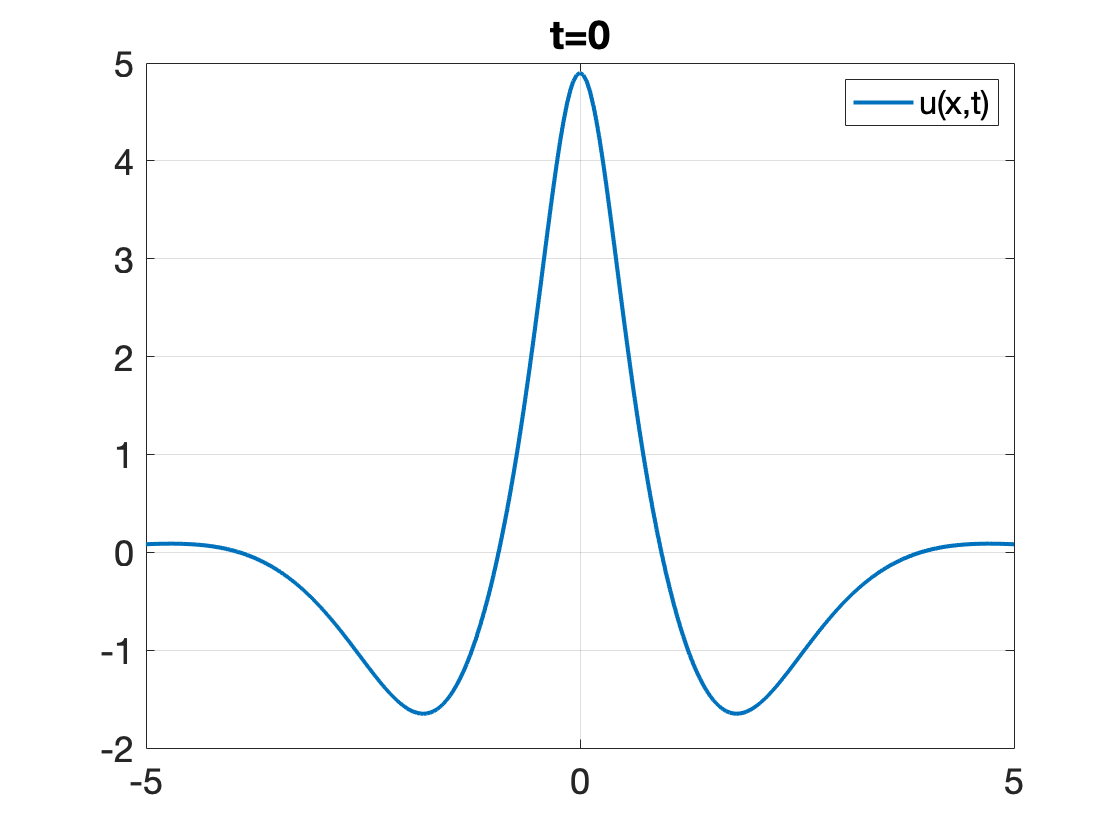}
\includegraphics[width=0.24\textwidth]{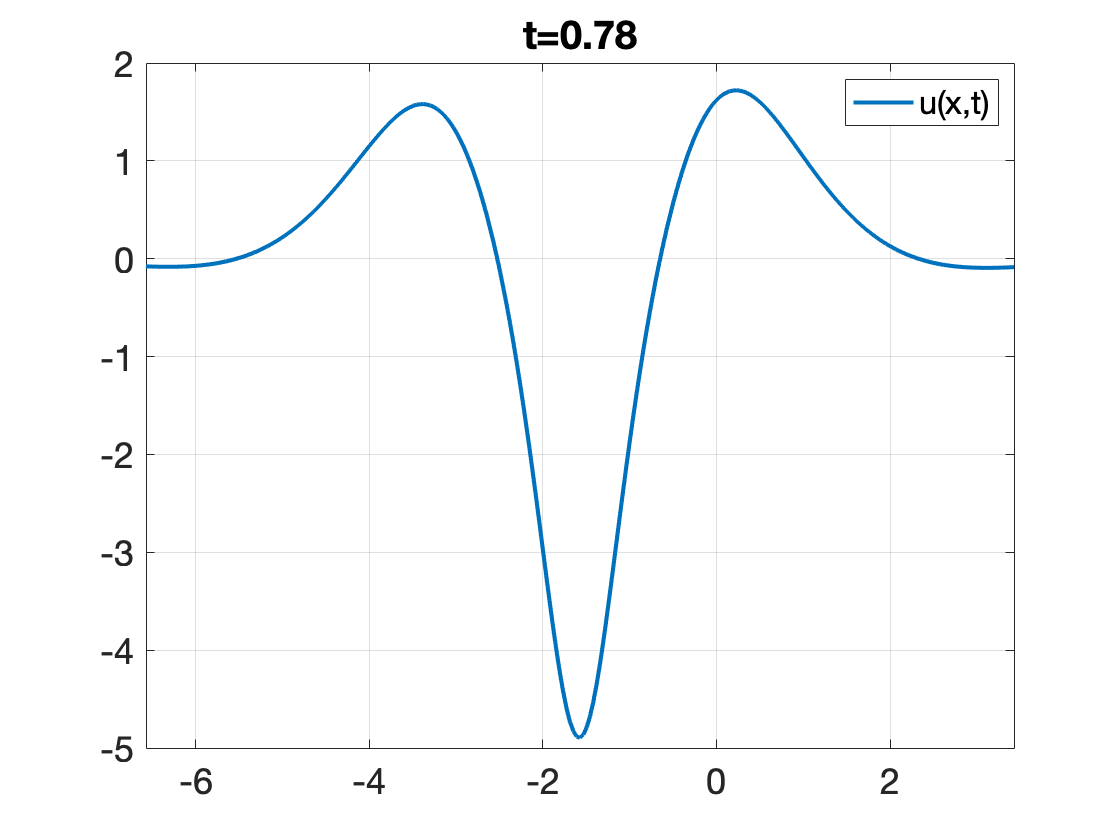}
\includegraphics[width=0.24\textwidth]{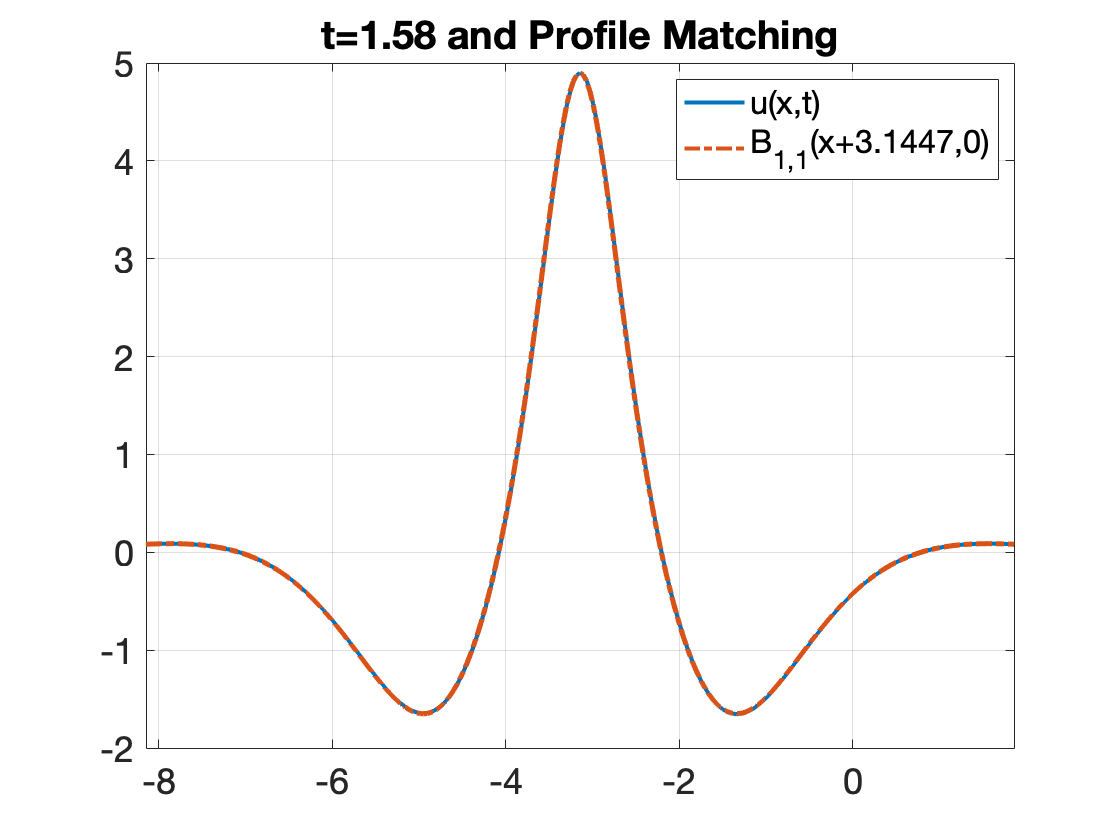}
\includegraphics[width=0.24\textwidth]{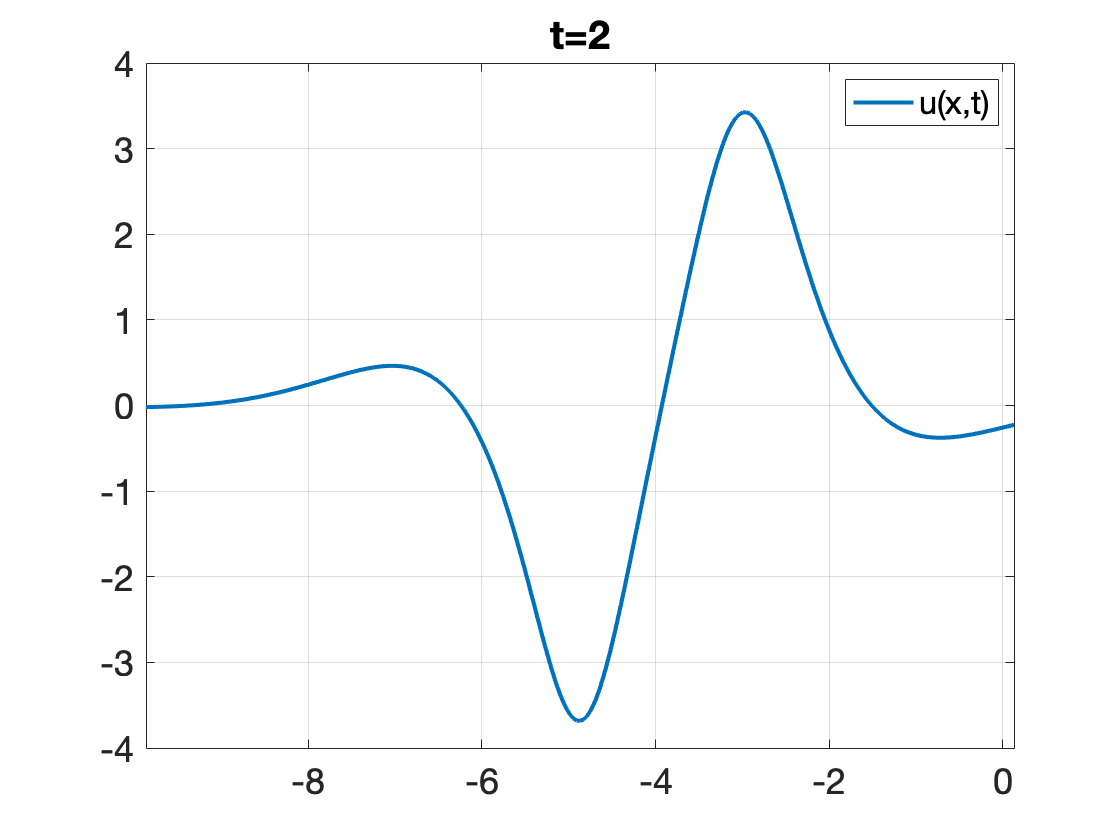}
\newline 
(a) \hspace{102.5pt} (b) \hspace{102.5pt} (c) \hspace{102.5pt} (d)
\caption{\label{B} {\small Time evolution of $u_0(x) = B_{1,1}(x,0)$ from \eqref{mKdVBS}. In (c), matching the numerical simulation of $u(x,t)$ at $t=1.58$ with the exact (shifted) breather $B_{1,1}(x-a,0)$, $a=-3.1447$.}}
\end{center}
\end{figure}

\noindent $\bullet$ \underline{Case $\alpha=1$, $\beta=1$.}
Our first breather example is given in Figure \ref{B}, where the solution with the initial data $u_0(x)=B_{1,1}(x,0)$ in plot (a) propagates under the mKdV flow to the left, since $\gamma=2>0$ (see \eqref{dg} and the paragraph after that explaining the directions), showing the oscillating (`breathing') pattern. 

\vspace{-.3cm}
Note the periodic behavior of the breather: the original shape of the breather at $t=0$ in subplot (a) is also seen at $t \cong 1.58$ in subplot (c), where the breather has oscillated through one cycle. Also note the half cycle at $t \cong 0.78$ (symmetric but negative shape), given in subplot (b). 
Since the velocity of the breather $\gamma=2$ is positive, indicating that the breather is moving \textit{to the left}, in subplot (c) of Figure \ref{B}, we match the simulated dynamics with the exact breather $B_{1,1}(x-a,0)$ shifted to the left by $a=-3.1447$. In subplot (d), a snapshot at $t=2$ is given to show other configurations of this breather within its envelope. 
\smallskip

\noindent $\bullet$ \underline{Case $\alpha=2$, $\beta=1$.} \indent
We change the parameter $\alpha$ from 1 to 2, thereby increasing $\gamma$ from $2$ to $11$ (hence, the breather moves faster to the left than in the previous case).
\begin{figure}[htb!]
\begin{center}
\includegraphics[width=0.24\textwidth]{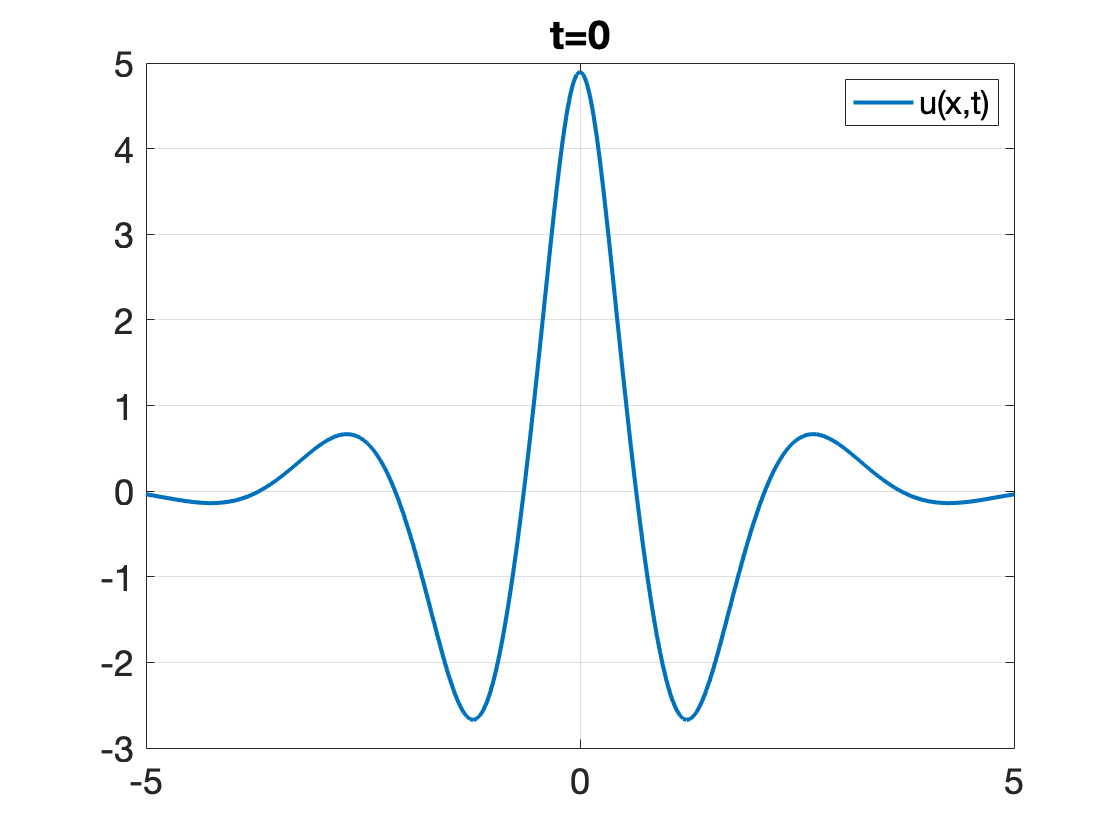}
\includegraphics[width=0.24\textwidth]{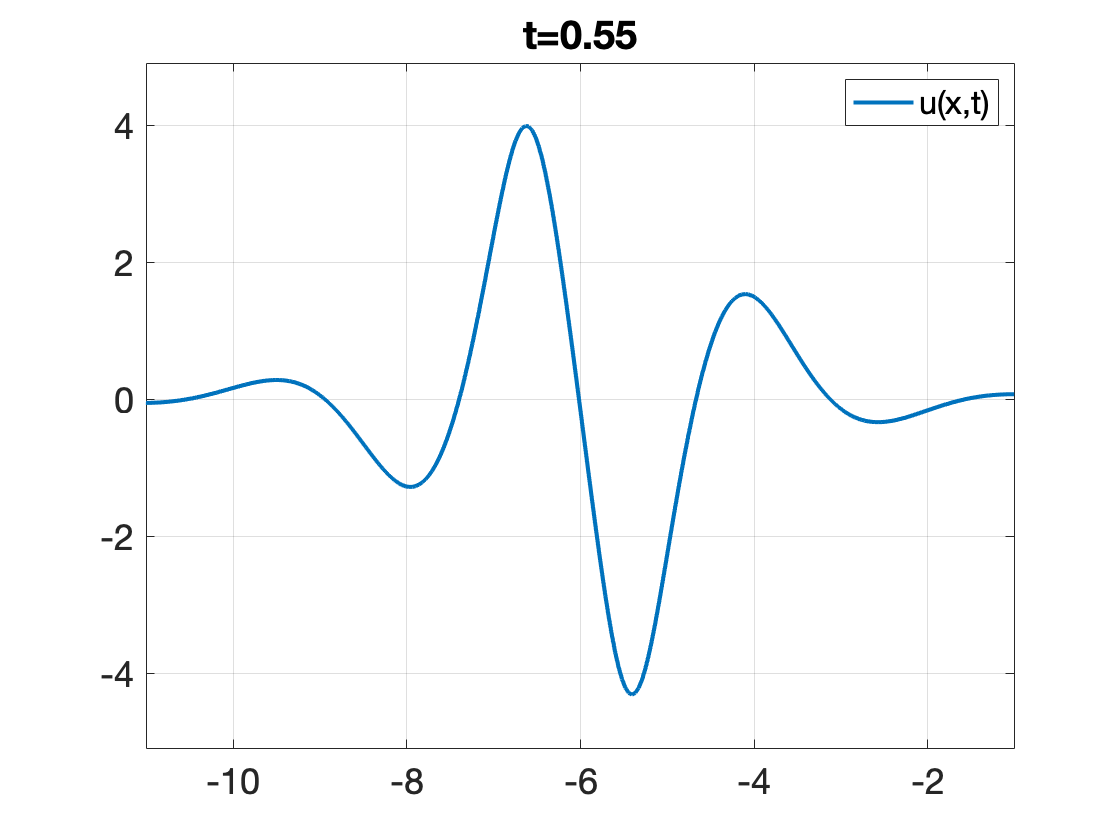} \hspace{6pt} 
\raisebox{2mm}{\includegraphics[width=0.195\textwidth]
{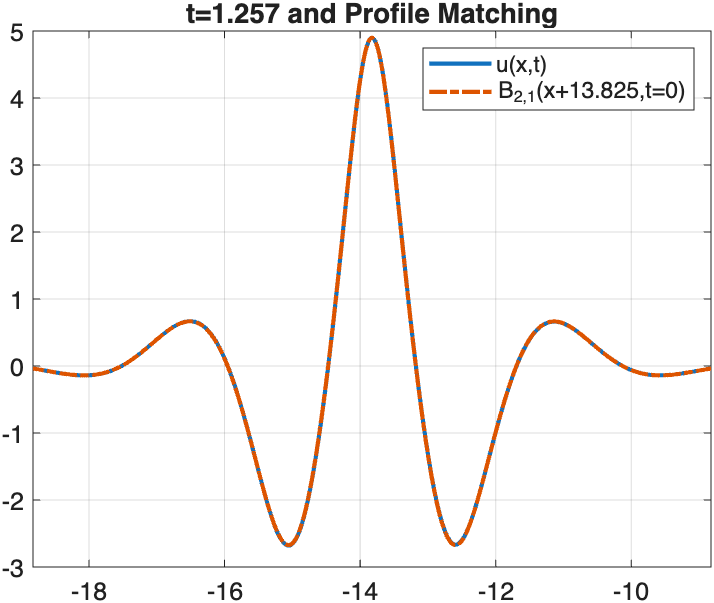}} \hspace{6pt}
\includegraphics[width=0.24\textwidth]{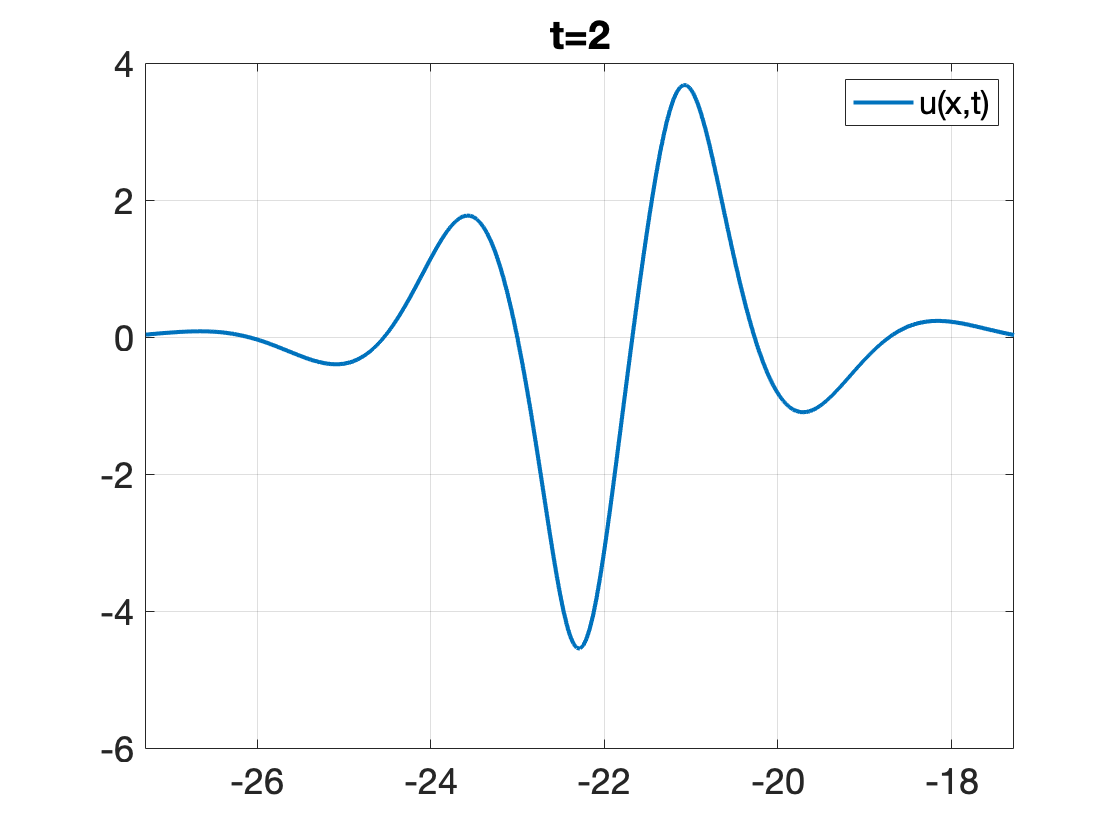}
\newline 
(a) \hspace{102.5pt} (b) \hspace{102.5pt} (c) \hspace{102.5pt} (d)
\caption{\label{B2} {\small Time evolution of $u_0(x) = B_{2,1}(x,0)$ from \eqref{mKdVBS}. In (c), matching the numerical simulation of $u(x,t)$ at $t=1.27$ with the exact (shifted) breather $B_{2,1}(x-a,0)$, $a=-13.825$.} }
\end{center}
\end{figure}
In Figure \ref{B2}, we plot several snapshots of the time evolution of the breather solution with the initial condition $u_0(x) = B_{2,1}(x,0)$.
Note that here $\delta = 1$, which is larger than the previous value $\delta=-2$, and this produces more oscillations within the envelope. Comparing Figures \ref{B} and \ref{B2}, we note that the breather oscillates faster within the envelope and travels farther distance-wise during the same time (compare the locations in subplot (d) of Figure \ref{B} and \ref{B2}).

The time periodicity can also be observed, the original shape as in (a) appears again at $t \cong 1.27$ in subplot (c), just as with $\alpha=1$ (but here after 4 cycles), and the first period occurs at $t \cong 0.315$. In both Figure \ref{B} and \ref{B2}, the second parameter $\beta=1$, so one could observe the difference in breather behavior as it relates to changes in $\alpha$: for $B_{1,1}$ the oscillation speed within the envelope $\delta = -2$ and for $B_{2,1}$ it is increased to $\delta = 1$. 

Next, we compare breathers with the same $\alpha$ but different $\beta$. \smallskip

\begin{figure}[h!]
\begin{center}
\includegraphics[width=0.24\textwidth]{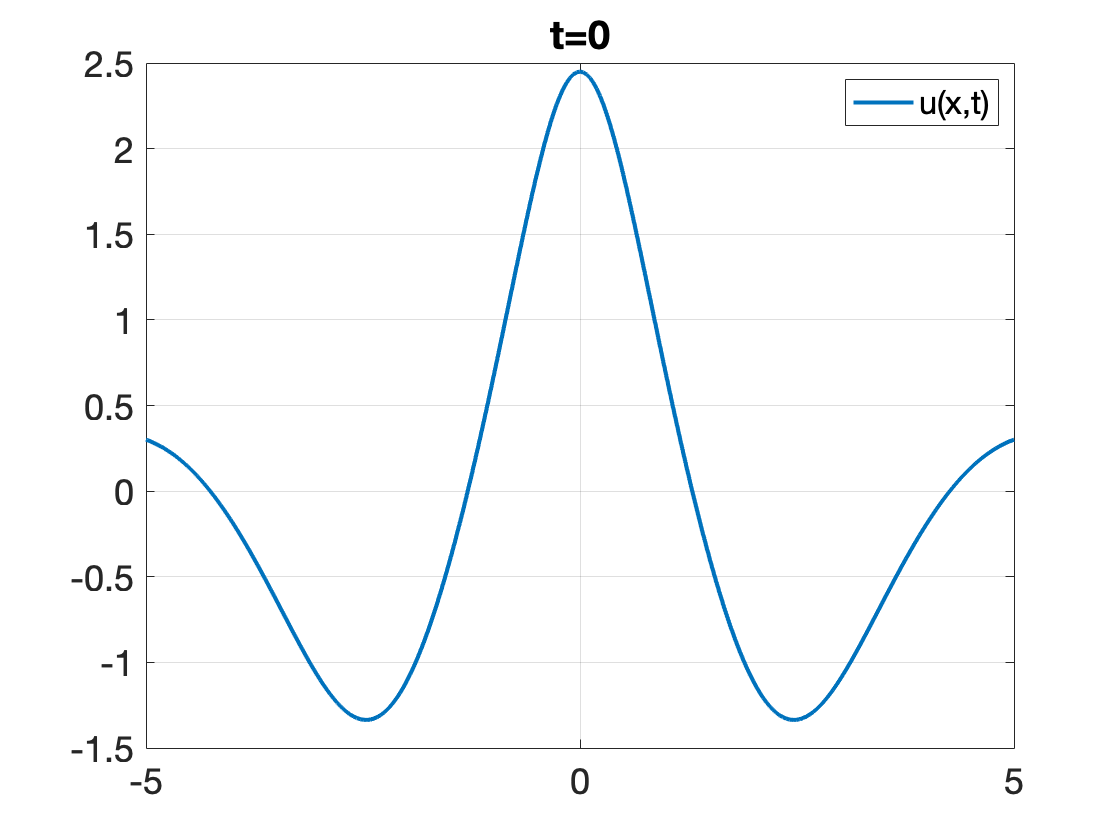}
\includegraphics[width=0.24\textwidth]{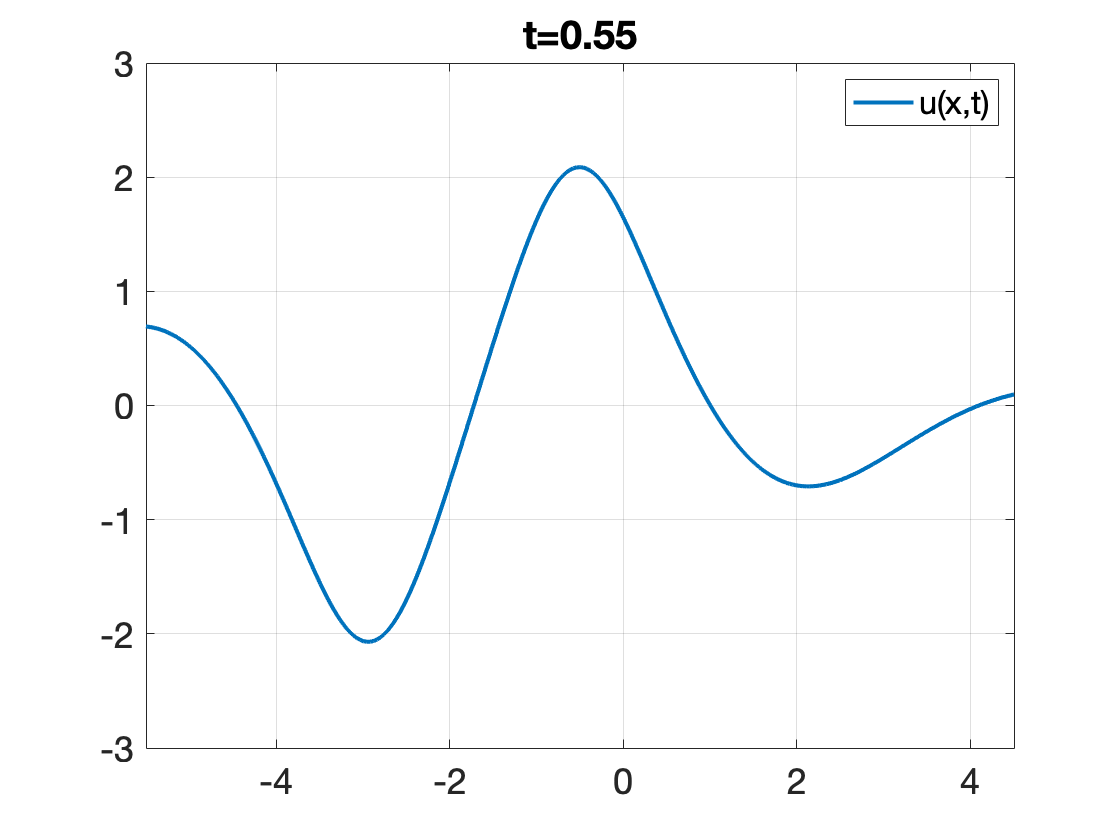}
\includegraphics[width=0.24\textwidth]{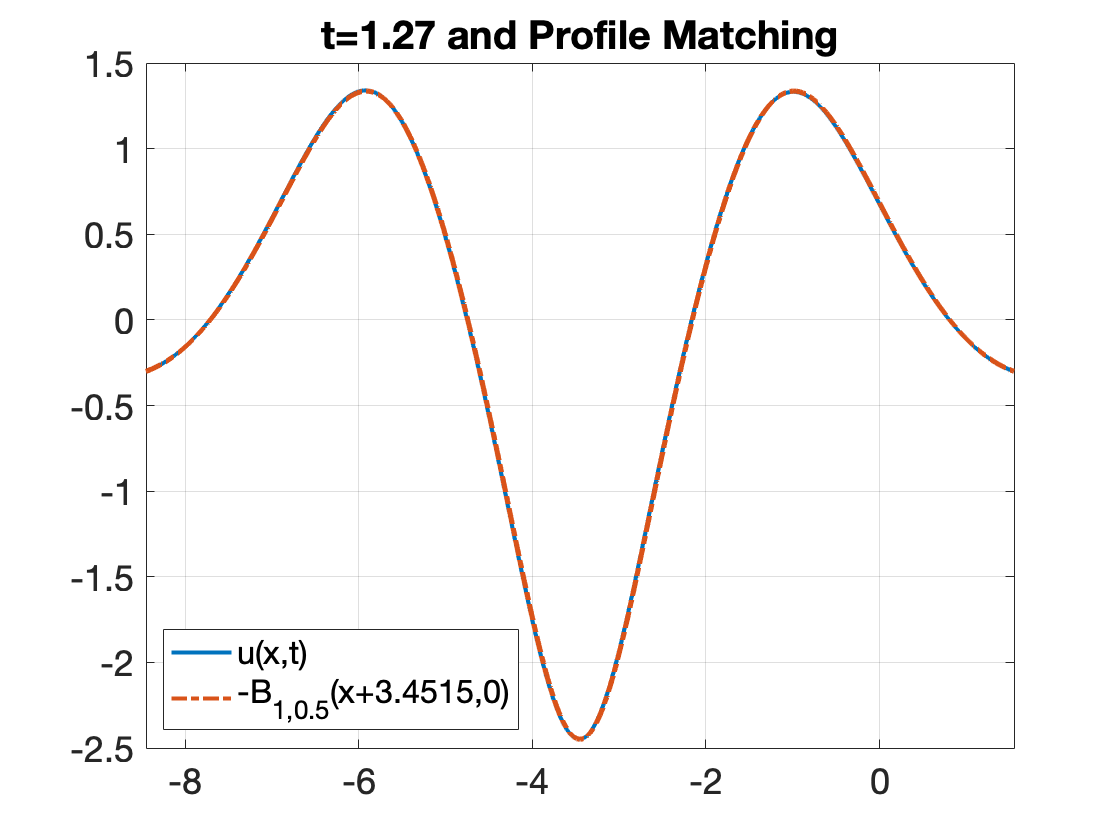}
\includegraphics[width=0.24\textwidth]{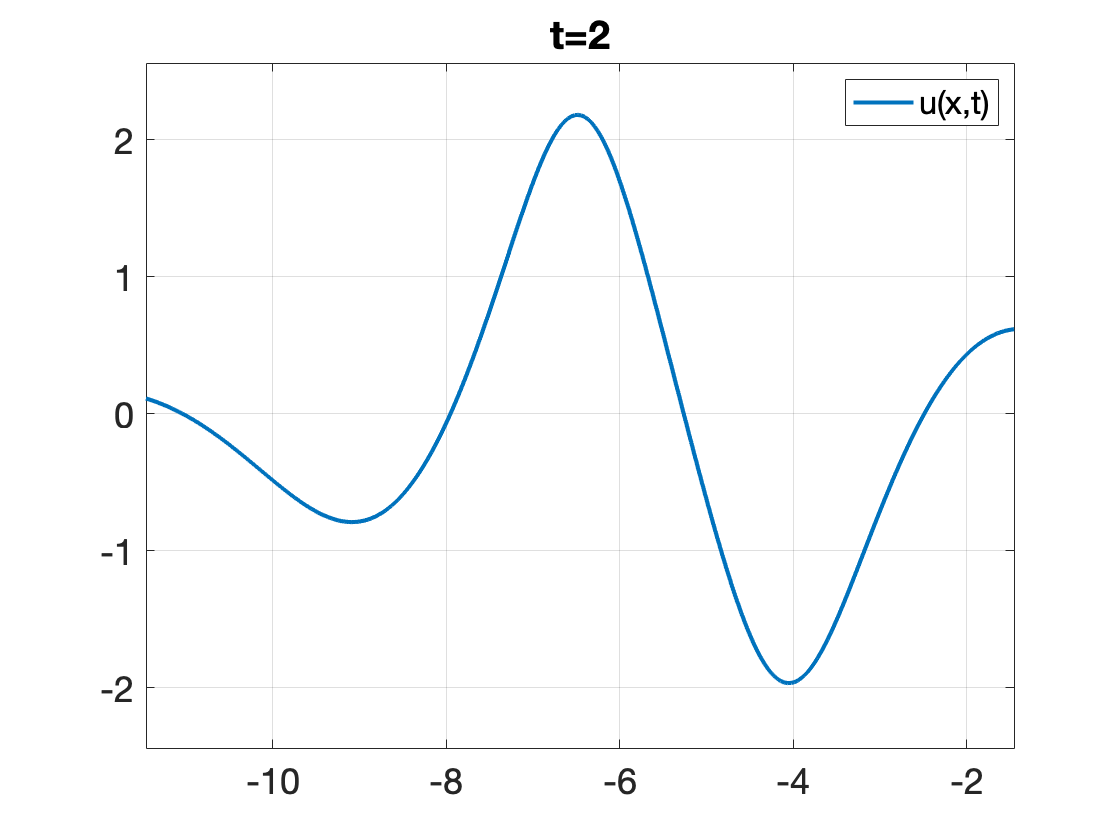}
\newline 
(a) \hspace{102.5pt} (b) \hspace{102.5pt} (c) \hspace{102.5pt} (d)
\caption{\label{B3} {\small Time evolution of $u_0(x) = B_{1,0.5}(x,0)$ from \eqref{mKdVBS}. In (c), matching $u(x,t)$ at $t=1.27$ with the exact (shifted) breather $B_{1,0.5}(x-a,0)$, $a=-3.4515$.} }
\end{center}
\end{figure}

\noindent $\bullet$ \underline{Case $\alpha=1$, $\beta=0.5$.}
\indent In Figure \ref{B3} the snapshots of the mKdV evolution of the initial condition $u_0(x) = B_{1,0.5}(x,0)$ are given; here, the value of $\beta$ is decreased to $1/2$, while $\alpha$ remains equal to 1, the same as in the first example. Comparing Figure \ref{B} with Figure \ref{B3}, one may notice that decreasing $\beta$ makes the shape of the breather wider or less tight around its initial center ($x=0$), and oscillations within its envelope are slower. 

With $\alpha=1$ and $\beta=0.5$, it follows that $\delta=\frac14$, which means that the rate of internal oscillations is slow. Thus, it takes longer in time for the solution to complete a full period. Note that the simulation in Figure \ref{B3} ends at $t=2$, which is before the solution can complete a full period. Thus, we choose to match the profile at the half-period mark at $t\cong 1.27$ in subplot (c) of Figure \ref{B3}. By knowing the half-period, we can conclude that for the solution $B_{1,0.5}(x,t)$, a full period takes about $t \cong 2.54$ to complete and travels about $x\cong 6.9$ to the left. \smallskip

\noindent $\bullet$ \underline{Comparison of $L^\infty$ norms.} \indent In all of the shown cases, the matching of our simulated mKdV time evolution is nearly perfect with the exact breather profile in subplots (c), confirming the validity of our numerical scheme. 
\begin{figure}[ht!]
\begin{center}
\includegraphics[width=0.32\textwidth]{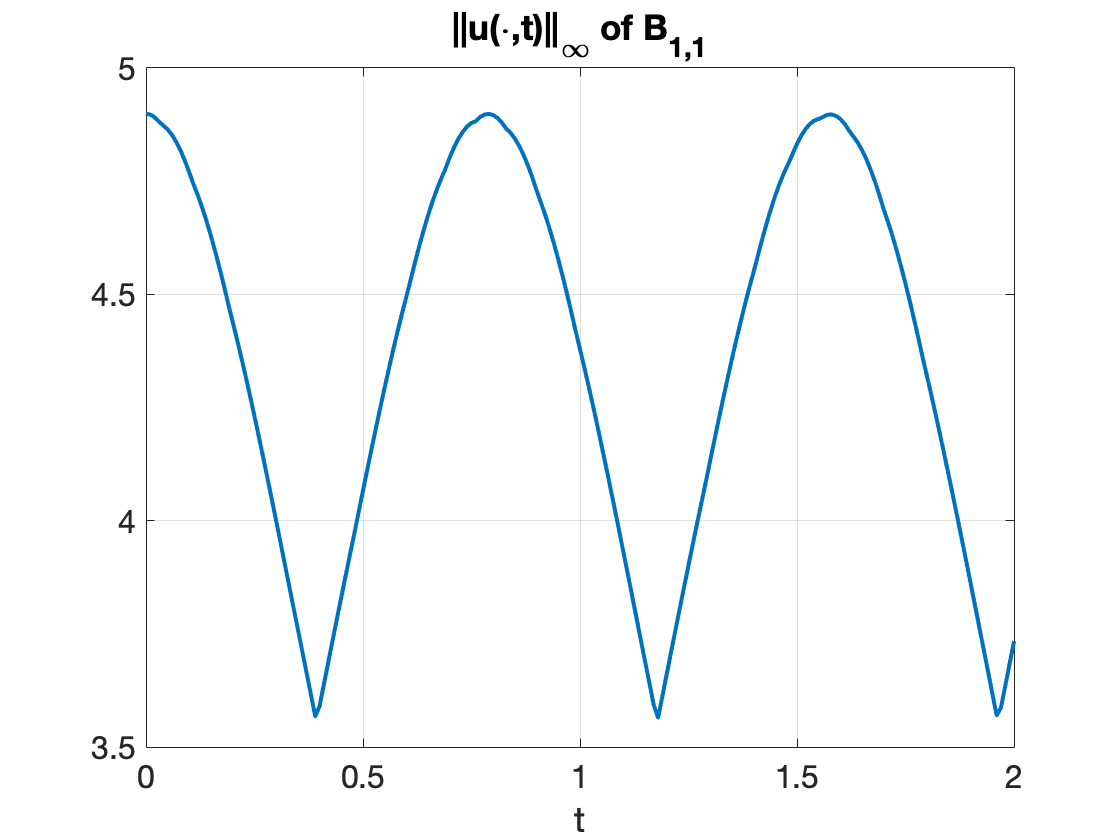} \hspace{15pt}
\includegraphics[width=0.2675\textwidth]{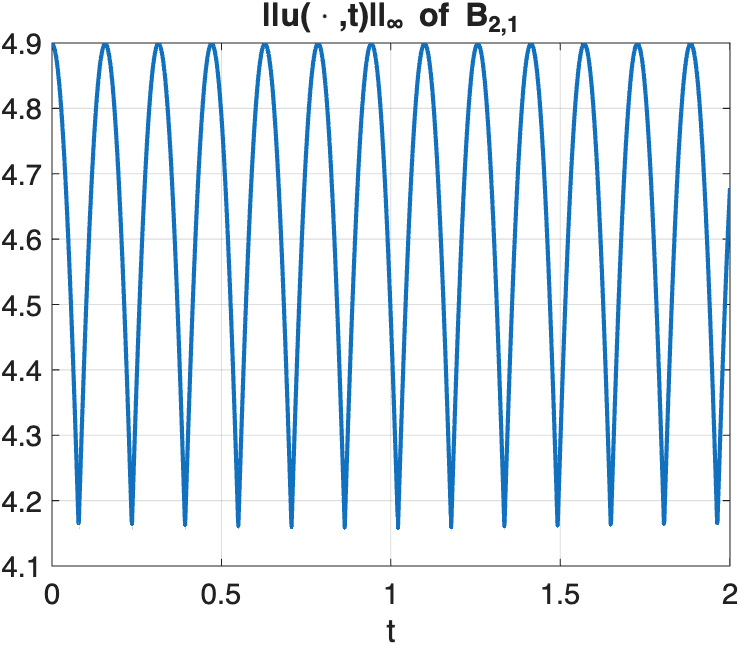} \hspace{15pt}
\includegraphics[width=0.32\textwidth]{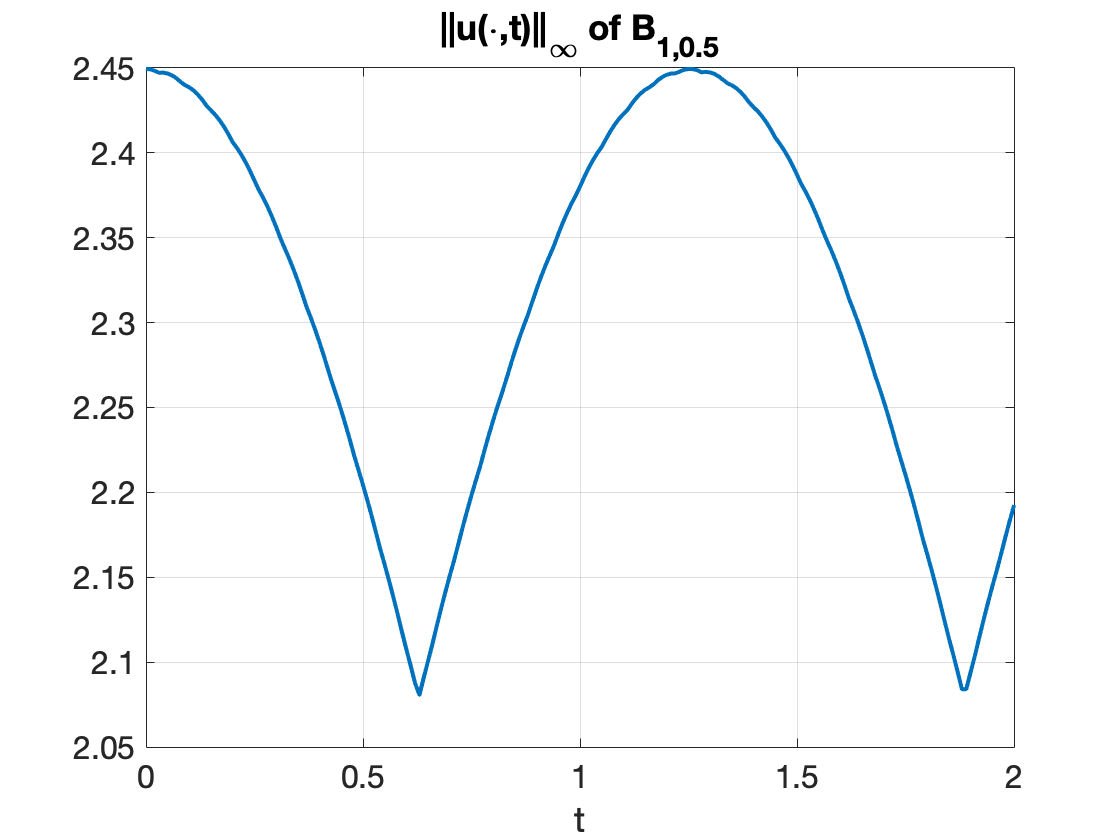}\\
\hspace{2pt} (a) \hspace{148.5pt} (b) \hspace{146pt} (c) 
\caption{\label{BsNorms} {\small {Time dependence of the $L^{\infty}$ norm of the numerically computed mKdV breathers: (a) $B_{1,1}(x,t)$, (b) $B_{2,1}(x,t)$, and (c) $B_{1,0.5}(x,t)$.}}}
\end{center} 
\end{figure}
Note that due to the symmetry of the breathers, their initial supremum norm
\begin{equation}\label{E:Linfty}
\|u(0)\|_{L^\infty} = \|B_{\alpha,\beta}(0,0)\| = 2\sqrt 6 \beta
\end{equation}
gives the initial maximum height of the breather and allows us to later compute the time period of the breather. Note that the initial height does not depend on the parameter $\alpha$. Thus, the initial maximum height of the first two breathers is $\|u(0)\|_{L^\infty} = 2\sqrt 6 \approx 4.90$, which can be seen as the peak in Figures \ref{B} and \ref{B2}, as well as in the $L^\infty$ norm in (a) and (b) of Figure \ref{BsNorms}. %\\

In Figure \ref{BsNorms}, time dependence of the $L^\infty$ norms of all three breathers from Figure \ref{B}, \ref{B2}, and \ref{B3} are shown. All three graphs in Figure \ref{BsNorms} have periodic behavior in time, one of {\it the defining features} of a breather, which we also exploit later when breathers are not given explicitly. Due to the difference in $\alpha$ and $\beta$ values between the three cases, we see different rates of oscillation, and checking the (second) maximum in each case, we confirm the period times given earlier (since the breather is symmetric in its shape around the $x$-axis, the first maximum in the $L^\infty$ norm gives the time when it is negative and looks exactly opposite to the initial shape; thus, taking double of that time gives the full period). 

The fastest oscillation in Figure \ref{BsNorms} is in subplot (b), corresponding to $B_{2,1}$, whose evolution is shown in Figure \ref{B2}. This behavior is expected given that this case has the highest $\alpha$ value of the 3 examples considered: a higher $\alpha$ means more internal oscillations and a larger value of $|\gamma|$, while a lower $\beta$ means a wider, lower-amplitude breather envelope. Due to the numerical resolution (in our case, Matlab), the middle plot was re-simulated with $dt=0.001$, $N=2^{16}$, and $L=100\pi$ to improve the resolution and have equal minima (for visual purposes).\smallskip

\begin{figure}[ht!]
\begin{center}
\raisebox{3.5mm}{\includegraphics[width=0.49\textwidth]{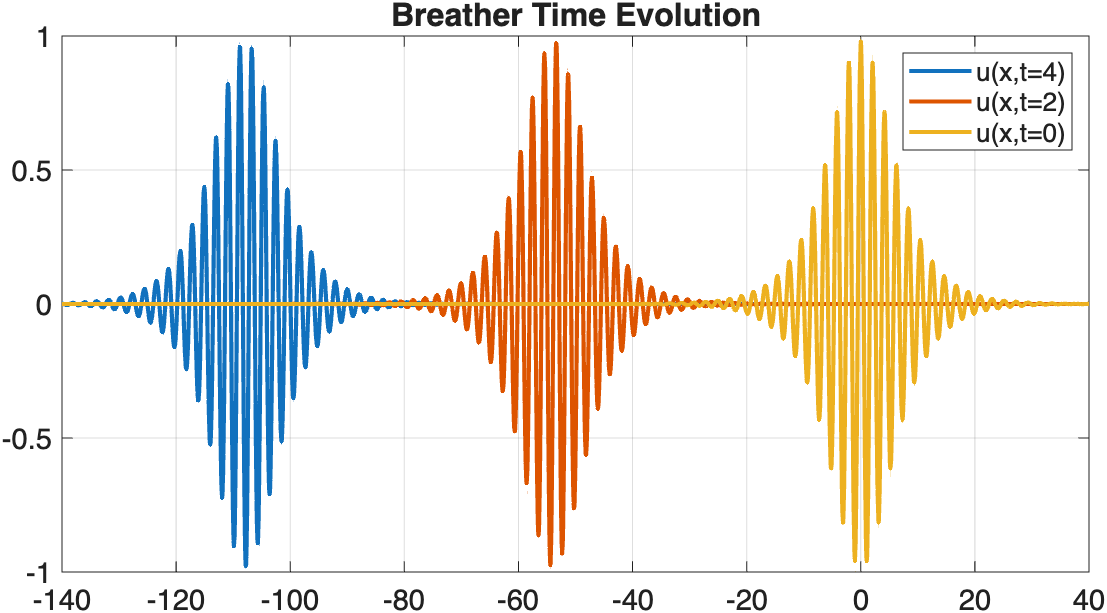}}
\includegraphics[width=0.40\textwidth]{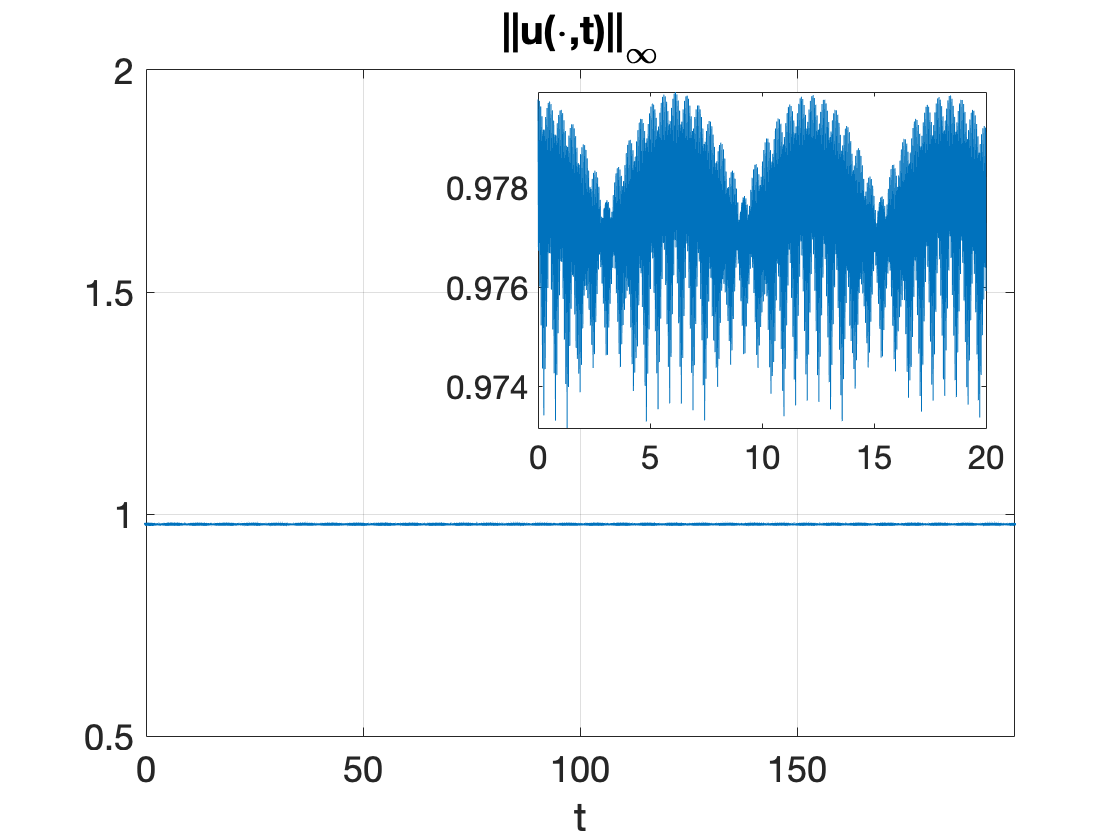}
\caption{\label{GKdVB5.1} Example of a constant $L^\infty$ norm in a breather with $u_0(x) = B_{3,0.2}(x,0)$: profiles at $t=0$, $t=2$, and $t=4$ (left), $L^\infty$ norm up to $t=200$ with the zoomed-in subplot (right).}
\end{center}
\end{figure}

\noindent $\bullet$ \underline{Breather with nearly constant $L^\infty$ norm.} \indent In Figure \ref{GKdVB5.1}, we further investigate the parameters $\alpha$ and $\beta$ by generating a case where the breather will result in a nearly constant $L^\infty$ norm. Here, we take a much higher $\alpha$ value than $\beta$ value, to be precise $B_{3,0.2}(x,0)$. In doing so, we choose a profile with more internal oscillations (and thus shorter pockets), and it also has a smaller envelope height compared to the previous examples. 
Since the breather profile at $t=0$ is symmetric with dense oscillations cascading down on both sides of the profile's center and the surrounding peaks are nearly as tall in amplitude, this evolution produces an $L^\infty$ norm that looks numerically constant, see the right subplot of Figure \ref{GKdVB5.1}. 
Therefore, by looking only at the $L^\infty$ norm, it would {\it not be possible to identify a soliton from a breather}. 

Taking a closer look at the $L^{\infty}$ norm, although it seems to be constant around $0.97$ on a larger $y$-axis interval $[0.5,2]$, zooming-in to a significantly smaller interval (changes in the third decimal) reveals the oscillatory behavior of the breather that resembles the $L^{\infty}$ norms in the previous Figure \ref{BsNorms}. The phenomenon we see in Figure \ref{GKdVB5.1} is important to keep in mind for identifications of breathers as the $L^{\infty}$ norm may not show the periodicity (especially in much larger values of $\alpha$ than $\beta$). 

\section{Interactions of multiple mKdV coherent structures}\label{interactionsmKdV}

\indent Since the mKdV equation is a completely integrable system, it is important to investigate the behavior of its solutions under forced interactions. We investigate breathers as they interact with each other and also with solitons. In the following examples, we show that breathers remain stable before, during, and after interactions with other solutions (although, of course, during the interactions, they go through deformations). Since soliton-soliton interactions are well known and studied, e.g., see examples of interactions in \cite{frrsy2022} for the generalized KdV equations, we omit those cases. Instead, we include examples of soliton-breather, breather-breather, multiple breathers, and multiple breathers-soliton interactions. Recent works \cite{Sem2022}, \cite{Sem2023}, \cite{CS2023} showed that multi-breathers are unique and orbitally stable provided they travel in the positive direction. With our convention, this corresponds to physical velocity $-\gamma>0$, or $\gamma<0$. Here, we show examples of multi-breathers and breather-soliton interactions with negative physical velocities, equivalently $\gamma>0$. The computational parameters for this section are $N=2^{13}$, $L=14\pi$ (for a finite domain $[-L,L]$), and $dt=0.01$.
\smallskip

Complete integrability is associated with known {\it analytical} features such as the existence of a Lax pair, \cite{Lax1968}, or having infinitely many conserved quantities; but how do completely integrable systems present themselves {\it numerically}? In each case of this section, the simulations are consistent with the elastic behavior expected from complete integrability of the mKdV equation. Specifically, within the numerical resolution used here, the simulations exhibit {\it no effects of radiation} or dispersion coming from the interactions. If the mKdV model were not completely integrable, or in non-integrable perturbation models, by contrast, one would generally observe small-scale dispersion and radiation generated from interactions.
\smallskip

\begin{figure}[ht!]
\begin{center}
\includegraphics[width=0.32\textwidth]{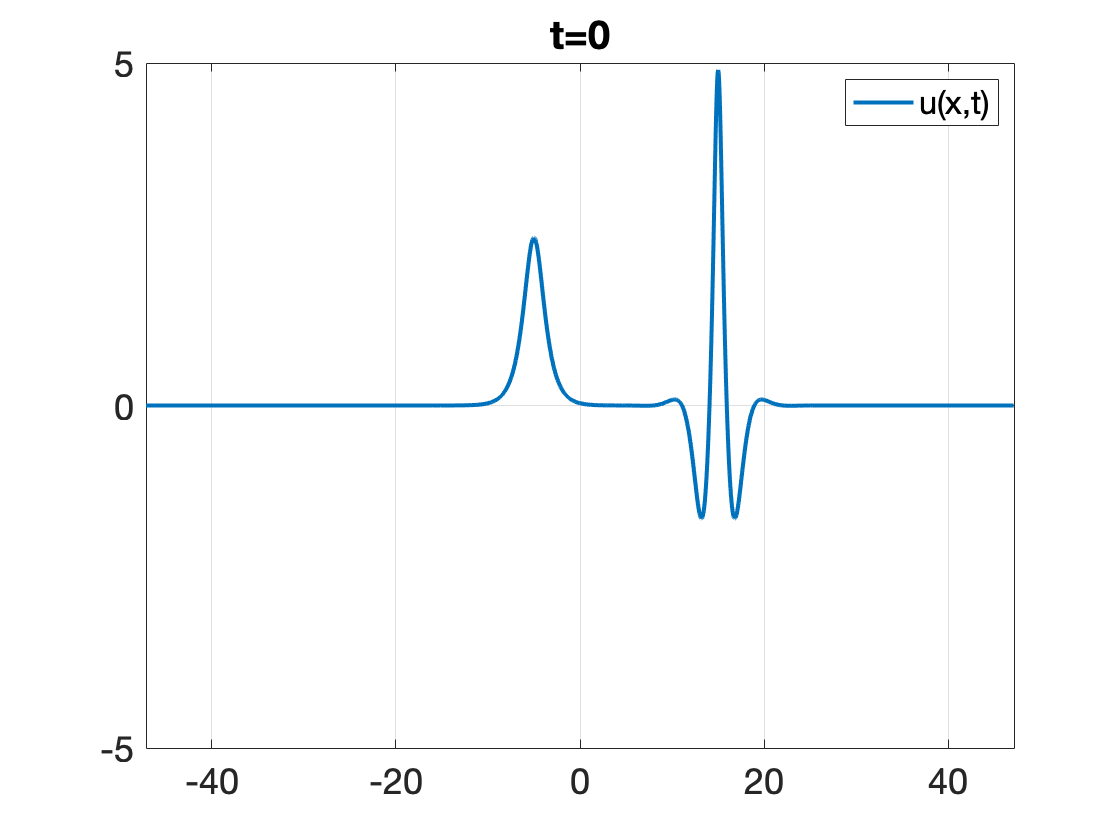}
\includegraphics[width=0.32\textwidth]{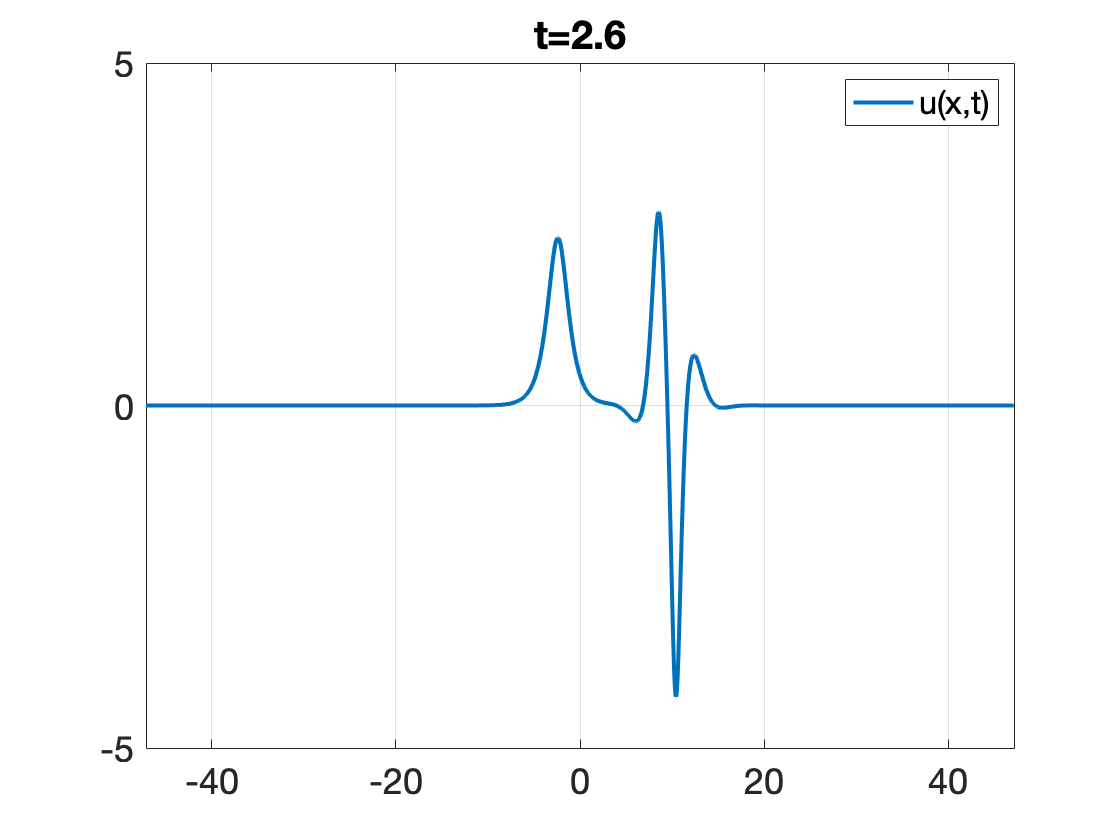}
\includegraphics[width=0.32\textwidth]{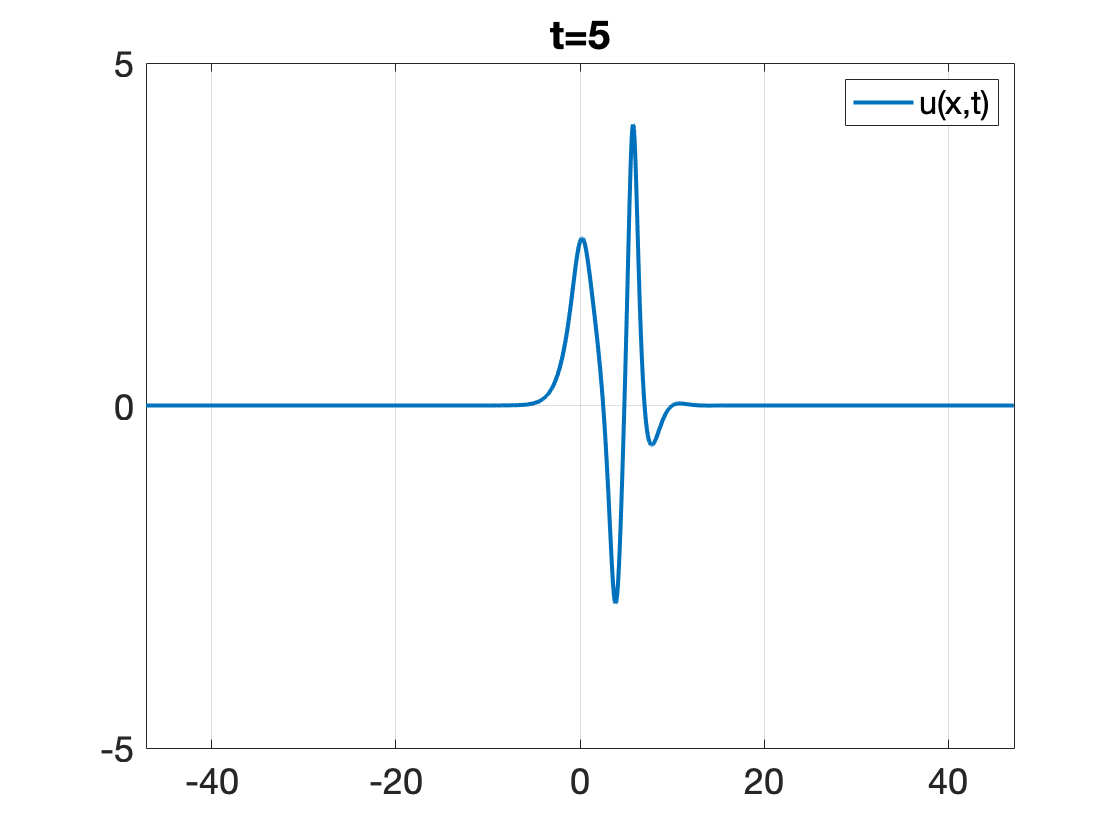}
\includegraphics[width=0.32\textwidth]{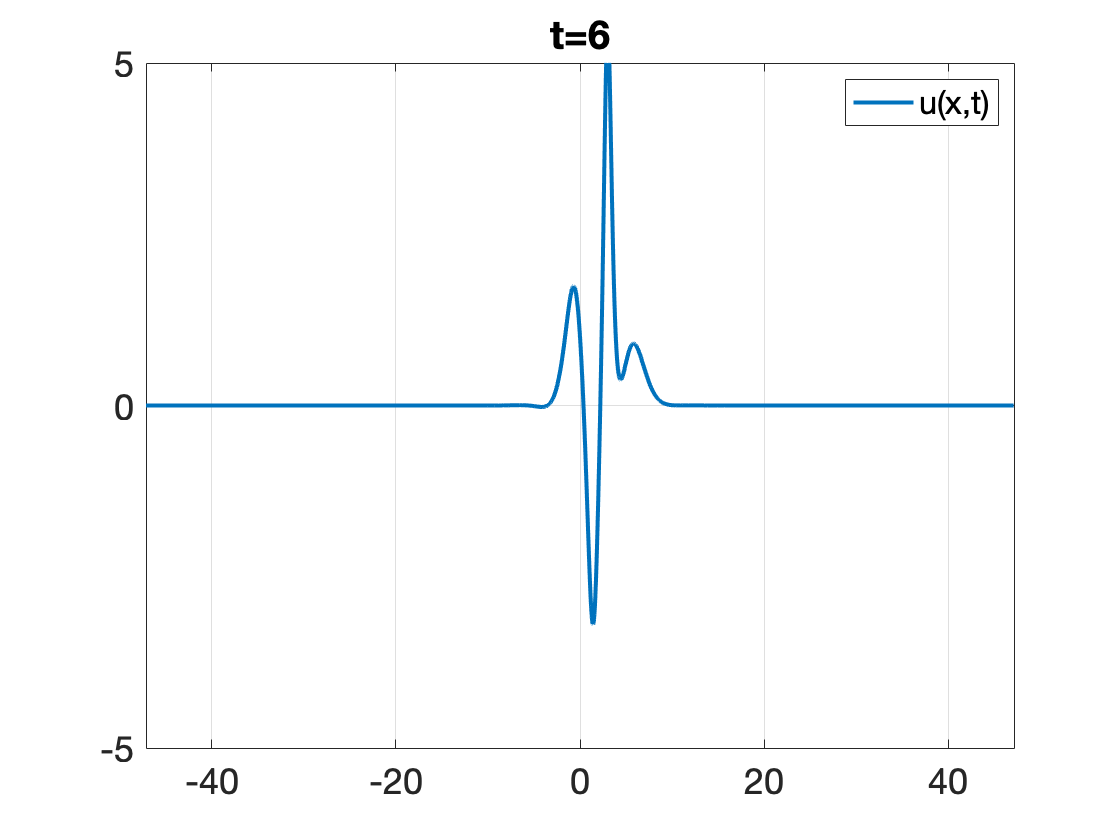}
\includegraphics[width=0.32\textwidth]{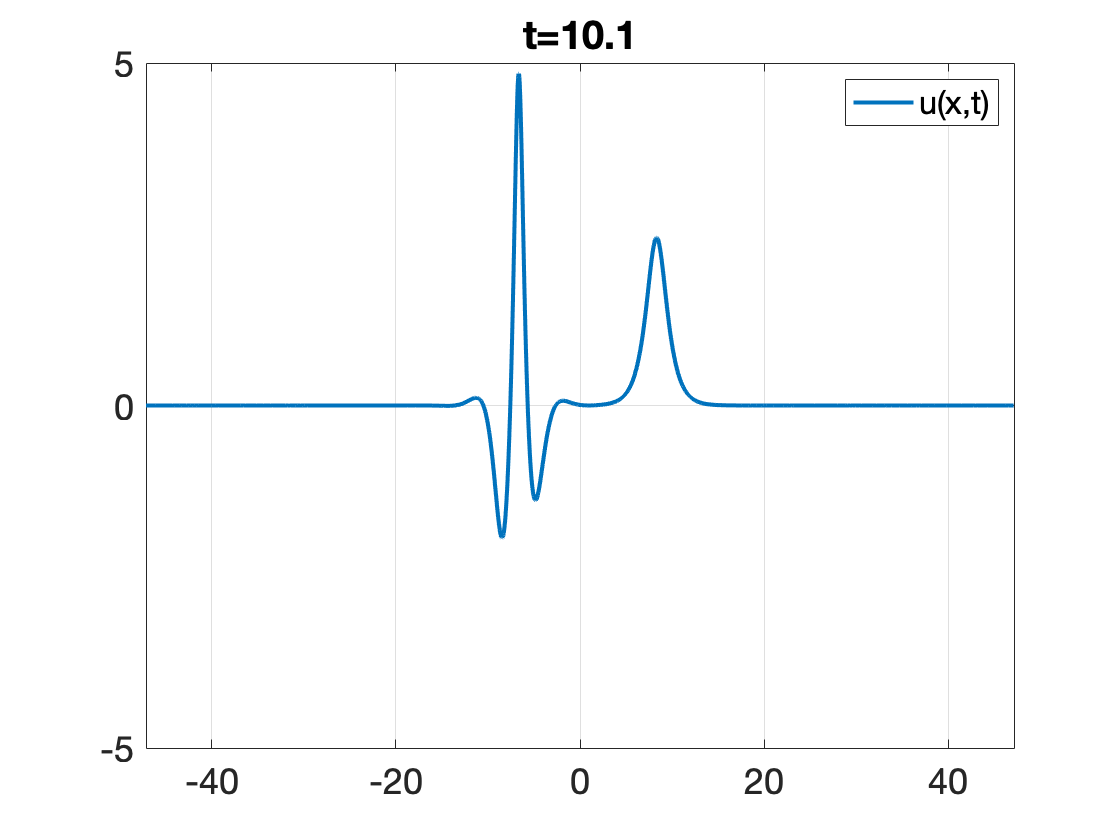}
\includegraphics[width=0.32\textwidth]{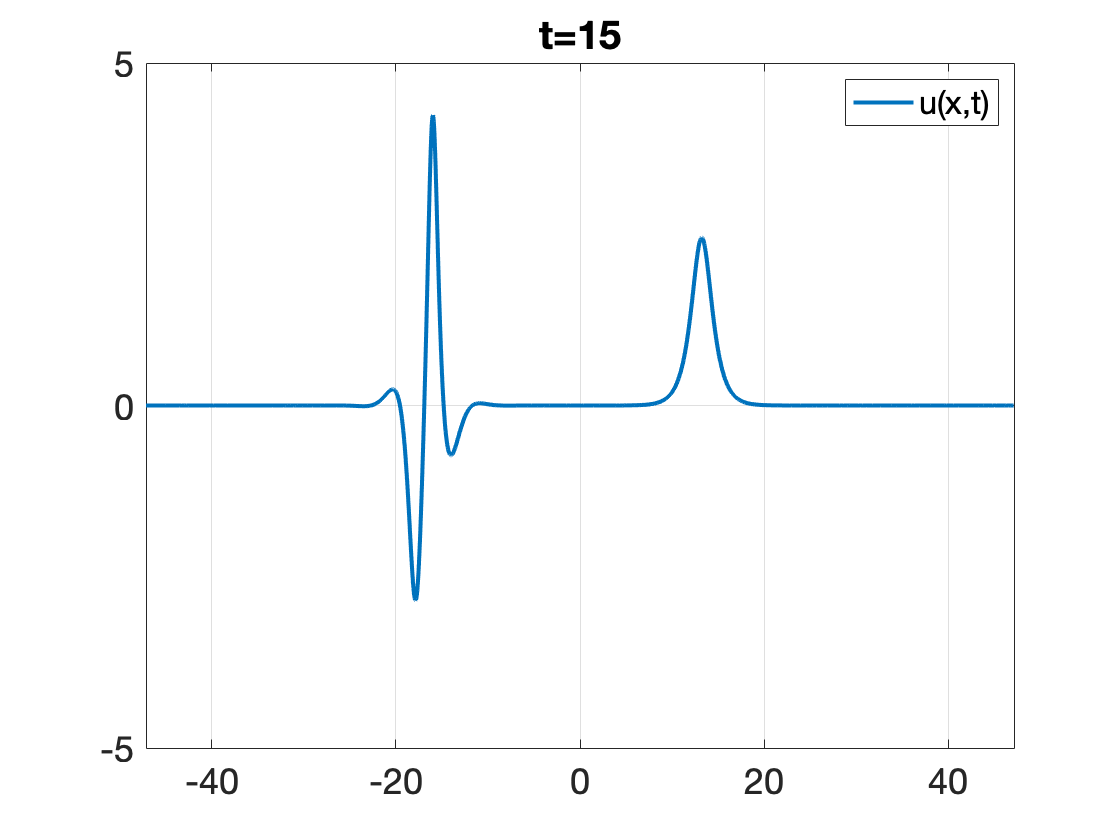}
\caption{\label{BS} Interaction of a soliton and a breather in mKdV: time evolution of $u_0(x) = Q_1(x+5)+B_{1,1}(x-15,0)$ from $t=0$ to $t=15$.}
\end{center}
\end{figure}
\subsection{Breather-Soliton Interaction:}\indent In Figure \ref{BS}, we simulate the interaction between a soliton and a breather. Given that solitons always travel right, we chose $\alpha$ and $\beta$ so that the breather travels left, inducing the interaction of two solutions. Thus, we take the soliton $Q_1$ from \eqref{E:mKdV-S-rescale} and shift it by 5 spaces to the left, and shift the breather $B_{1,1}(x,0)$ by 15 spaces to the right, making the initial condition $$u_0(x)=Q_1(x+5)+B_{1,1}(x-15,0).$$
In the time evolution, we see the collision between the soliton and the breather. From around $t=4.5$ to $t=6$, we see the breather collide with the soliton, distorting both solution profiles, then returning to their original shapes as the breather separates from the left of the soliton and continues traveling left while the soliton continues traveling to the right. During the collision, the height of the breather increases momentarily. After the two solutions separate, both the breather and soliton return to their original size and shape, with no visible radiation from the collision at the numerical resolution used, in agreement with the elastic interactions expected for the integrable mKdV flow. 
\smallskip

{\bf Remark.} We mention that our shown simulations are non-periodic; however, it is possible to run them with a `wrapping-around' effect (i.e., on a periodic domain, which is often done in computations). In that case, the soliton would reappear on the left and the breather on the right, and would interact with each other again. The interactions can be repeated multiple times if the simulation is set to run for a longer time. In the case of multiple interactions, one will still observe that {\it no} radiation or dispersion comes from the repeated collisions and interactions.

\noindent \subsection{Breather-Breather Interaction.} 
\begin{figure}[h!]
\begin{center}
\includegraphics[width=0.32\textwidth]{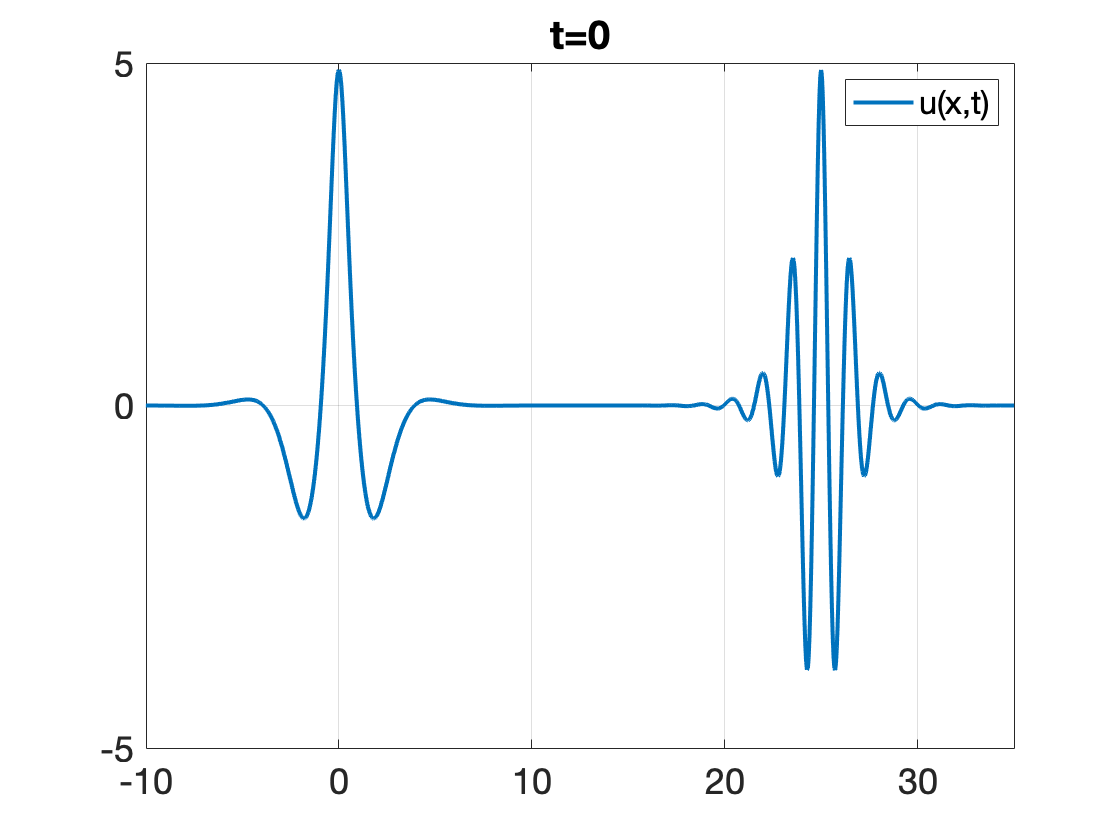}
\includegraphics[width=0.32\textwidth]{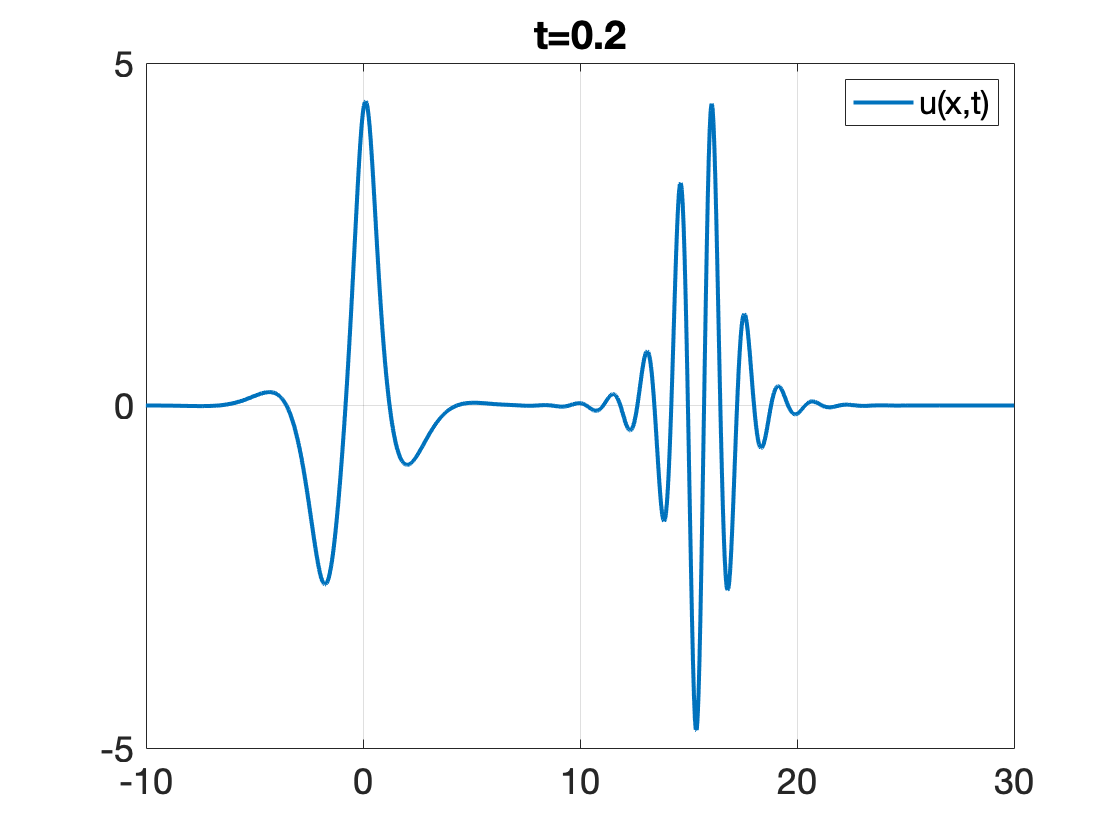}
\includegraphics[width=0.32\textwidth]{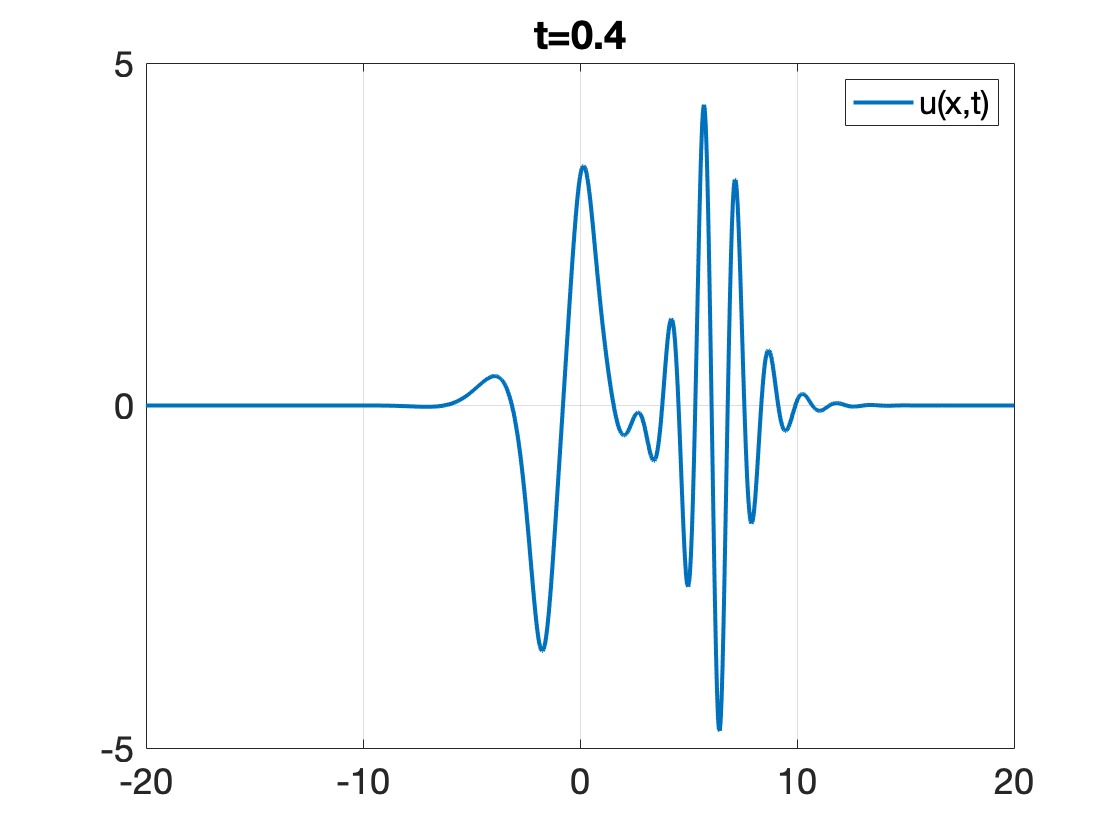}
\includegraphics[width=0.32\textwidth]{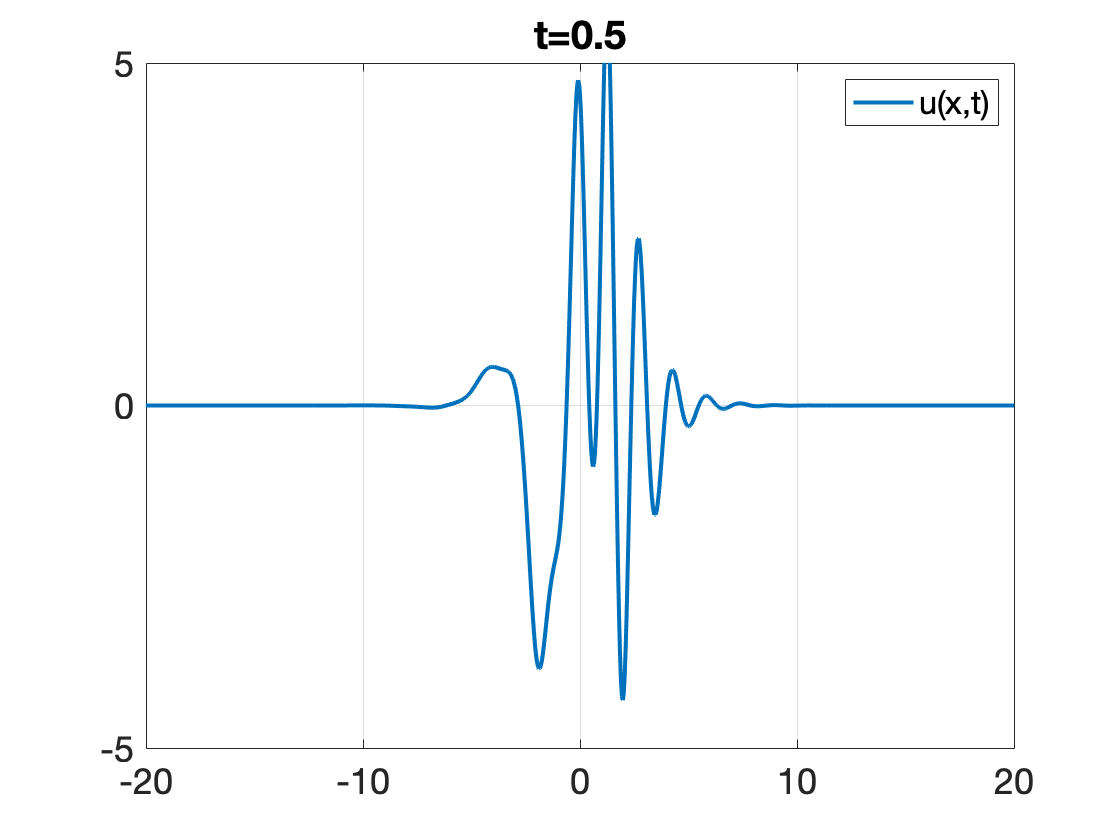}
\includegraphics[width=0.32\textwidth]{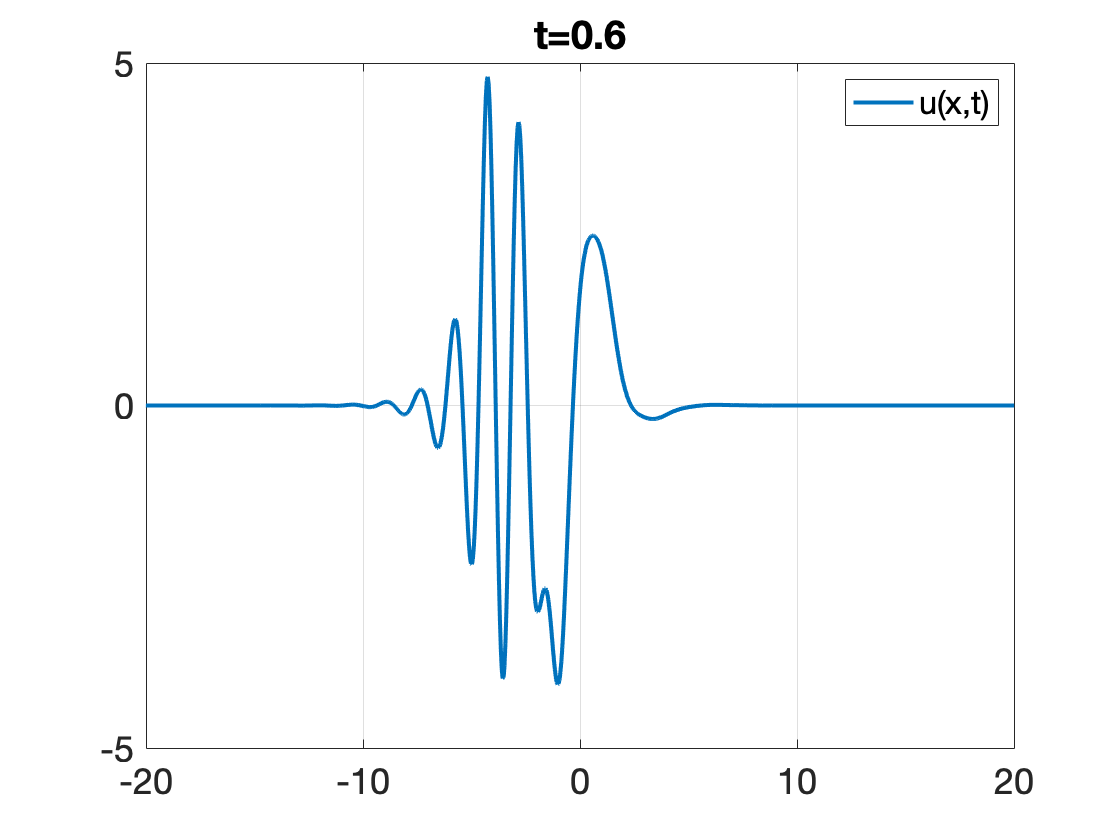}
\includegraphics[width=0.32\textwidth]{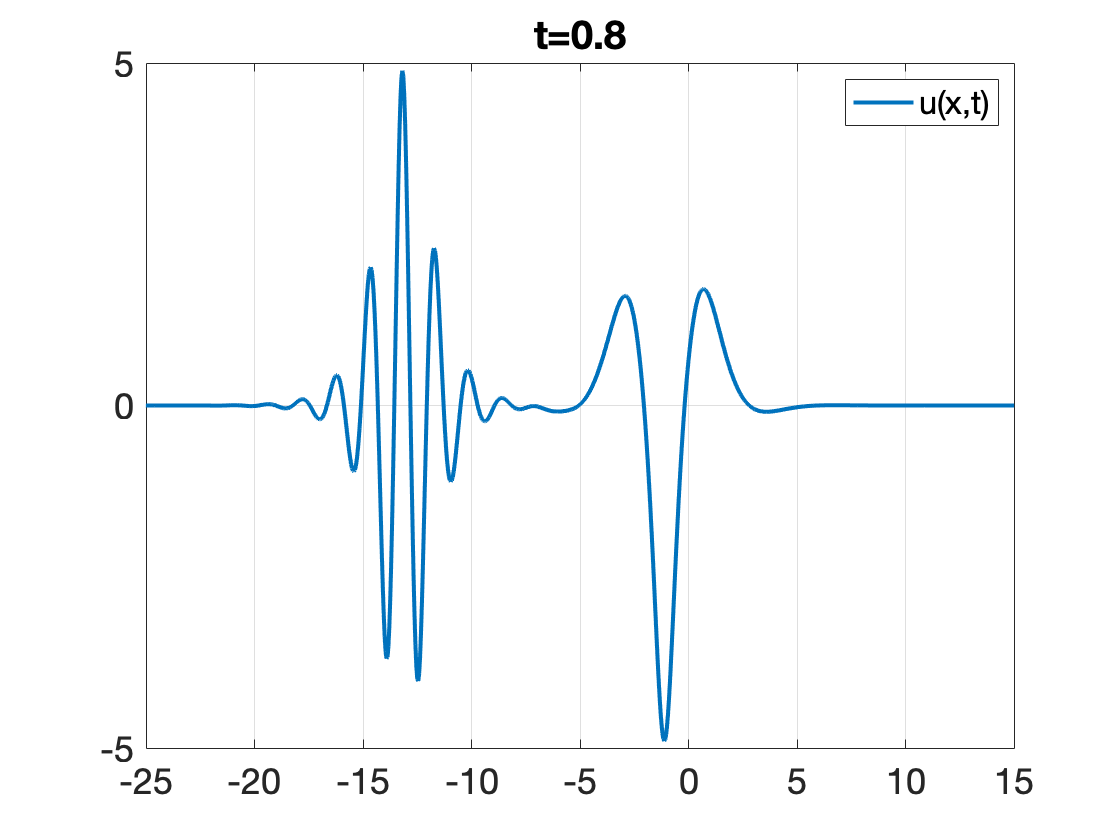}
\caption{\label{BB1} Interaction of two breathers both traveling left in mKdV: time evolution of $u_0(x) = B_{1,1}(x,0)+B_{4,1}(x-25,0)$.}
\end{center}
\end{figure}

\indent Next, we simulate the interaction between two breathers. 
In Figure \ref{BB1}, we have two breathers with different $\alpha$-values but the same $\beta$-value to further illustrate the effect of the parameters on the profile of the initial breather and its behavior. The two breathers, $B_{1,1}$ and $B_{4,1}$, are sufficiently separated in space, the second breather being shifted to the right by 25, 
$$
u_0(x)=B_{1,1}(x,0)+B_{4,1}(x-25,0).
$$ 
The time evolution plots in Figure \ref{BB1} showcase the different speeds of the two breathers due to their difference in $\alpha$, and hence different values for $\delta$ and $\gamma$. Note that both breathers in Figure \ref{BB1} travel to the left. The breather with larger $\alpha=4$, initialized at $x=25$, travels to the left and collides with the breather initialized at $x=0$. The collisions occur around $t=0.5$ and continue up to $t=0.6$. We observe no radiation or dispersion coming from the interaction and the breathers separate. At $t=0.8$, $B_{4,1}(x,t)$ is located around $x=-14$, while $B_{1,1}(x,t)$ is located around $x=-1$. Since the $\beta$ values of both breathers are equal to $1$, we observe that both breathers have the same width and amplitude. Then, the different $\alpha$ values produce a different number of internal oscillations of the breathers' profile and the speed at which both breathers travel left. 

\begin{figure}[h!]
\begin{center}
\includegraphics[width=0.32\textwidth]{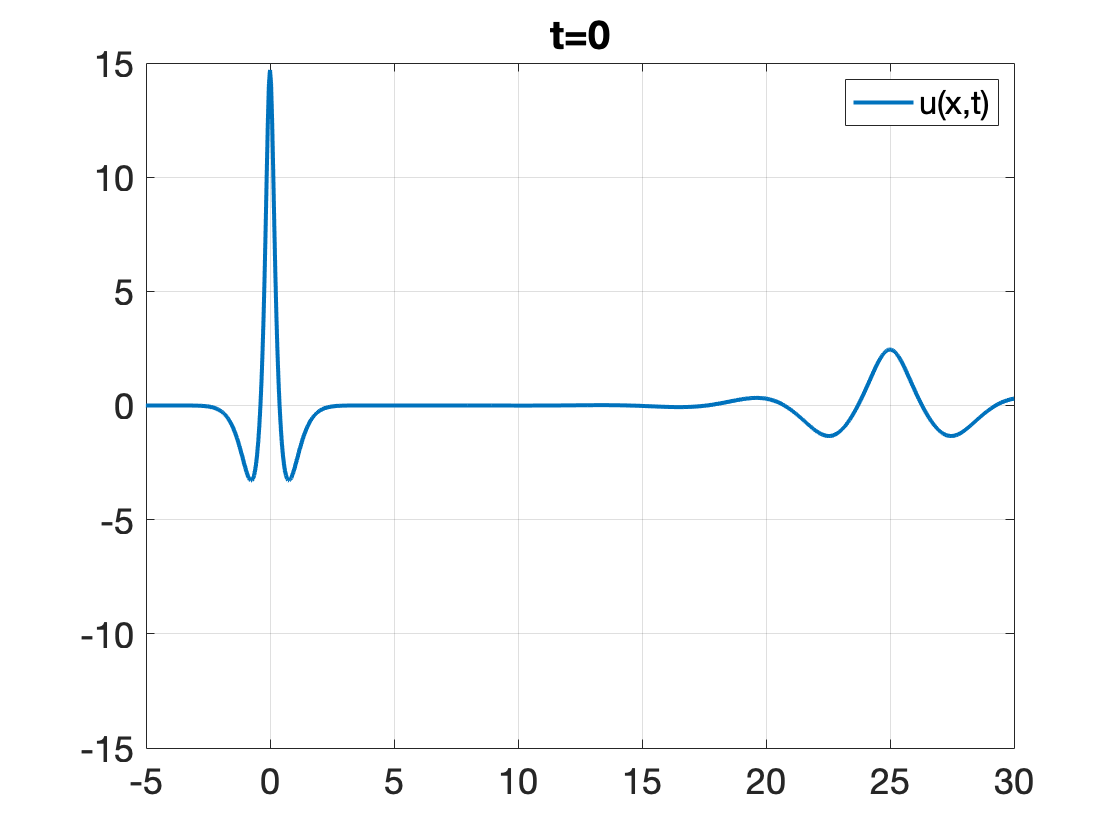}
\includegraphics[width=0.32\textwidth]{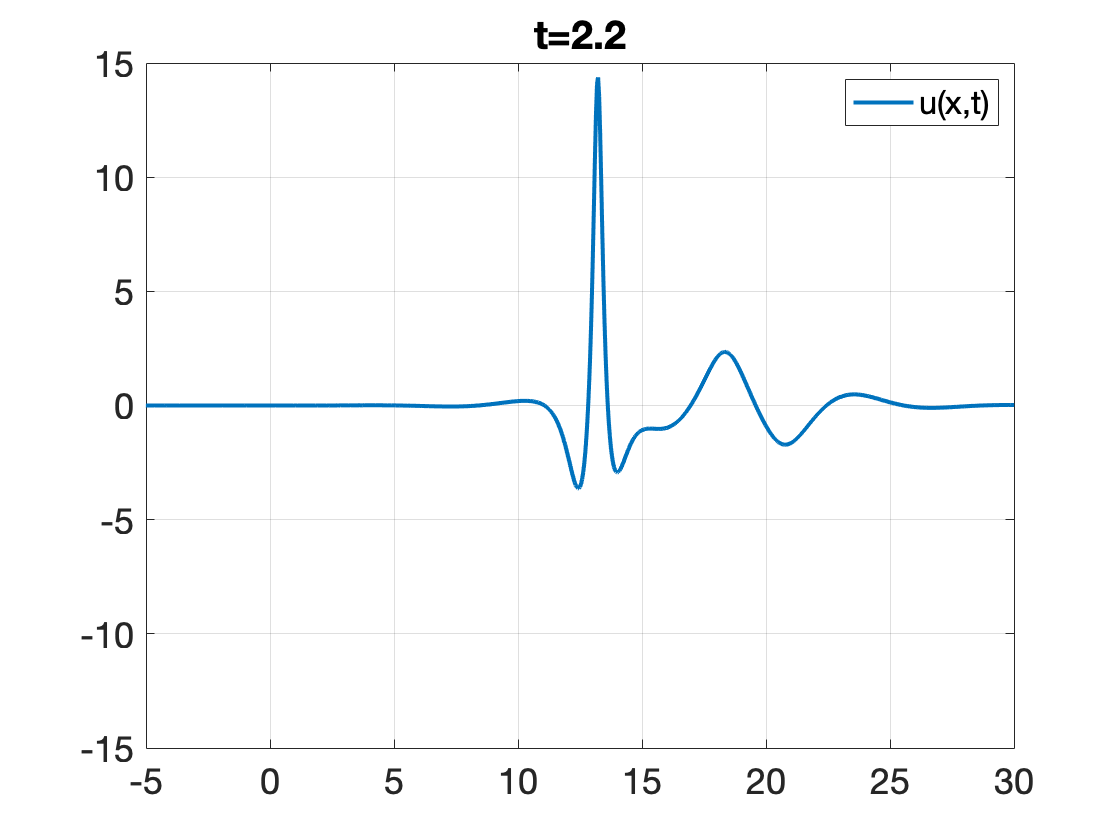}
\includegraphics[width=0.32\textwidth]{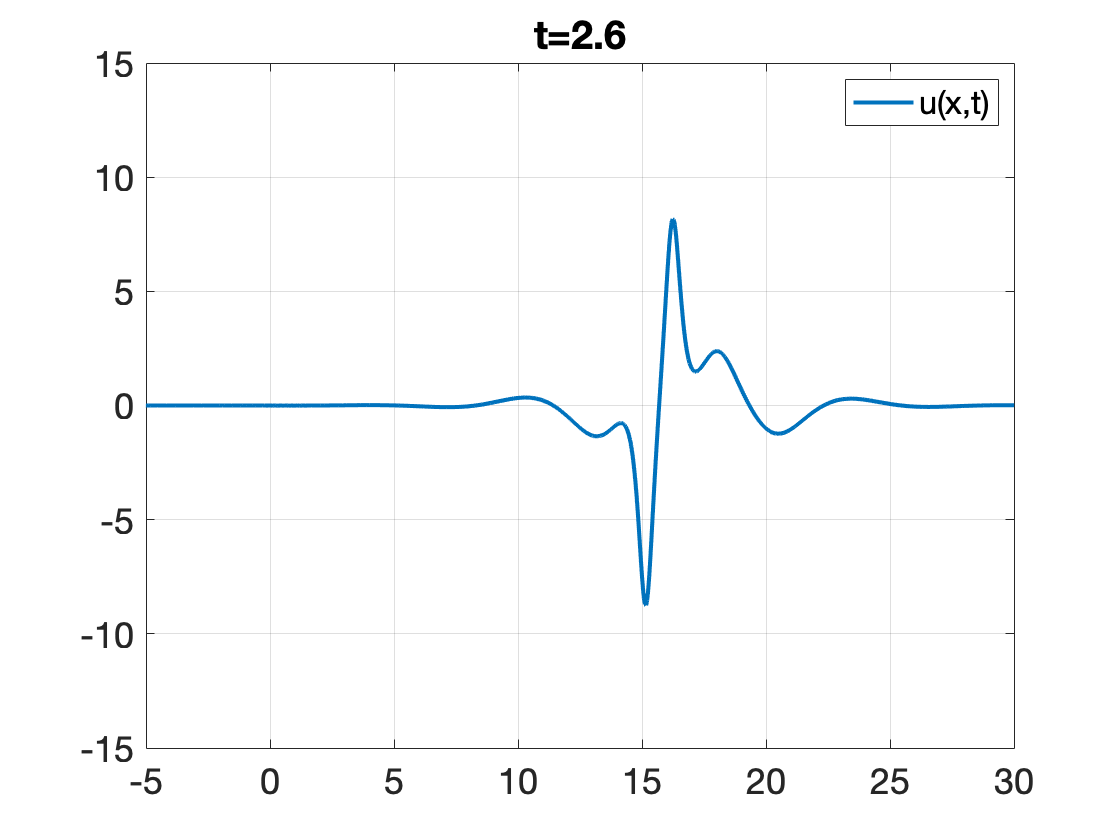}
\includegraphics[width=0.32\textwidth]{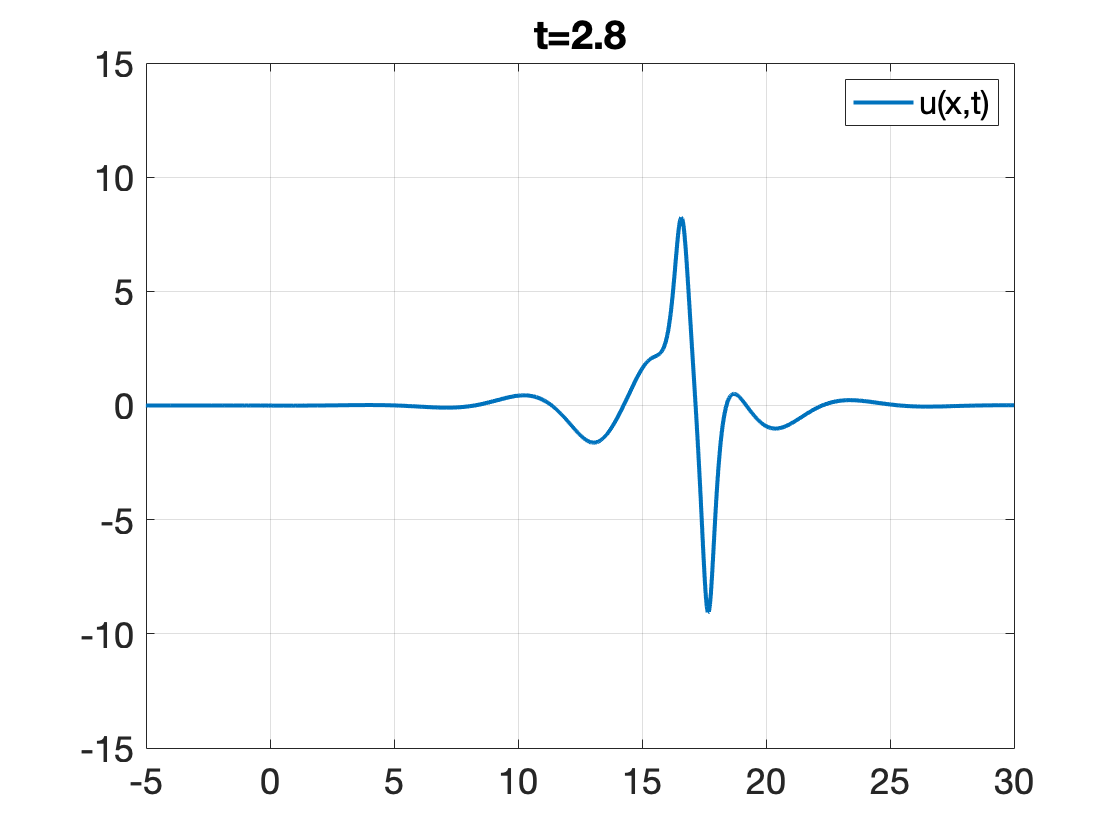}
\includegraphics[width=0.32\textwidth]{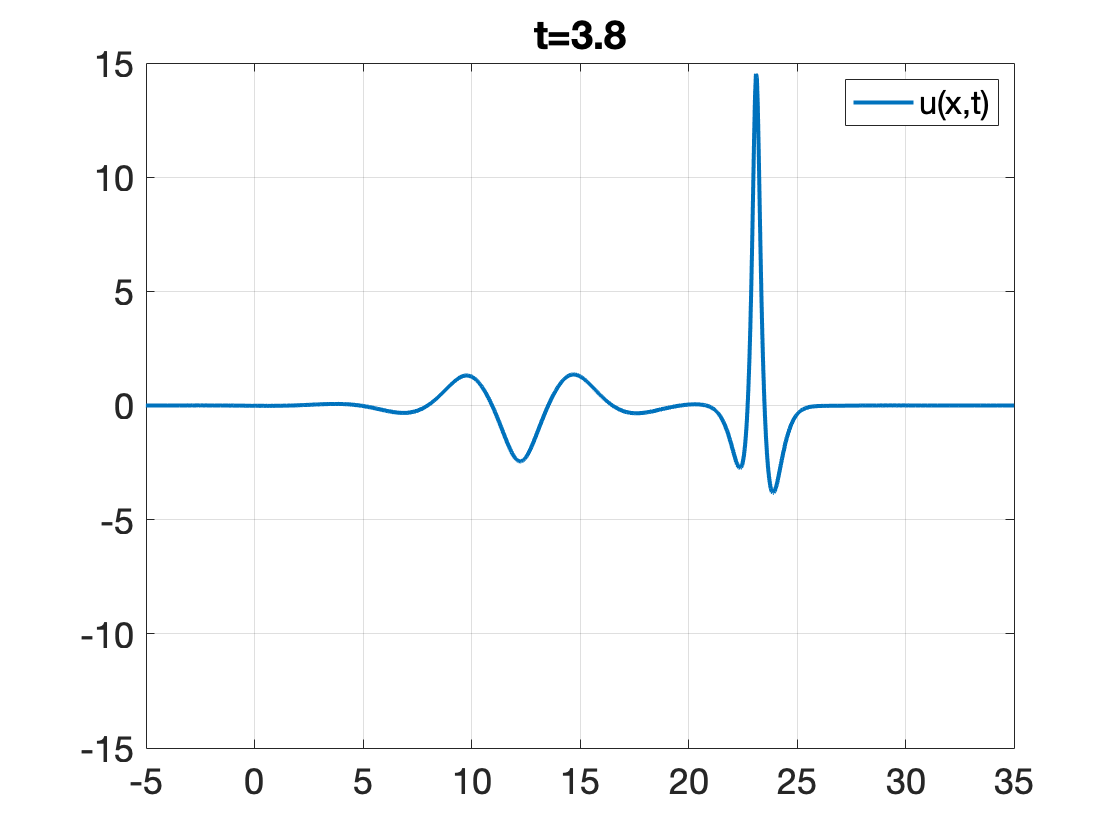}
\includegraphics[width=0.32\textwidth]{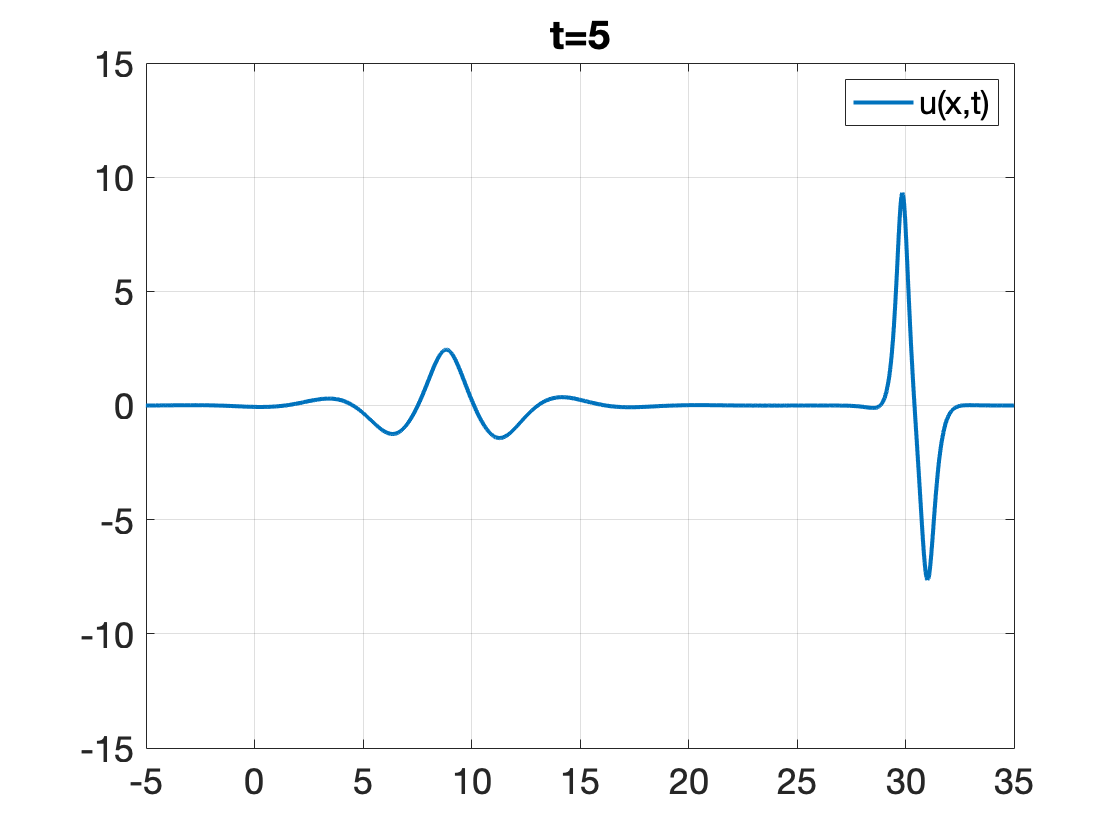}
\caption{\label{BB2} Interaction of two breathers traveling in different directions in mKdV: time evolution of $u_0(x) = B_{1,3}(x,0)+B_{1,0.5}(x-25,0)$.}
\end{center}
\end{figure}

In Figure \ref{BB2}, we simulate the interaction of two breathers of different $\beta$ values and the same $\alpha$ values. We take $\alpha=1$ and $\beta=3$ for the first breather, and $\alpha=1$ and $\beta=0.5$ for the second breather, with the second breather shifted to the right by 25; thus, producing the initial condition 
$$u_0(x)=B_{1,3}(x,0)+B_{1,0.5}(x-25,0).$$
The two breathers are initialized apart to observe the behaviors of both solutions before and after the interaction. Figure \ref{BB2} shows that the breather initialized at $x=0$ travels to the right while the second breather initialized at $x=25$ travels to the left. For the chosen values of $\beta=3$ and $\alpha=1$, we get $\gamma<0$, which means that the first breather at $x=0$ for $t=0$ travels to the right, while $\beta=0.5$ and $\alpha=1$ yield $\gamma>0$, so the second breather at $x=25$ for $t=0$ travels to the left.\smallskip
The breathers collide from around $t=2.6$ to $t=2.8$, momentarily distorting the shapes of both breathers' profiles. Then at $t=2.8$ one can observe that the breather with $\beta=3$ emerges from the right-hand side of the breather with $\beta=0.5$. By $t=3.8$, both breathers have sufficiently separated, returned to their initialized shape and size, and no radiation formed as a result of the interaction.\smallskip

In summary, Figures \ref{BS}, \ref{BB1}, and \ref{BB2} provide numerical evidence that breathers and solitons maintain their original shapes before and after interactions (their speeds and internal oscillations are also preserved after interaction) in the mKdV equation. We have simulated various other configurations and this holds for any interaction between breathers of different $\alpha$ and $\beta$ values as well as between breathers and different solitons. 
In the next subsection, we demonstrate that this holds true for {\it multiple} breather interactions, traveling at different speeds and with different internal oscillations. We also note that no radiation occurs in these examples.

\subsection{Triple Breathers Interaction} 
\quad Here, we show an additional example involving a three-breather interaction, complementing the previous cases. 
\begin{figure}[h!]
\begin{center}
\includegraphics[width=0.28\textwidth]{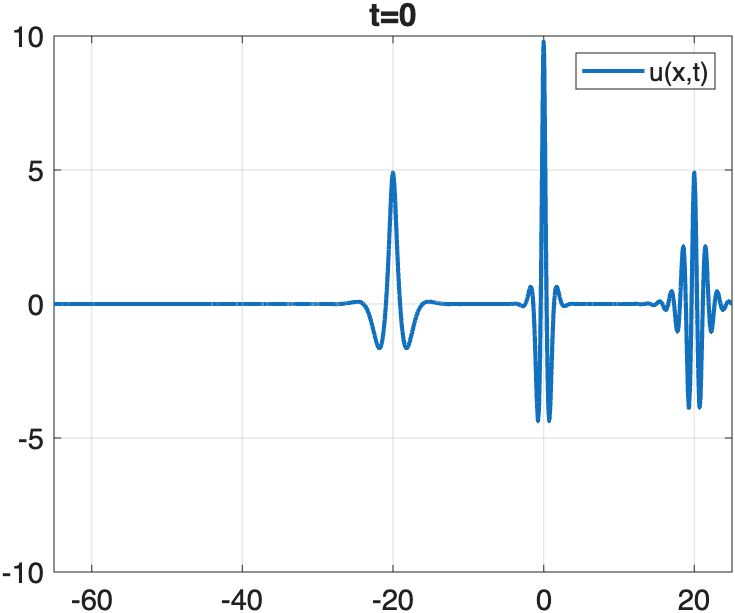} \hspace{20pt}
\includegraphics[width=0.28\textwidth]{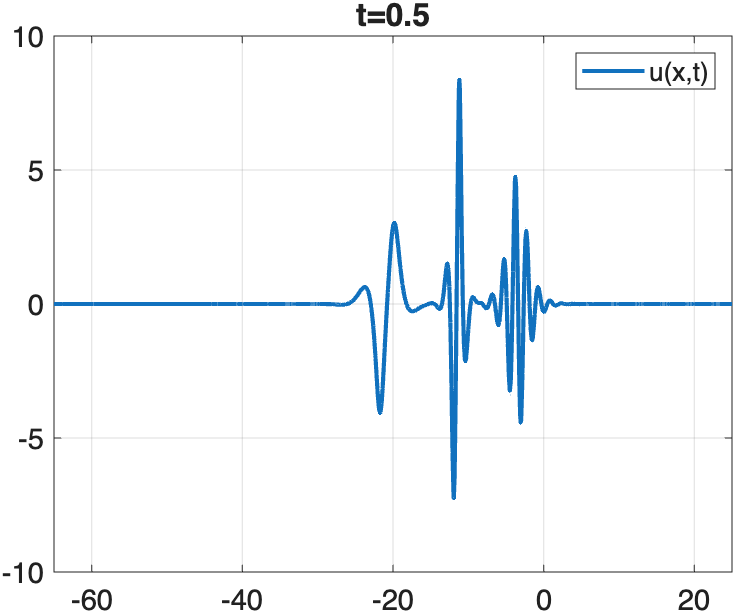} \hspace{20pt}
\includegraphics[width=0.28\textwidth]{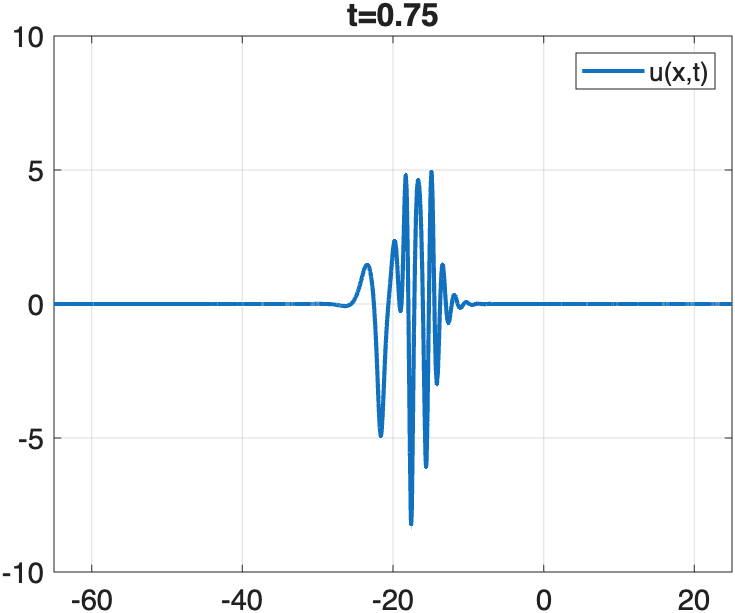} \\
\vspace{10pt}
\includegraphics[width=0.28\textwidth]{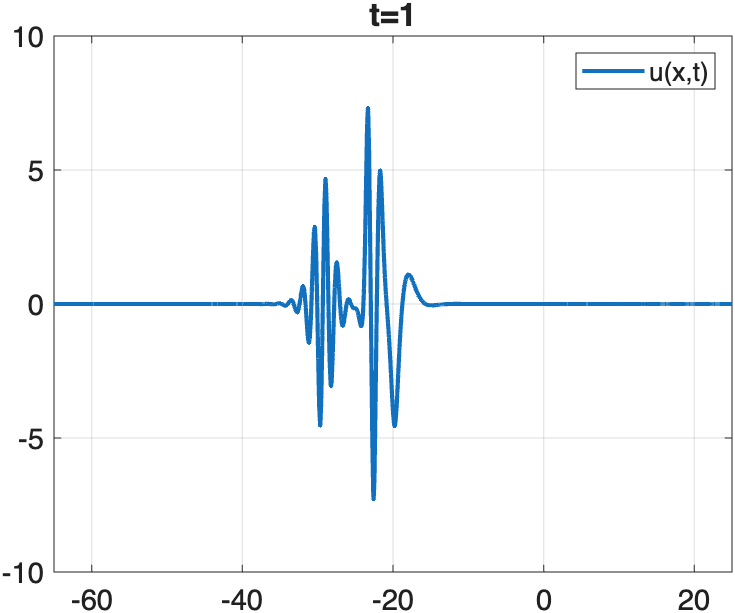} \hspace{20pt}
\includegraphics[width=0.28\textwidth]{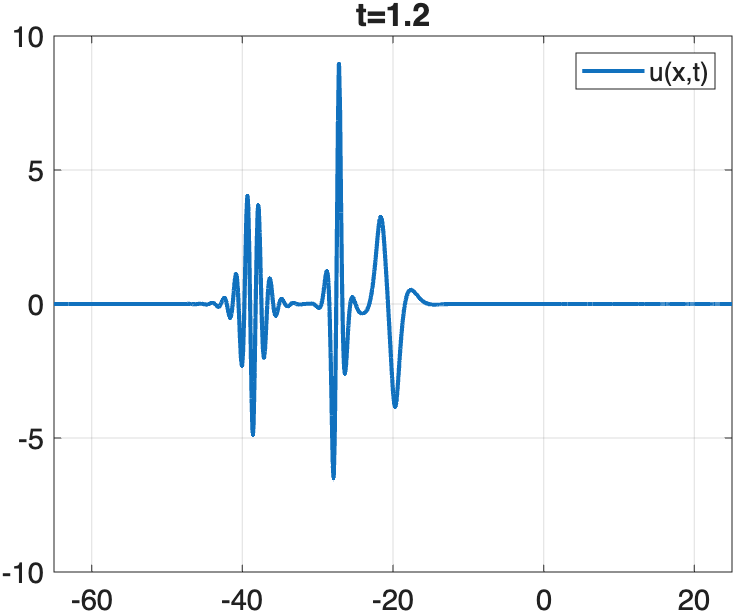} \hspace{20pt}
\includegraphics[width=0.28\textwidth]{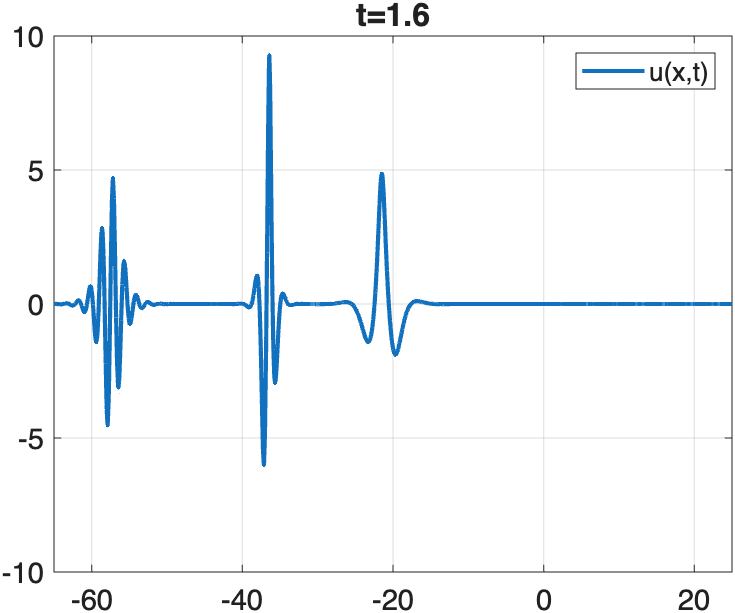} %\\
\caption{\label{BBB} Interaction of three breathers (all traveling in the left direction) in mKdV: time evolution of $u_0(x) = B_{1,1}(x+20,0)+B_{3,2}(x,0)+B_{4,1}(x-20,0)$.}
\end{center}
\end{figure}
In Figure \ref{BBB}, we simulate the interaction of three different breathers with different $\alpha$ and $\beta$ values shifted away from each other at $t=0$. Thus, our initial condition is 
$$
u_0(x) = B_{1,1}(x+20,0)+B_{3,2}(x,0)+B_{4,1}(x-20,0).
$$ 
Note that all three breathers travel to the left with different speeds due to the difference in their parameters. 

The leftmost breather initialized at $x=-20$ is the slowest moving left, while the rightmost breather initialized at $x=20$ is the fastest moving left. By $t=0.5$, notice that the center and rightmost breathers travel left and collide with the leftmost breather by $t=0.75$. At the point of collision, all three breathers interact and their profiles are distorted. By $t=1$, the former rightmost breather, with $\alpha=4$ and $\beta=1$, is separating from the left of the collision, followed by the former center breather, with $\alpha=3$ and $\beta=2$. Again, there is no radiation coming out from the interaction of the three breathers. By $t=1.2$, two of the three faster left-traveling breathers have separated from the slowest left-traveling breather. Lastly, at $t=1.6$, the breathers are separated from each other with profiles that match their original size and shape.

\subsection{Triple Breathers-Soliton Interaction}
\begin{figure}[h!]
\begin{center}
\includegraphics[width=0.28\textwidth]{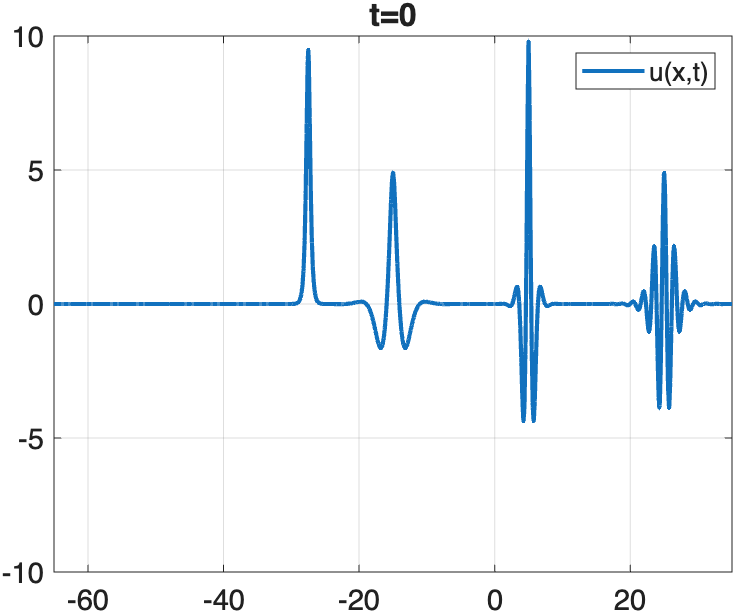} \hspace{20pt}
\includegraphics[width=0.28\textwidth]{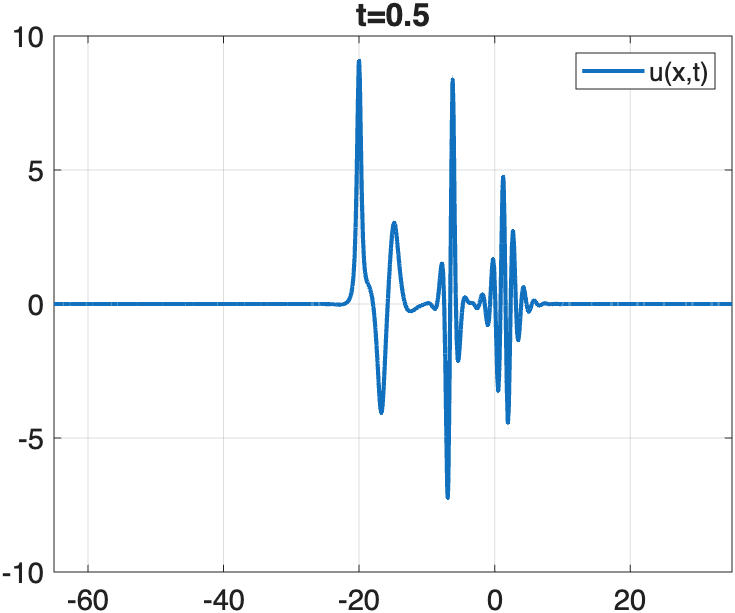} \hspace{20pt}
\includegraphics[width=0.28\textwidth]{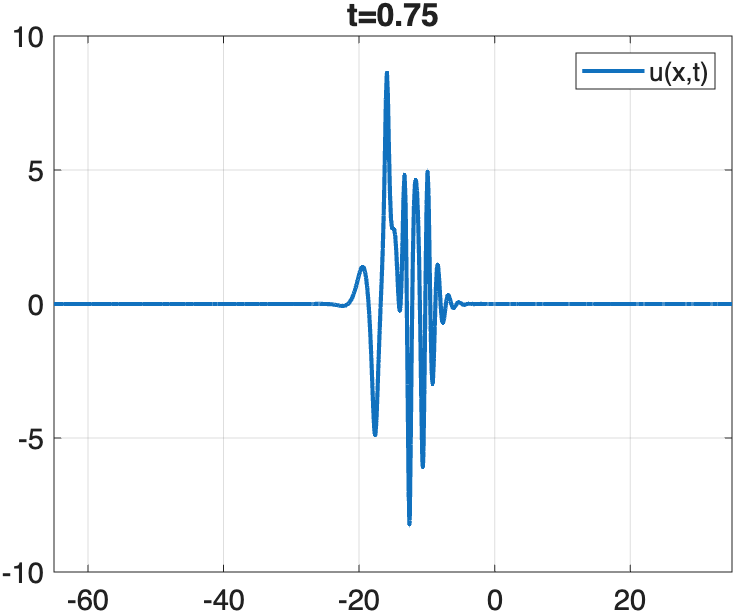} \\
\vspace{10pt}
\includegraphics[width=0.28\textwidth]{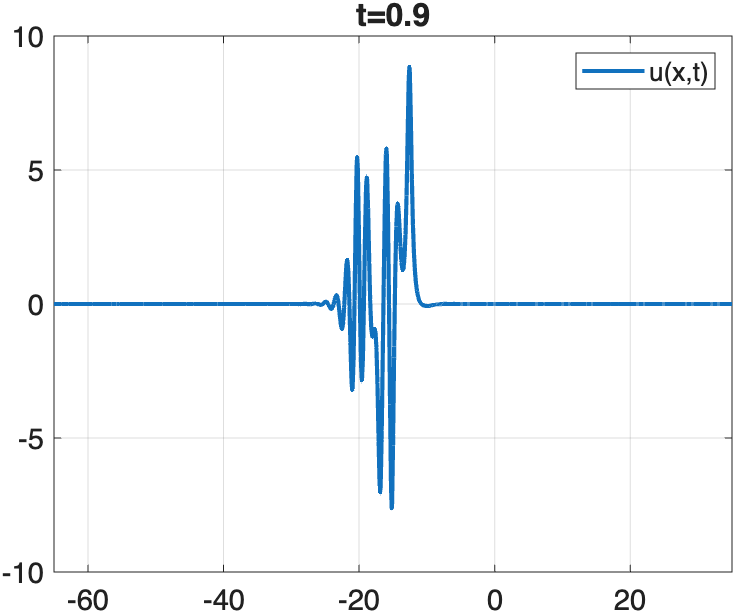} \hspace{20pt}
\includegraphics[width=0.28\textwidth]{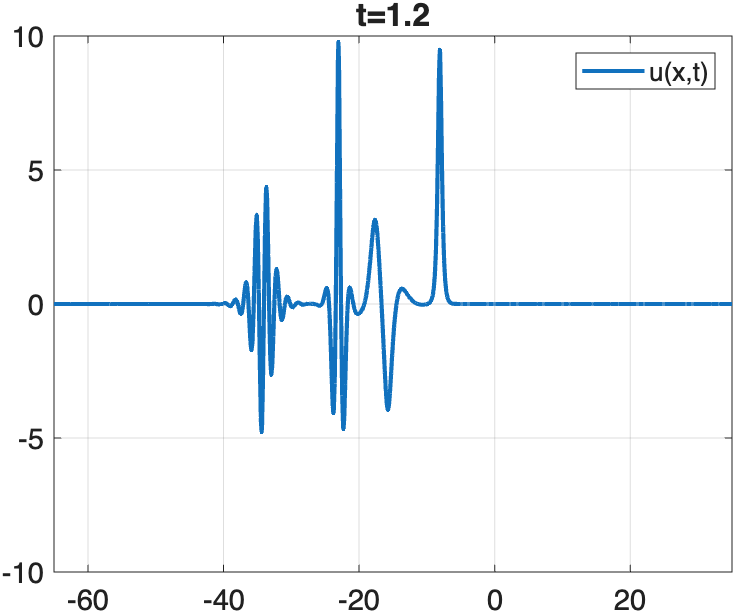} \hspace{20pt}
\includegraphics[width=0.28\textwidth]{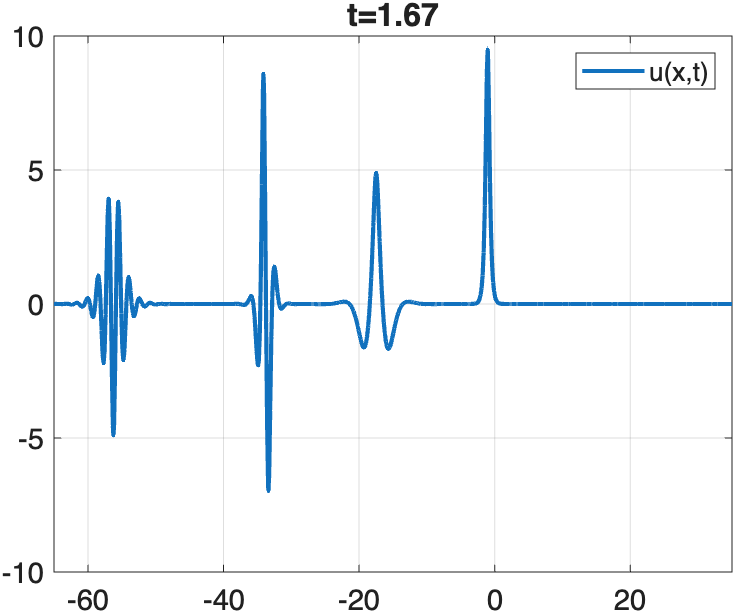}
\caption{\label{BBBS} Interaction of three breathers (all traveling in the left direction) and one soliton (traveling to the right) in mKdV: time evolution of $u_0(x) = Q_{15}(x+27.5)+B_{1,1}(x+15,0)+B_{3,2}(x-5,0)+B_{4,1}(x-25,0)$.}
\end{center}
\end{figure}

The last example of this section concerns the interaction of three breathers and a single soliton. In Figure \ref{BBBS}, we have the initial data of $$u_0(x) = Q_{15}(x+27.5)+B_{1,1}(x+15,0)+B_{3,2}(x-5,0)+B_{4,1}(x-25,0).$$ 
Since solitons are single-wave solutions traveling to the right, we initialize the soliton in the leftmost position at $x=-27.5$ with $c=15$ so that it travels fast enough to collide with the slowest left-traveling breather initialized at $x=-15$ at the same moment as the faster left-traveling breathers, initialized at $x=5$ and $x=25$, creating a simultaneous interaction of all solutions. \smallskip

In Figure \ref{BBBS}, the time evolution shows that all four solutions collide from around $x=-16$ and to the left, starting at a time around $t=0.75$. The separate components start to reappear (break off) from the mutual interaction around $t=0.9$: the soliton emerges to the right of the collision, while the fastest traveling breather emerges to the left. Note that in the subplots for $t=0.75$ and $t=0.9$, we observe no radiation or dispersion from the collision of these different coherent structures; later, when we discuss and measure radiation, this interaction example has none. By $t=1.2$, all the parts are practically separated from each other. Lastly, at $t=1.67$, the triple breather and single soliton solutions are adequately spaced apart, and one can observe that each solution profile has its original size and shape (in particular, as discussed above, we match to confirm the initial shapes; more about matching is discussed in the next section).

\section{Perturbations of mKdV breathers}\label{perturbations}

\indent In this section, we consider perturbations of breathers by multiplying the initial shape of a breather by a constant $A$, and, thus, modifying its initial {\it amplitude} as follows: 
\begin{equation}\label{E:AB}
u_0(x) = A \, B_{\alpha,\beta}(x,0).
\end{equation}
This allows us to study numerically the stability of breathers (including asymptotic behavior and stability), while also observing the radiation shedding to the left as it would be in the case of soliton perturbation. We simulate perturbations of the breather $B_{\alpha, \beta}$ \eqref{mKdV-B} separately for $A>1$ and $A<1$. To identify the breather that evolves and stabilizes from this perturbation (in particular, the parameters $\alpha, \beta$ may not be as close to the initial ones), we develop a technique that identifies breathers and allows us to match the numerical profile with the exact profile \eqref{mKdV-B}.  

\subsection{The computational approach}\label{S:CompTools}

Our method extracts the breather parameters $\alpha$, $\beta$, $\gamma$, and $\delta$ from numerical simulations and tracks their evolution in time to study asymptotic behavior.
Using the corresponding `final parameters', we identify the numerically computed breather and match it %with the known 
to the explicit breather \eqref{mKdV-B}. This provides numerical 
evidence for the asymptotic stability of a perturbed breather and, more broadly, evidence for
the soliton-breather resolution conjecture.
\smallskip

{\it Parameter tracking.}
More precisely, at each time step we advance the numerical solution and extract the breather parameters via a two-stage procedure:

(i) we use the known expressions for the mass \eqref{mBmass} and the energy \eqref{mBenergy} to obtain 
$\beta$ and $\gamma$ at each numerical time $t$, thereby producing $\beta(t)$ and $\gamma(t)$; 

(ii) we then solve for $\alpha$ and $\delta$ from \eqref{dg} at each time step, yielding $\alpha(t)$ and $\delta(t)$. Hence, we track all four parameters in time and 
plot their evolution 
(see, e.g., Figures \ref{abdg09} and \ref{1.1abdg}). 
\smallskip

{\it Notation.}
Notation-wise, the four parameters at the initial time $t=0$ are denoted by the corresponding Greek letters, while the corresponding parameters at the final computational time $T$ are denoted using prime notation; %, i.e., 
$$
\alpha(0)=\alpha \quad \mbox{and} \quad \alpha(T)=\alpha^\prime,
$$
When discussing each parameter as time-dependent, we write it as a function, e.g., $\alpha(t)$. 
Ideally (or in analytical proofs), $T = \infty$, but for numerical simulations, it is sufficient to have a finite value, for instance, at a time when the parameters are stabilized (or asymptotically level out). Therefore, we show that 
$$\alpha(t) \to \alpha^\prime \quad \mbox{as} \quad t \to T, 
$$ 
which is a numerical approximation of $t \to \infty$. 
\smallskip

{\it Initial jump of parameters.}
Another important point we would like to emphasize is that, when considering the perturbed amplitude breather in the initial condition, i.e., $u_0(x) = A\, B_{\alpha, \beta}(x,0)$, while the breather is specified by parameters $\alpha$ and $\beta$, the values extracted from the simulation may jump significantly at the very beginning of the simulation, since the scaling by $A$ changes the mass and energy.
Therefore, if we use the mass and energy of the initial condition to determine $\alpha(t)$ and $\beta(t)$ at $t=0$ (e.g., $M[u_0] \equiv M[A B_{\alpha,\beta}] = M[B_{\alpha^*, \beta^*}]$ 
for some $\alpha^*$ and $\beta^*$), then these starred values $\alpha^*, \beta^*$ can significantly differ from the initial values $\alpha,\, \beta$, 
since the perturbation immediately changes them % parameter values 
and these extracted values are typically closer to the values that the solution asymptotically approaches; this can be seen in Figure \ref{abdg09} for $\alpha(t)$. Later (at a `final' computational time, or another time of interest), these tracked parameters are used to {\it perform profile matching} of the evolved solution %initial data 
to an exact breather given via the explicit 
formula \eqref{mKdVBS}, see examples in Figures \ref{0.9m} and \ref{1.1m}.
\smallskip

{\it Localization of breathers.}
Note that the mass and energy quantities of the initial data are calculated at $t=0$ on the entire domain $[-L,L]$. For accurate breather identification, % however, 
in the case of perturbed breathers, we compute these quantities on a {\it localized domain} determined by the predicted breather's shape and location. 
To do so, we compute the values of local mass and energy, and thus identify the parameters of the `final state' breather more accurately. Furthermore, this localization inherently allows us to quantify the mass and energy of the shedding radiation (in other words, we use the mechanism of decoupling of coherent states from radiation) throughout our simulations; see examples in Figures \ref{0.9m} and \ref{1.1rm}. In Figure \ref{abdg09} we show an example %is provided 
comparing the difference in a breather's parameters when the mass and energy are calculated on a global versus a local domain. Reasoning note: the localized-window selection and the decoupling of coherent structures from radiation are numerical procedures here; the paper should state the window-selection rule when reproducibility is needed. 
\smallskip

We now discuss different perturbations of mKdV breathers.

\subsection{Case \texorpdfstring{$A < 1$}{A < 1}:}
\begin{figure}[h!]
\begin{center}
\includegraphics[width=0.32\textwidth]{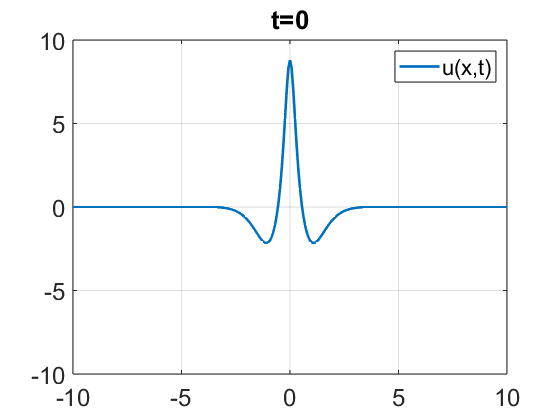}
\includegraphics[width=0.32\textwidth]{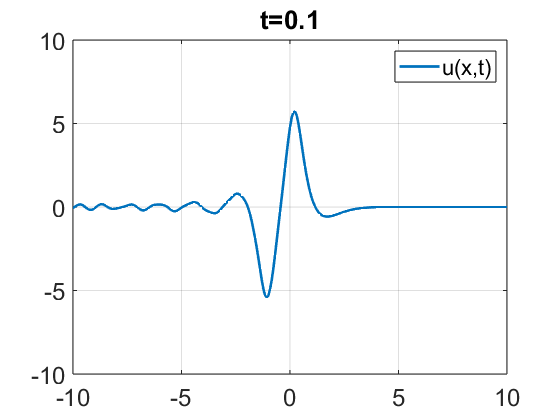}
\includegraphics[width=0.32\textwidth]{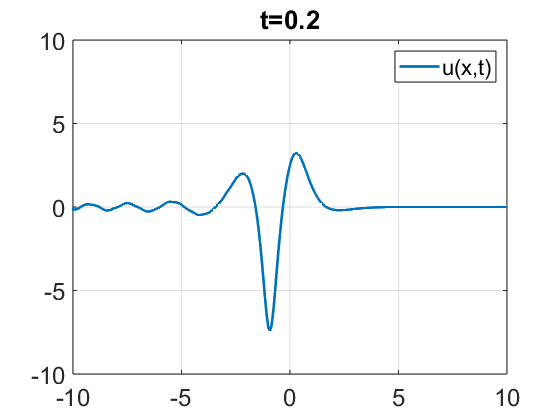}
\includegraphics[width=0.32\textwidth]{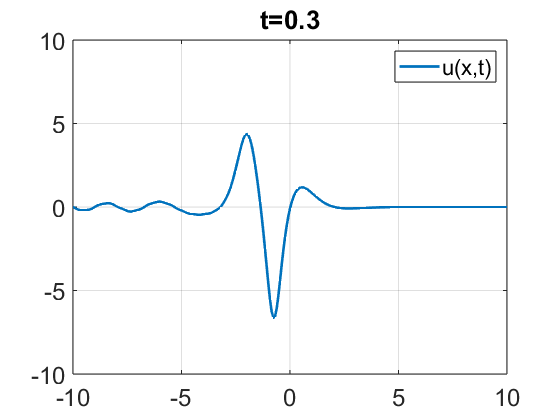}
\includegraphics[width=0.32\textwidth]{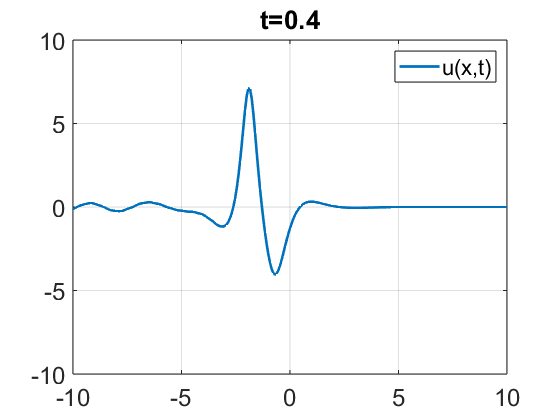}
\includegraphics[width=0.32\textwidth]{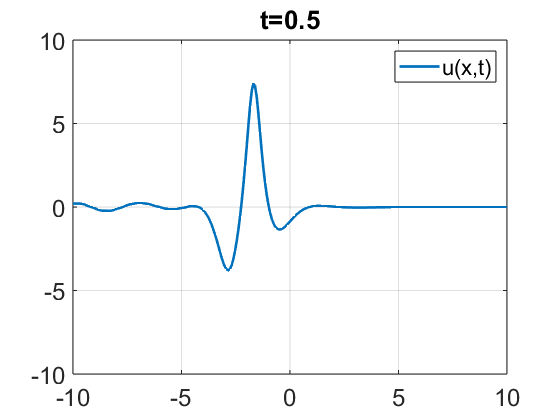}
\caption{\label{0.9B} Time evolution of $u_0(x) = 0.9B_{1,2}(x,0)$.}
\end{center}
\end{figure}

\begin{figure}[h!]
%\begin{centering}
\raisebox{3mm}
{\includegraphics[width=0.33\textwidth]{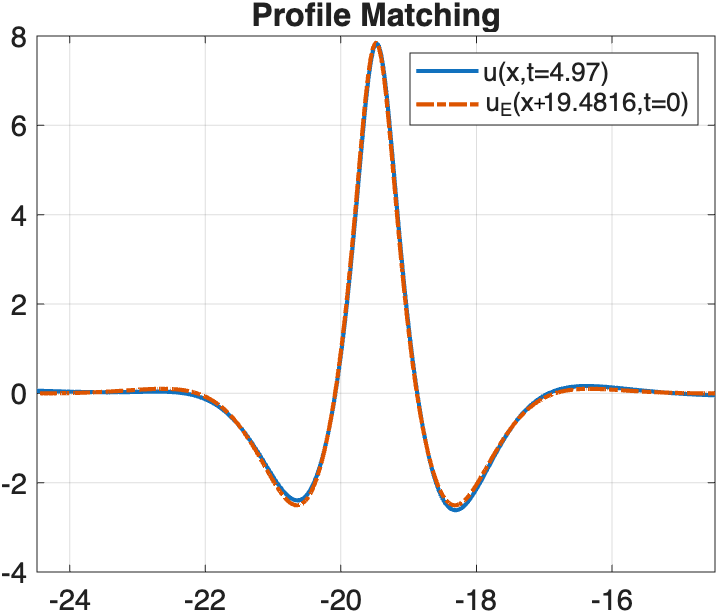}}
{\includegraphics[width=0.31\textwidth,height=0.3\textwidth]{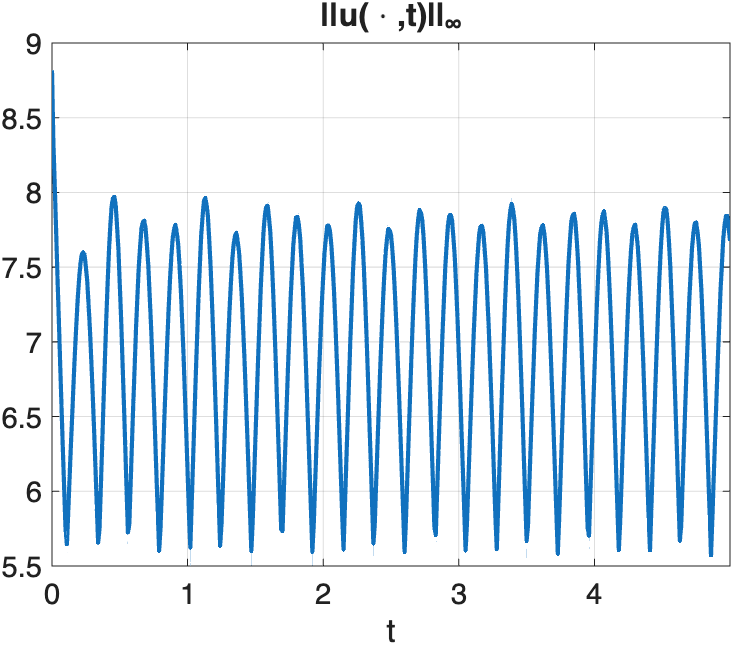}}
\includegraphics[width=0.33\textwidth,height=0.3\textwidth]{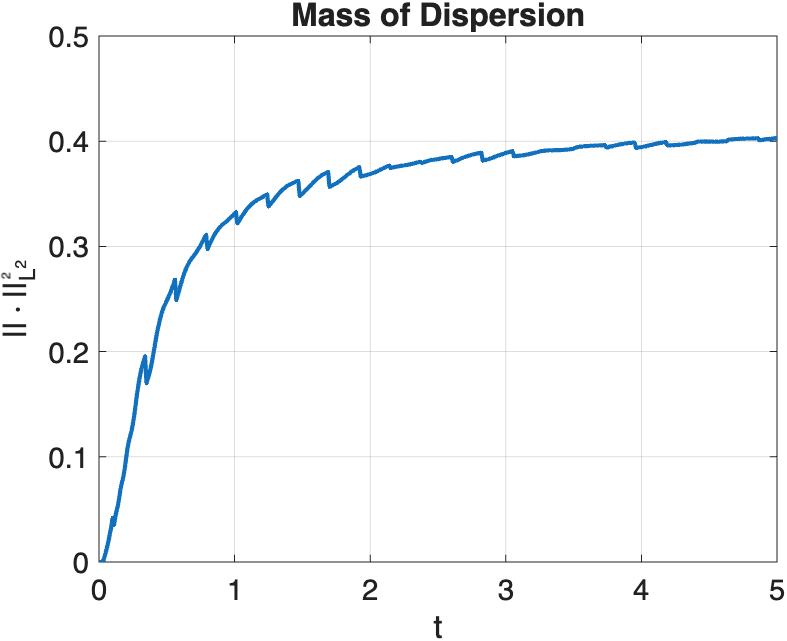}
\caption{\label{0.9m} {\small Time evolution $u(x,t)$ of the initial condition $u_0(x) = 0.9B_{1,2}(x,0)$: matching at $t=4.97$ with an exact breather $u_E(x)=B_{1.4703,1.6032}(x+19.4816,0)$ (left); $L^{\infty}$ norm %for $u_0 = 0.9B_{1,2}(x,0)$ 
up to $t=5$ (middle); quantifying the amount of radiation via the $L^2$ norm to $t=5$ (right). }}
\end{figure}

We first consider amplitudes with $A<1$. The computational parameters in this section are $L=100\pi$, $N=2^{15}$, and $dt=0.0025$. The time evolution $u(x,t)$ of $$u_0(x)=0.9B_{1,2}(x,0)$$ is shown in Figure \ref{0.9B}. The perturbed breather starts to shed some radiation to the left and then stabilizes with a height of approximately 7.75. Observe that between $t=0.1$ and $t=0.4$, the radiation is dispersing off the breather to the left as the breather also slowly moves \textit{to the left}. This behavior will result in an $L^{\infty}$ plot with internal and external oscillations, meaning that the peak amplitudes will alternate in heights, having a wave-like pattern. The $L^{\infty}$ norm for this example is the middle subplot of Figure \ref{0.9m}. % (top-right subplot). 

\begin{figure}[h!]
\begin{centering}
\includegraphics[width=0.295\textwidth]{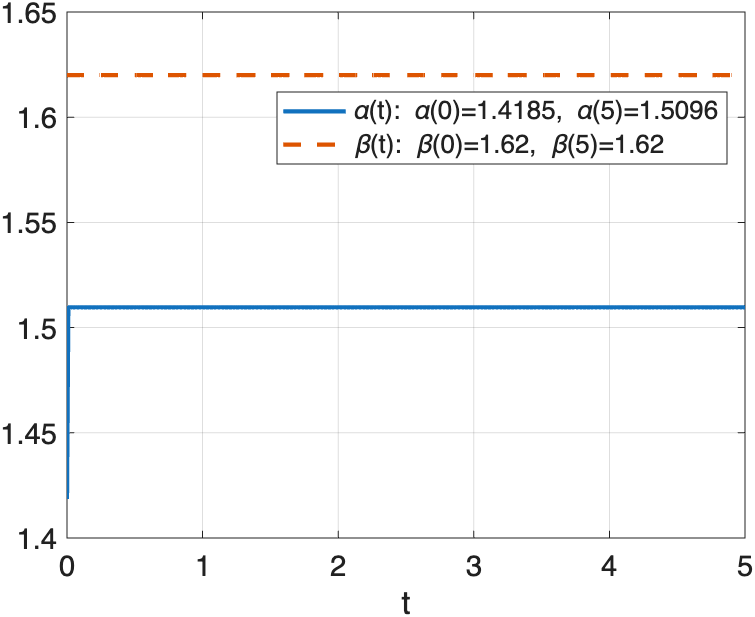} \hspace{20pt}
\includegraphics[width=0.285\textwidth]{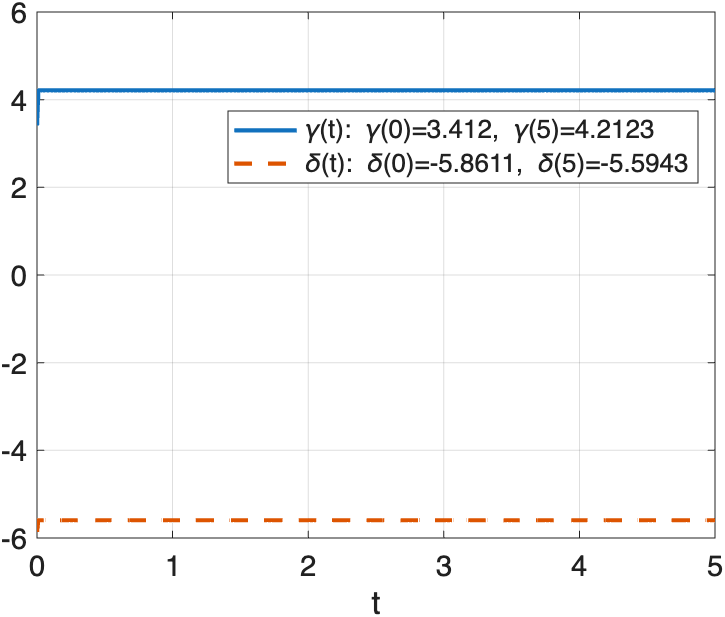} \\
\includegraphics[width=0.295\textwidth]{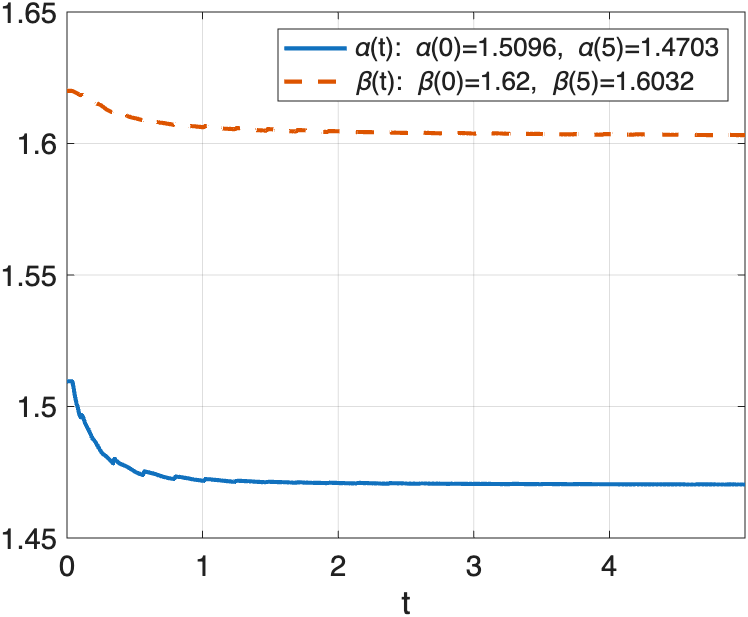} \hspace{20pt}
\includegraphics[width=0.285\textwidth]{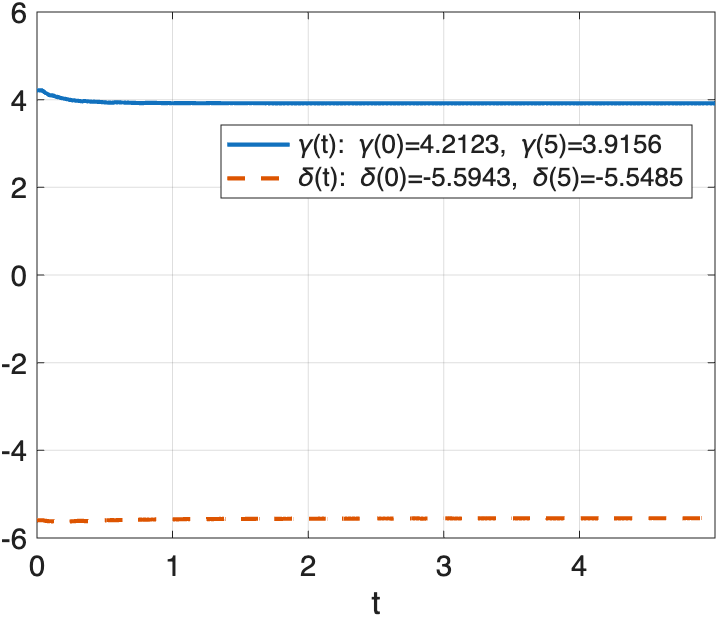} 
\caption{\label{abdg09} {\small Global tracking (top row) and local tracking (bottom row) of $\alpha$, $\beta$, $\delta$, and $\gamma$ values in time for the evolution of $u_0(x) = 0.9B_{1,2}(x,0)$. } }
\end{centering}
\end{figure}

Using our computational approach, %the iD method, 
we calculate $\beta$ and $\gamma$ described in Section \ref{S:CompTools}. To compare localized vs. non-localized results, the parameters of the perturbed breather are tracked during the evolution via two methods: on a local and on a global domain. In Figure \ref{abdg09}, we plot the parameters $\alpha(t), \beta(t), \gamma(t), \delta(t)$, observing that they stabilize sufficiently quickly 
approaching a possible final asymptotic state. We compare the accuracy of the tracked parameters by: first, calculating the mass and energy of the solution from either a local domain or a global one; secondly, using those values of mass and energy, we extract the parameter values and then, thirdly, we plot them from the {\it entire} domain (top row of Figure \ref{abdg09}) and from the {\it local} domain (bottom row of Figure \ref{abdg09}). Calculating locally means computing the mass and energy of the evolving solution separately from the radiation throughout the entire simulation (which is possible, since the radiation splits from the breather and travels faster to the left than the breather). 
In the top row of Figure \ref{abdg09}, all 4 parameters are nearly constant (since mass and energy are conserved quantities) except for the initial jumps. This poorly captures the change in parameter values and the identification of the final breather (plot not shown, since it is worse than the matching we show for the locally extracted parameters in the left plot of Figure \ref{0.9m}); 
the bottom plots of the parameters approach their asymptotic values more smoothly (and furthermore, as we show later, produce a very good matching of profiles).
In the bottom row of Figure \ref{abdg09}, the values of the asymptotic parameters are $\alpha^\prime=1.4703$, $\beta^\prime=1.6032$, $\gamma^\prime=3.9156$, and $\delta^\prime=-5.5485$. The starting and final values for $\alpha(t)$ differ by $-0.04$, for $\beta(t)$ by $-0.02$, for $\gamma(t)$ by $-0.3$, and for $\delta(t)$ by $0.05$. This means that the perturbed breather stabilizes by shedding radiation, which changes the number of oscillations in the profile, the breather pocket size, the breather speed, and the internal oscillation rate. Note that due to the decrease of the initial amplitude by 10\%, i.e., $A=0.9$, the initial $\alpha=1$ is not equal to $\alpha(0)=1.5096$, in fact, $\alpha<\alpha(0)$, and the initial $\beta=2$ $> \beta(0) =1.62$. We mention that the same behavior of the initial $\beta>\beta(0)$ and the initial $\alpha<\alpha(0)$ is observed in all cases with a perturbation of $A<1$. 

To further confirm that our calculations of $\alpha^\prime$ and $\beta^\prime$ (via local domain) are accurate with respect to the asymptotic profile, the `final' state (at the time of computation) of the perturbed breather is matched with an exact %the asymptotic 
breather, based on these converged $\alpha^\prime$ and $\beta^\prime$ values. We show that the evolution of the perturbed breather at $t=4.97$ matches the exact breather $B_{\alpha^{\prime}, \beta^{\prime}}(x,t)$ with $\alpha^{\prime}=1.4703$ and $\beta^{\prime}=1.6032$, shifted $19.4816$ to the right, see left of Figure \ref{0.9m}. (The non-ideal matching is due to the short final computational time at which the matching is done in this example; it is sufficient to illustrate the matching and asymptotic convergence, and the matching improves when the simulation is run for a longer time.)
This suggests that the $\alpha^\prime$ and $\beta^\prime$ values, calculated with the localized mass and energy approach, are accurate in determining the asymptotic state of the perturbed breather. %\ 

The time evolution of the $L^{\infty}$ norm (denoted as $\lVert u(\cdot, t) \rVert_{\infty}$) is calculated 
to confirm that the solution is truly an asymptotically stable breather. The $L^{\infty}$ norm tracks the maximum height of the breather at each time $t$ to determine whether the breather decays (or radiates away) or remains stable over the simulated time interval. 
In Figure \ref{0.9B}, the breather is traveling in the same direction as the dispersion, which causes the $L^{\infty}$ norm plot to exhibit slight perturbations in its peaks; overall, the oscillatory pattern is between $5.5$ and $8$.

In the right of Figure \ref{0.9m}, the {\it mass of the radiation} dispersed is plotted as a function of time. Here, we list the initial mass, via \eqref{mBmass}, and numerically computed mass on a domain $[-L,L]$ (recalling that the initial mass is computed before the time evolution begins), together with the local mass (the value of which we give at time $t=5$):
$$ \|u_0\|^2_{L^2(\mathbb R)} = M[0.9 B_{1,2}] = 0.9^2 48 = 38.88, \quad \|u_0\|^2_{L^2([-L,L])}=38.88 \;\; \text{ and } \;\; \|u(t=5)\|^2_{L^2_{loc}}=38.477,
$$ 
where %$L$ indicates entire domain and 
$loc$ indicates the local domain (the same local domain used to calculate the parameters during the simulation). Since the perturbed breather immediately moves to the left, the mass of the radiation, as a function of time, steadily grows and plateaus around $0.4$ by $t=3$. Note that the sharpness of the curve is due to the structure and `breathing' pattern of the breather (each sharp turn of the curve in the bottom-left subplot corresponds to each trough in the $L^\infty$ norm of the top-right subplot of Figure \ref{0.9m}). Taking the difference of the initial mass and the local mass at each time step we can quantify the mass of the dispersion as shown on the right of Figure \ref{0.9m}, which converges to $0.4030$ at $t=5$. We conclude that the mass of the dispersion at that (final computational) time percentage-wise is about $1.04\%$ of the initial breather mass. 
\smallskip

Summarizing the $A<1$ case, we typically see that such perturbations scale down the initial data profile and increase the number of internal oscillations, consistent with the results of $\beta>\beta(0)$ and $\alpha<\alpha(0)$. Figure \ref{0.9B} illustrates that the breather starts immediately shedding the radiation to the left as it is trying to approach its stable shape. 
Figures \ref{abdg09} and \ref{0.9m} confirm that the stable state is being reached asymptotically. Naturally, one would like to confirm that this behavior holds true for all initial data with $A<1$. For all such cases we indeed found that the initial value $\beta$ is greater than $\beta(0)$, i.e., $\beta >\beta(0)$ and initial $\alpha<\alpha(0)$. 
Further decrease in $A$ shows 
that one would need to take larger $\beta$ to combat the decrease in the height of a breather needed to 
numerically distinguish the breather from its radiation. We numerically found that $A=0.6$ with $\beta=4$ is approximately a good value for the threshold: this means that numerical simulation of cases with $A<0.6$ would require $\beta>4$, otherwise, the resulting breather and its evolution may be difficult to distinguish from its radiation (as it would not be possible to separate them). 

\subsection{Case \texorpdfstring{$A > 1$}{A > 1}:} We now investigate perturbations with values of $A$ greater than 1, in particular, we present $A =1.1, 1.2$ and $1.3$.
\smallskip

%\subsubsection
$\blacklozenge$ \underline{$A=1.1$:} ~
For comparison of the properties in this case, we fix $\alpha=1$ and take $\beta=1, 1.1$, and $1.2$. 
We choose to vary $\beta$ to show 
breather evolutions that travel left, travel right, or remain relatively stationary at the origin. Thus, we consider
the initial conditions 
$$
u_0(x)=1.1B_{1,1}(x,0), \ u_0(x)=1.1B_{1,1.1}(x,0), \ \text{and } u_0(x)=1.1B_{1,1.2}(x,0).
$$
The computational parameters in this section are $L=100\pi$, $N=2^{15}$, and $dt=0.0025$.
\begin{figure}[h!]
\begin{center}
\includegraphics[width=0.25\textwidth]{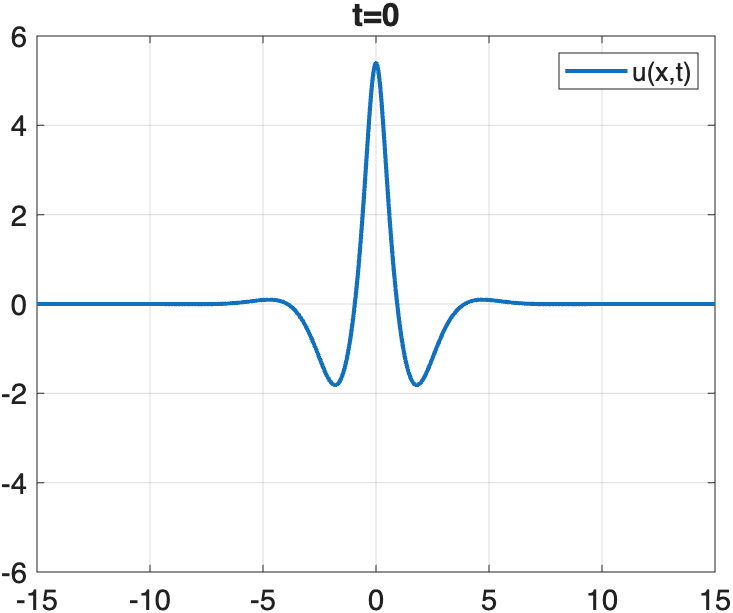} \hspace{20pt}
\includegraphics[width=0.25\textwidth]{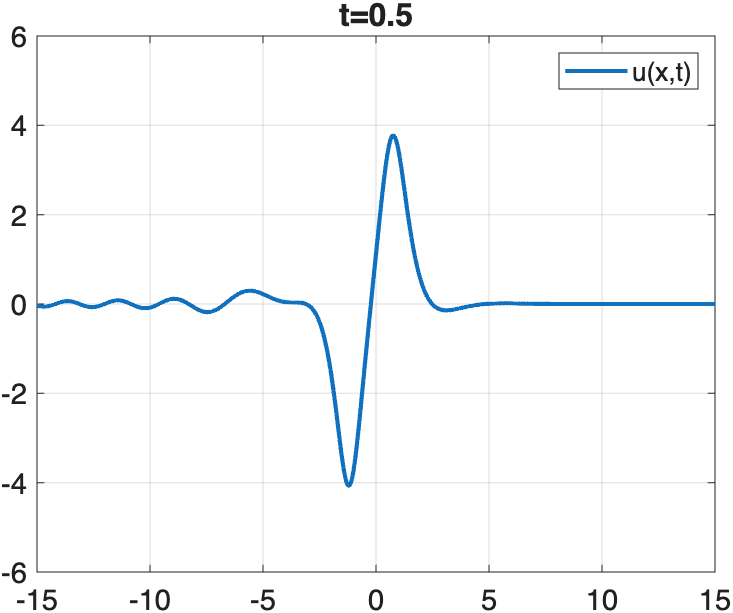} \hspace{20pt}
\includegraphics[width=0.25\textwidth]{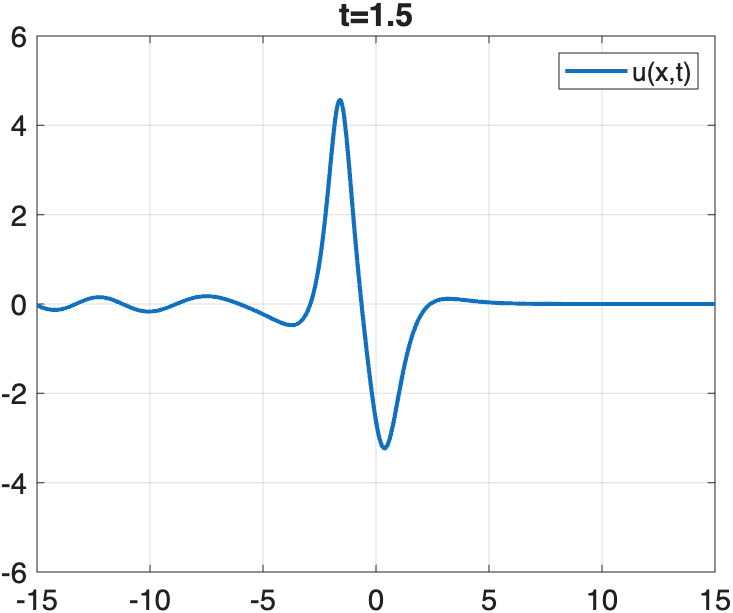}\\ \vspace{10pt}
\includegraphics[width=0.25\textwidth]{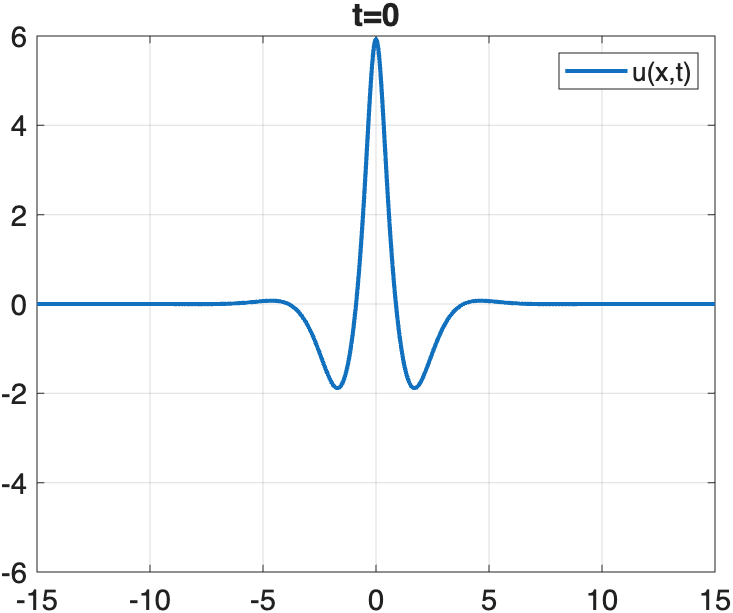} \hspace{20pt}
\includegraphics[width=0.25\textwidth]{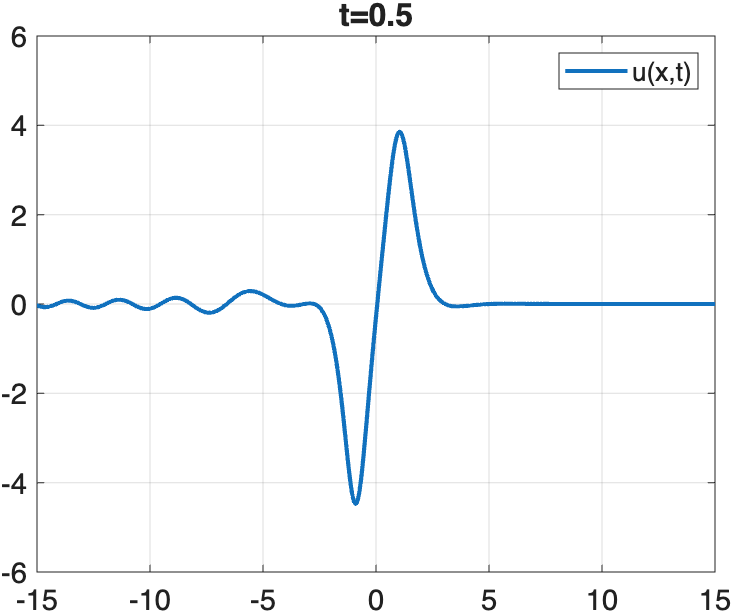} \hspace{20pt}
\includegraphics[width=0.25\textwidth]{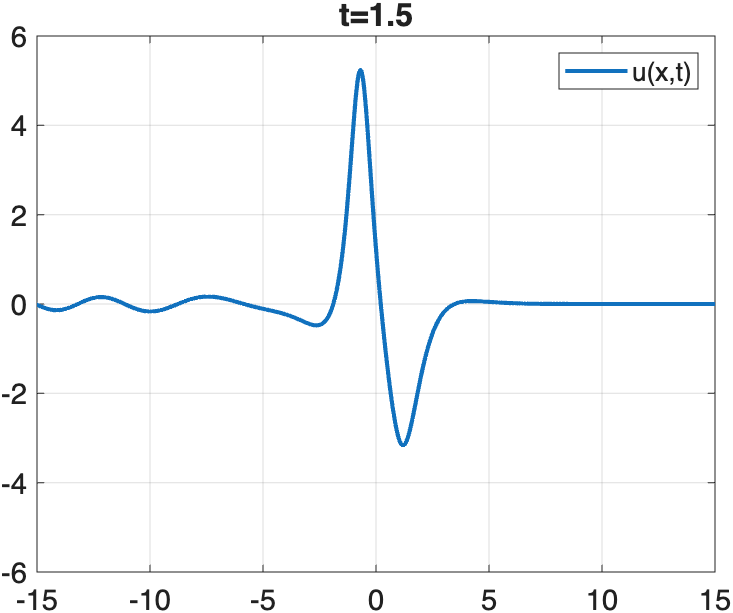}\\ \vspace{10pt}
\includegraphics[width=0.25\textwidth]{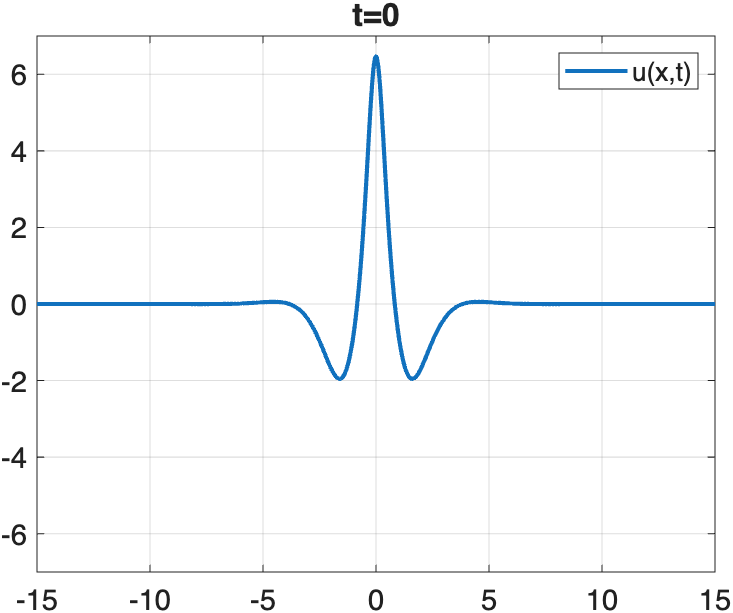} \hspace{20pt}
\includegraphics[width=0.25\textwidth]{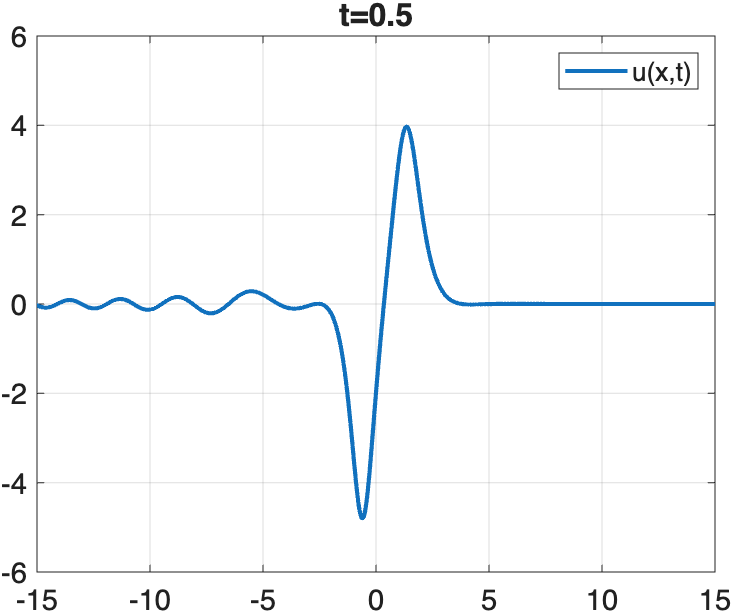} \hspace{20pt}
\includegraphics[width=0.25\textwidth]{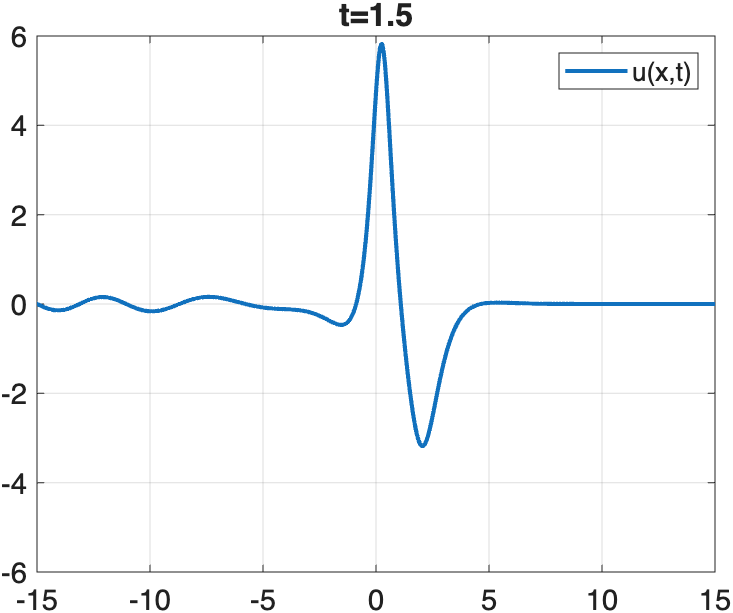}
\caption{\label{1.1B}{\small Time evolution of the perturbed breather with %initial conditions when 
$u_0(x) = 1.1B_{1,\beta}(x,0)$ and $\beta = 1$ (top row), $\beta = 1.1$ (center row), and $\beta = 1.2$ (bottom row).} }
\end{center}
\end{figure}

In Figure \ref{1.1B}, all initial evolutions start shedding off radiation to the left, while trying to find their asymptotic states. Comparing the profiles from the top row to bottom row, the breathers incrementally increase in their height. Then, at the times $t=0.5$ and $t=1.5$, all evolving breathers exhibit similar profile shape, though notice slight difference in breather sizes and positions at $t=1.5$. 

\begin{figure}[h!]
\begin{centering}
\includegraphics[width=0.26\textwidth]{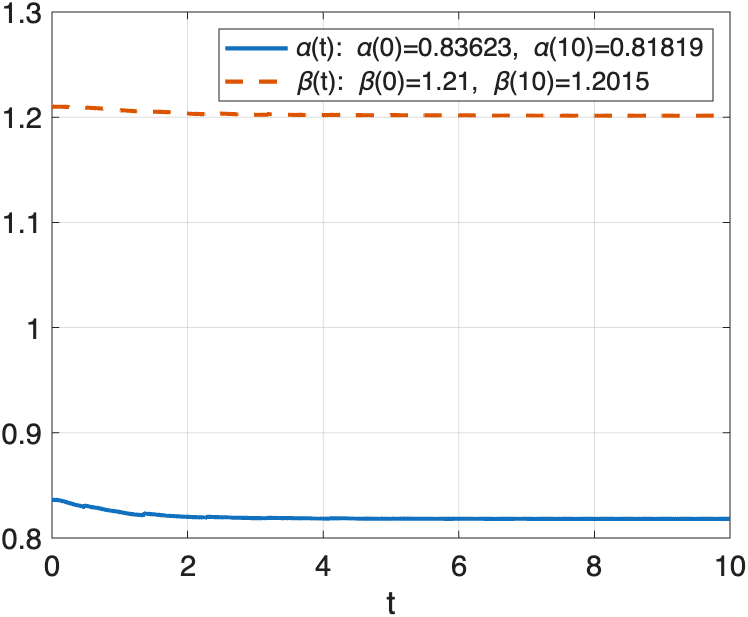} \hspace{20pt}
\includegraphics[width=0.26\textwidth]{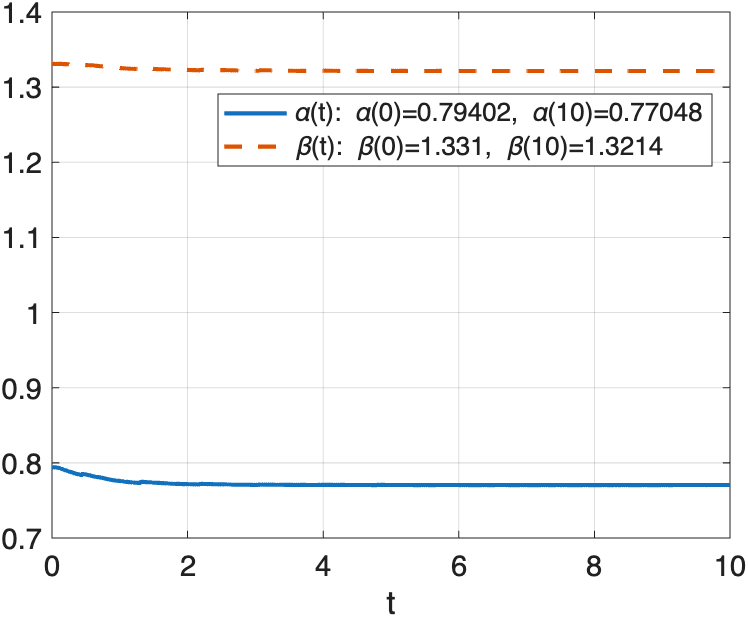} \hspace{20pt}
\includegraphics[width=0.26\textwidth]{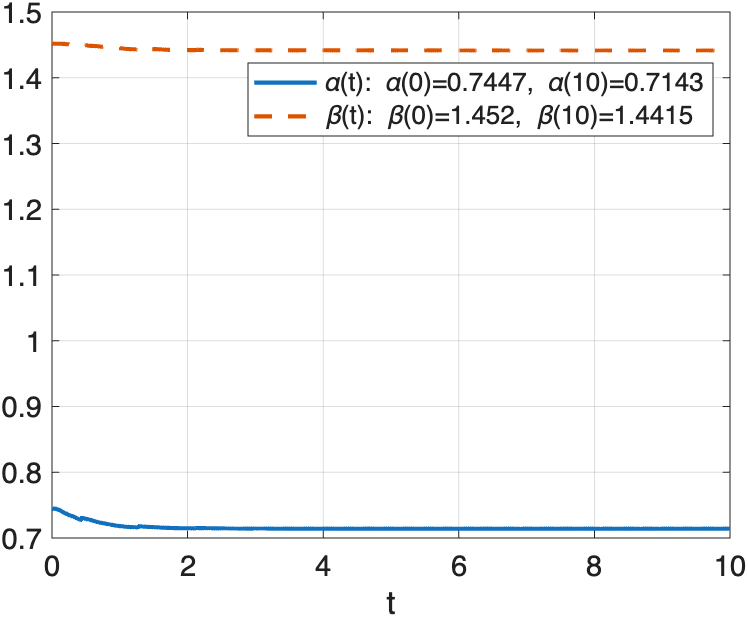}\\ \vspace{5pt} \hspace{-1.5pt}
\includegraphics[width=0.255\textwidth]{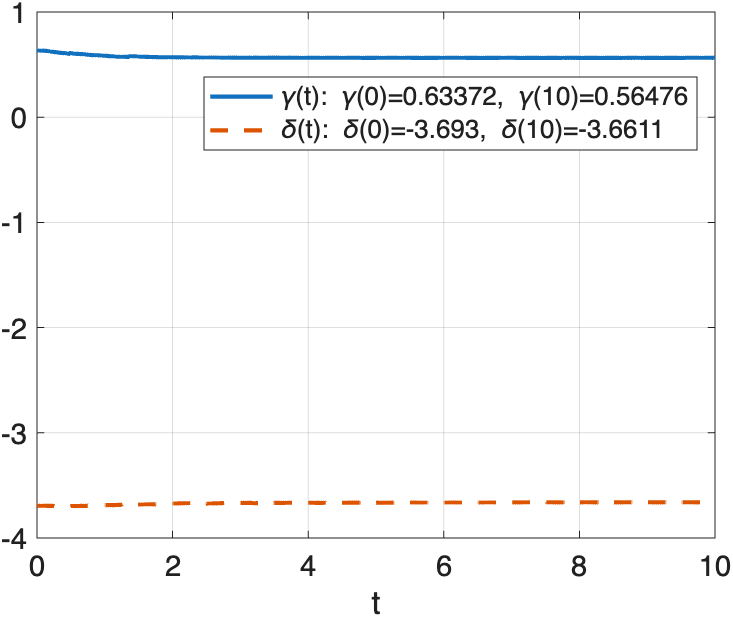} \hspace{22pt}
\includegraphics[width=0.255\textwidth]{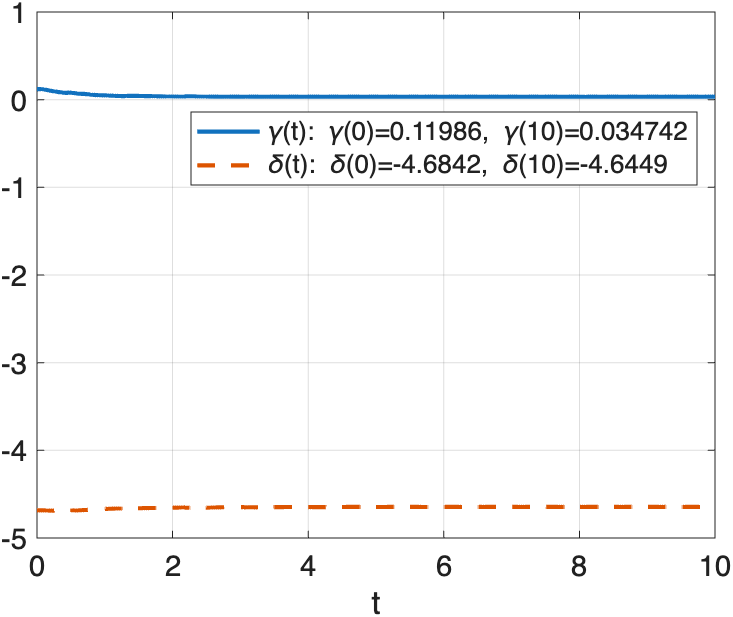} \hspace{22pt}
\includegraphics[width=0.255\textwidth]{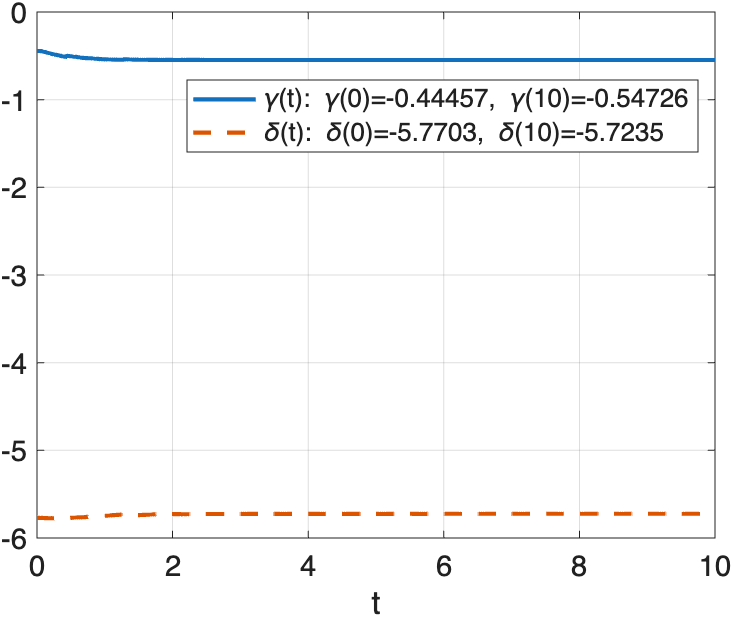} 
\caption{\label{1.1abdg} {\small Tracking of $\alpha(t)$, $\beta(t)$, $\delta(t)$, and $\gamma(t)$ for time evolution of $u_0(x) = 1.1B_{1,\beta}(x,0)$ with $\beta = 1$ (left column), $\beta = 1.1$ (center column), and $\beta = 1.2$ (right column).}
}
\end{centering}
\end{figure}

For all three evolutions (with the initial values $\beta=1, 1.1$ and $1.2$ at $t=0$), the time evolution of parameters $\alpha(t)$, $\beta(t)$, $\gamma(t)$, and $\delta(t)$ is in Figure \ref{1.1abdg} and the values at the final time are listed in Table \ref{T:1}:
\begin{comment}
{\small
\begin{center}
\begin{tabular}{| c || c | c | c | c |} 
 \hline
 \multicolumn{5}{|c|}{\qquad Asymptotic Values (from initial $A=1.1$, $\alpha=1$)} \\
 \hline
 Initial $\beta$ & $\alpha^\prime$ & $\beta^\prime$ & $\gamma^\prime$ & $\delta^\prime$\\ 
 \hline
 1 & 0.81819 & 1.2015 & 0.564760 & -3.6611 \\ 
 \hline
 1.1 & 0.77048 & 1.3214 & 0.034742 & -4.6449 \\
 \hline
 1.2 & 0.71430 & 1.4415 & -0.547260 & -5.7235\\
 \hline
\end{tabular}
%\caption{Asymptotic Values (from initial $A=1.1$, $\alpha=1$)}
% \label{T:1}
\end{center} 
}

{\small
\begin{table}[ht]
\centering
\begin{tabular}{| c || c | c | c | c |} 
 \hline
 \multicolumn{5}{|c|}{\qquad Asymptotic Values (from initial $A=1.1$, $\alpha=1$)} \\
 \hline
 Initial $\beta$ & $\alpha^\prime$ & $\beta^\prime$ & $\gamma^\prime$ & $\delta^\prime$\\ 
 \hline
 1 & 0.81819 & 1.2015 & 0.564760 & -3.6611 \\ 
 \hline
 1.1 & 0.77048 & 1.3214 & 0.034742 & -4.6449 \\
 \hline
 1.2 & 0.71430 & 1.4415 & -0.547260 & -5.7235\\
 \hline
\end{tabular}
\caption{Asymptotic Values (from initial $A=1.1$, $\alpha=1$)}
\label{T:1}
\end{table}
}
\end{comment}
{\small
\begin{table}[ht]
\centering
\begin{tabular}{| c || c | c | c | c |} 
 \hline
 Initial $\beta$ & $\alpha^\prime$ & $\beta^\prime$ & $\gamma^\prime$ & $\delta^\prime$\\ 
 \hline
 1 & 0.81819 & 1.2015 & 0.564760 & -3.6611 \\ 
 \hline
 1.1 & 0.77048 & 1.3214 & 0.034742 & -4.6449 \\
 \hline
 1.2 & 0.71430 & 1.4415 & -0.547260 & -5.7235\\
 \hline
\end{tabular}
\vspace{-6pt}
\caption{Asymptotic Values from initial $A=1.1$, $\alpha=1$.}
\label{T:1}
\end{table}
}

%\vspace{-6pt}
As the initial $\beta$ value increases, the asymptotic value of corresponding $\alpha^\prime$ decreases and $\beta^\prime$ increases. The corresponding $\gamma^\prime$ values are interesting with this example, showing that these perturbed breathers can evolve as left-traveling states (i.e., $\gamma^\prime>0$, \textit{case with $\beta=1$}), remain relatively stationary at $x=0$ (i.e., $\gamma^\prime\cong 0$, \textit{case with $\beta=1.1$}), or right-traveling states (i.e., $\gamma^\prime< 0$, \textit{case with $\beta=1.2$}). 
\smallskip

With the parameter functions in Figure \ref{1.1abdg}, the change in each function helps us interpret the effects of amplitude perturbations on breather dynamics as they all converge to their stable states. The perturbed breather sheds radiation, which effectively decreases the number of internal oscillations and increases the breather envelope size. Notice that for $A>1$, we have $\beta<\beta(0)$ and $\alpha>\alpha(0)$, opposite of the results in the previous section for $A<1$.

\begin{figure}[h!]
\begin{center}
\includegraphics[width=0.27\textwidth]{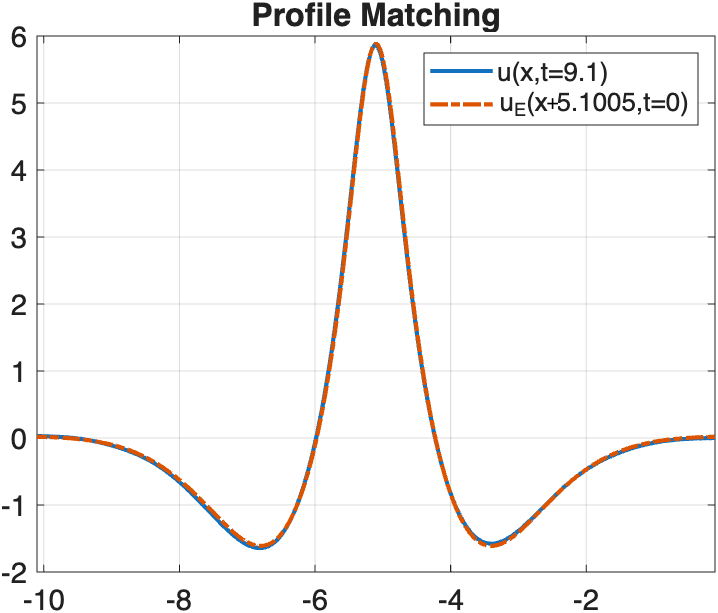} \hspace{20pt}
\includegraphics[width=0.27\textwidth]{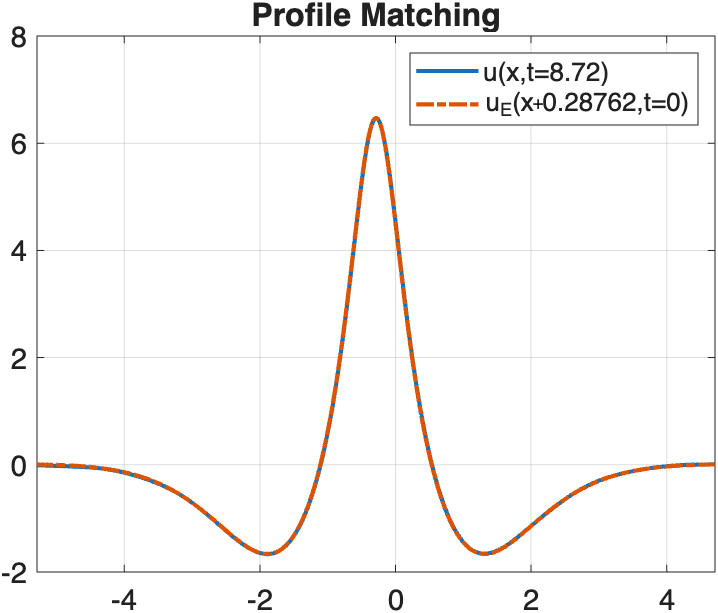} \hspace{20pt}
\includegraphics[width=0.27\textwidth]{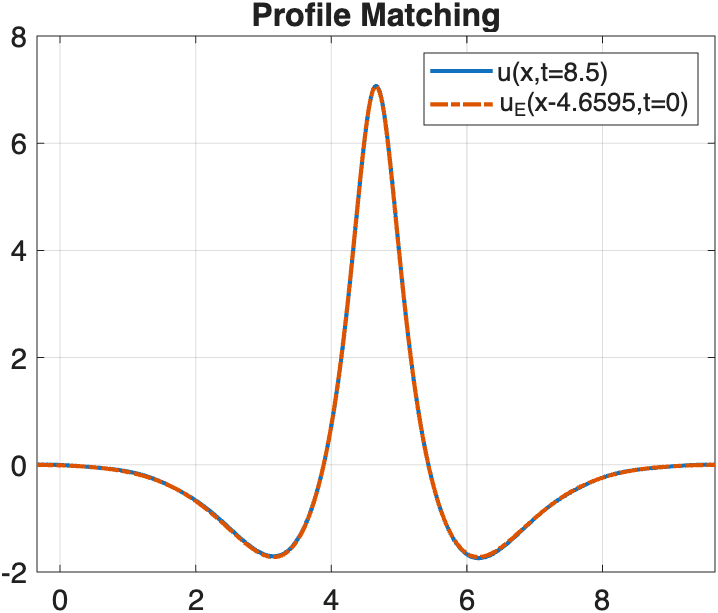}
\\\hspace{4pt} (a) \hspace{140pt} (b) \hspace{140pt} (c)
\caption{\label{1.1m} {\small Asymptotic profiles matching for $u_0(x)=1.1B_{1,\beta}(x,0)$ evolving via mKdV to $B_{\alpha^\prime,\beta^\prime}(x-c,0)$ for different $\beta$: time evolution of $u_0$ with (a) $\beta=1$
%(a) $u_0(x) = 1.1B_{1,1}(x,0)$ 
at $t=9.1$ matched to $u_E(x) =B_{0.81819,1.2015}(x+5.1005,0)$; (b) with $\beta=1.1$ 
%$u_0(x) = 1.1B_{1,1.1}(x,0)$ 
at $t=8.72$ to $u_E(x) =B_{0.77048,1.3214}(x+0.28762,0)$; 
(c) with $\beta=1.2$
%$u_0(x) = 1.1B_{1,1.2}(x,0)$ 
at $t=8.5$ to $u_E(x) =B_{0.7143,1.4415}(x-4.6595,0)$.} }
\end{center}
\end{figure}

Using the asymptotic values of $\alpha^\prime$ and $\beta^\prime$ from Figure \ref{1.1abdg}, the evolved breathers are matched with their asymptotic states $B_{\alpha^\prime,\beta^\prime}(x,0)$ after a certain number of cycles is completed. %The specific asymptotic values used for the profile matching are listed in the previous table. 
When $\beta=1$, the asymptotic breather is shifted by $5.1005$ to the left to make the profile match at $t=9.1$. For $\beta=1.1$, there is a smaller shift of $0.28762$ to the left (due to the small value of $\gamma$) to match at $t=8.72$. Lastly, for $\beta=1.2$, the shift is $4.6595$ to the right to match at $t=8.5$. In Figure \ref{1.1m}, the matching of these breathers confirms 
that the perturbations asymptotically lead to a `nearby' breather (since numerically we cannot do {\it infinitesimally small} perturbations, which would then show that the final state is very close to the shift of the original breather as was proved in the orbital stability \cite{am2013}, see also \cite{Sem2022}); this, in a sense, confirms 
the asymptotic stability (as well as the accuracy of the found $\alpha^\prime$ and $\beta^\prime$ values for the asymptotic profiles) of mKdV breathers. 

Table \ref{T:2} lists values corresponding to the global mass of the initial breather, the local mass of the evolving breather, the mass of the dispersion, and the percentage of initial mass loss to the dispersion. 
\vspace{-.3cm}
\begin{comment}
{\small
\begin{center}
\begin{tabular}{| c || c | c | c | c |} 
 \hline
 \multicolumn{5}{|c|}{Quantifying Radiation (for initial $AB_{1,\beta}(x,0)$ with $A=1.1$)} %, $\alpha=1$)} 
 \\
 \hline
 Initial & Total mass & Local Mass & Mass of Radiation & Percentage\\ 
 %\hline
 $\beta$ & $\|u\|^2_{L^2(\mathbb R)}$ & $\|B_{\alpha^\prime,\beta^\prime}\|^2_{L^2_{loc}}$ & 
 $\|u\|^2_{L^2(\mathbb R)} - \|B_{\alpha^\prime,\beta^\prime}\|^2_{L^2_{loc}}$ & of radiation mass\\ 
 \hline
 1 & 29.0400 & 28.8352 & 0.2048 & $0.71\%$ \\ 
 \hline
 1.1 & 31.9440 & 31.7144& 0.2296 & $0.72\%$ \\
 \hline
 1.2 & 34.8480 & 34.5960 & 0.2520 & $0.72\%$ \\
 \hline
\end{tabular}
\end{center} 
}
\end{comment}
{\small
\begin{table}[ht]
\centering
\begin{tabular}{| c || c | c | c | c |} 
 \hline
 Initial & Total mass & Local Mass & Mass of Radiation & Percentage\\ 
 $\beta$ & $\|u\|^2_{L^2(\mathbb R)}$ & $\|B_{\alpha^\prime,\beta^\prime}\|^2_{L^2_{loc}}$ & 
 $\|u\|^2_{L^2(\mathbb R)} - \|B_{\alpha^\prime,\beta^\prime}\|^2_{L^2_{loc}}$ & of radiation mass\\ 
 \hline
 1 & 29.0400 & 28.8352 & 0.2048 & $0.71\%$ \\ 
 \hline
 1.1 & 31.9440 & 31.7144& 0.2296 & $0.72\%$ \\
 \hline
 1.2 & 34.8480 & 34.5960 & 0.2520 & $0.72\%$ \\
 \hline
\end{tabular}
\vspace{-6pt}
\caption{Quantifying Radiation for initial condition $AB_{1,\beta}(x,0)$ with $A=1.1$.}
\label{T:2}
\end{table}
}

As the $\beta$ value grows by $0.1$, the mass of the dispersed radiation grows by approximately $0.025$. The percentage of initial mass loss for all $\beta$ values is about $0.72\%$, so the change in initial $\beta$ values does not have a significant effect on the final percentages, however, it does affect the initial and local mass of the breather, since larger $\beta$ values mean larger mass and amplitude (with a tighter envelope), and hence a larger-amplitude profile.

\begin{figure}[h!]
\begin{center}
\includegraphics[width=0.29\textwidth]{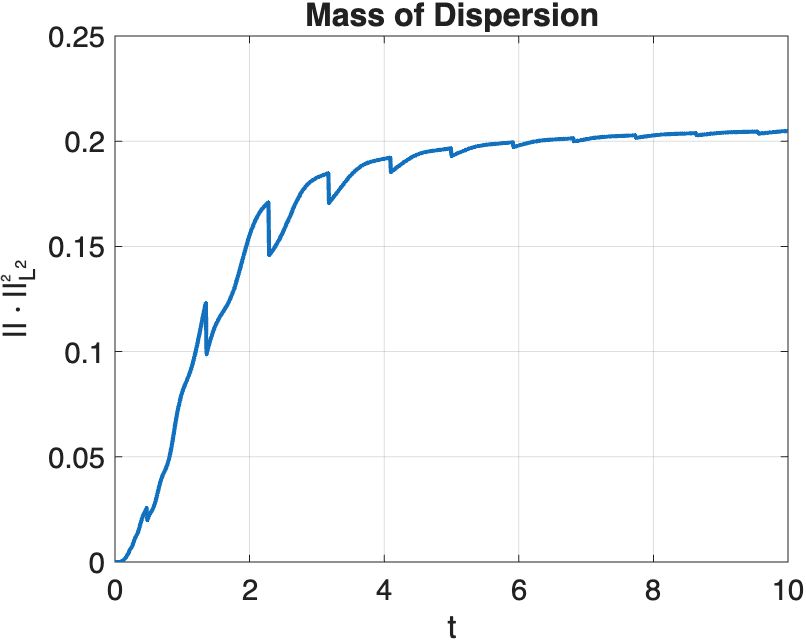} \hspace{15pt}
\includegraphics[width=0.29\textwidth]{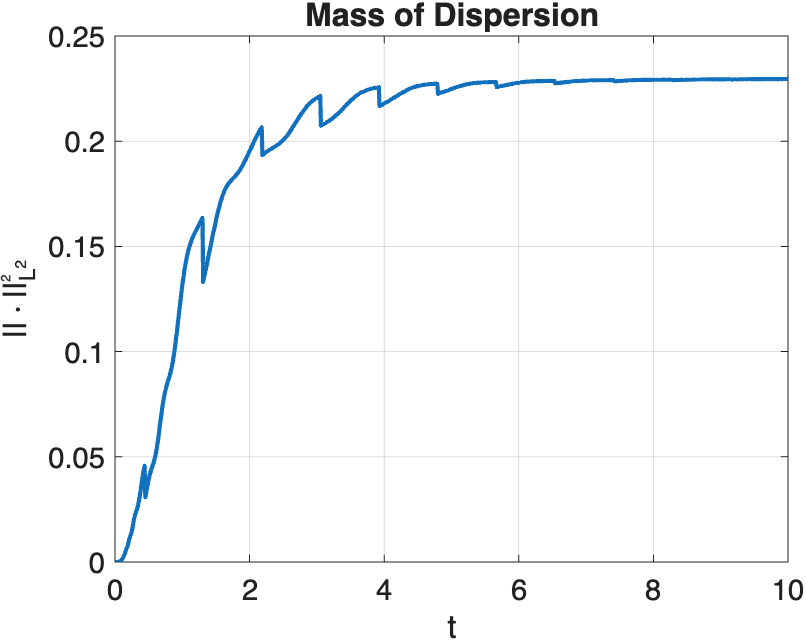} \hspace{15pt}
\includegraphics[width=0.29\textwidth]{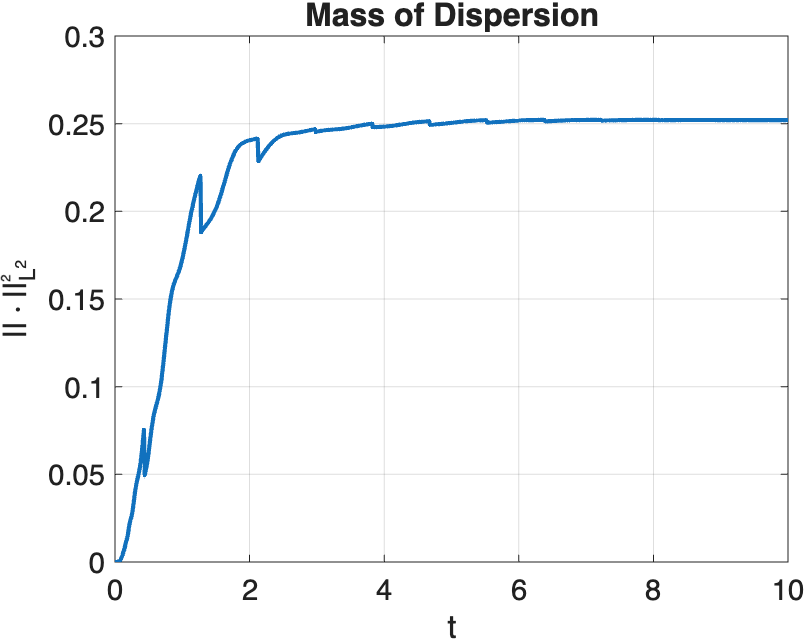}
\\\hspace{15pt} (a) \hspace{146pt} (b) \hspace{146pt} (c)
\caption{\label{1.1rm} Tracking mass of dispersion/radiation coming out to the left of the time evolution of perturbed breather $u_0(x) = 1.1B_{1,\beta}(x,0)$: (a) $\beta = 1$, %$u_0(x) = 1.1B_{1,1.1}(x,0)$ 
(b) $\beta=1.1$, %and %$u_0(x) = 1.1B_{1,1.2}(x,0)$ 
(c) $\beta=1.2$.}
\end{center}
\end{figure}

In Figure \ref{1.1rm}, the mass of dispersion coming from the left of the breathers during the time evolutions is plotted as a function of time. We note that the dips along the curve are an artifact of how the local domain is determined, and each dip corresponds to each trough of the respective $L^\infty$ norm.

\begin{figure}[h!]
\begin{center}
\includegraphics[width=0.29\textwidth]{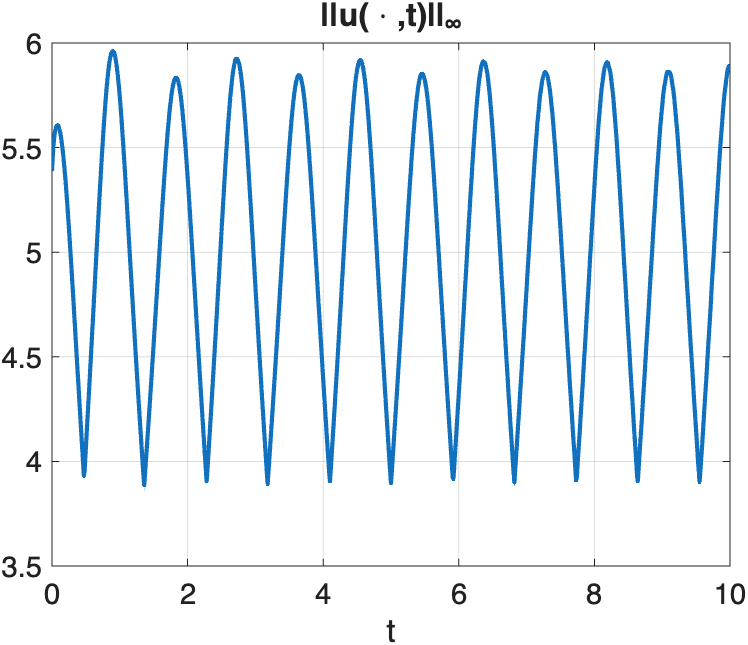} \hspace{10pt}
\includegraphics[width=0.29\textwidth]{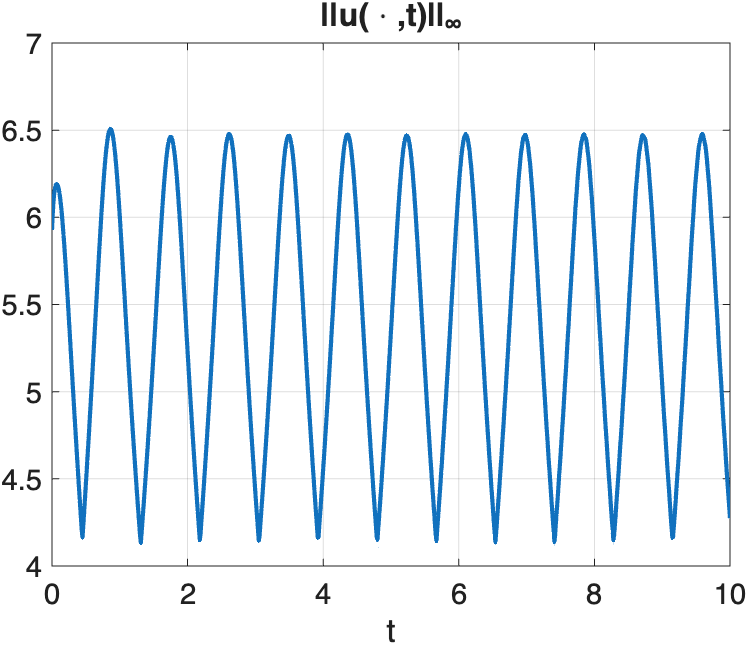} \hspace{10pt}
\includegraphics[width=0.29\textwidth]{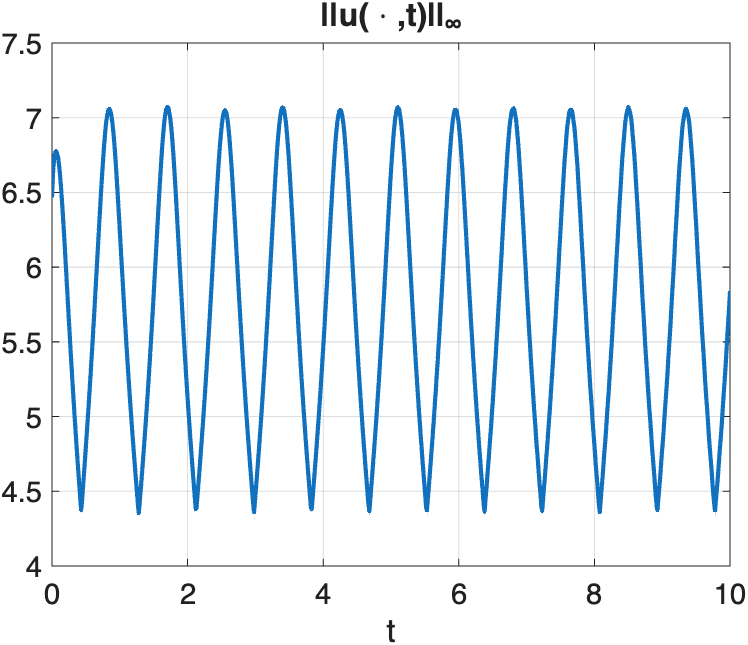}
\\\hspace{4pt} (a) \hspace{141pt} (b) \hspace{141pt} (c)
\caption{\label{1.1L}$L^{\infty}$ norms up to the time $t=5$ of the time evolution for $u_0(x) = 1.1B_{1,\beta}(x,0)$: (a) $\beta=1$, (b) $\beta=1.1$, (c) $\beta=1.2$.}
\end{center}
\end{figure}

In Figure \ref{1.1L}, the $L^{\infty}$ norms are used to confirm asymptotic stability of the perturbed breathers in Figure \ref{1.1B}. 
The periodic structure and oscillatory pattern form by the time $t=1$ in all cases. The $L^{\infty}$ norms differ by height: (a) shows an amplitude of about $6$, (b) has an amplitude of $6.5$, and (c) shows an amplitude of about $7.1$. This difference in amplitudes of the $L^{\infty}$ norm is due to the increasing $\beta$ values. Observe that the peaks of the $L^{\infty}$ norm from subplot (a) are alternating in amplitude. This is a numerical artifact due to the breather traveling in the same direction as the radiation. A longer simulation shows that the peaks of subplot (a) eventually level out, representing the radiation and solution separating or the radiation sufficiently spreading until it is undetectable in the $L^\infty$ norm. For subplots (b) and (c), the separation occurs quicker due to the $\gamma$ values (and thus, breathers are nearly stationary or traveling to the right). 

Thus, Figures \ref{1.1abdg} - \ref{1.1L} confirm the asymptotic stability of mKdV breathers in this example with $A=1.1$. 
Next, we investigate even larger perturbations of the initial data. 
\smallskip

%\subsubsection
$\blacklozenge$ \underline{$A=1.2$:}~
Here, we analyze the behavior of initial data of the form $$
u_0(x)= 1.2 \,B_{1,\beta}(x,0), 
$$ 
that is, the amplitude factor $A=1.2$, while fixing $\alpha=1$, and varying $\beta$. The numerical parameters in this example are $L=200\pi$, $N=2^{16}$, and $dt=0.0025$. 

\begin{figure}[h!]
\begin{center}
\vspace{8pt}
\includegraphics[width=0.22\textwidth]{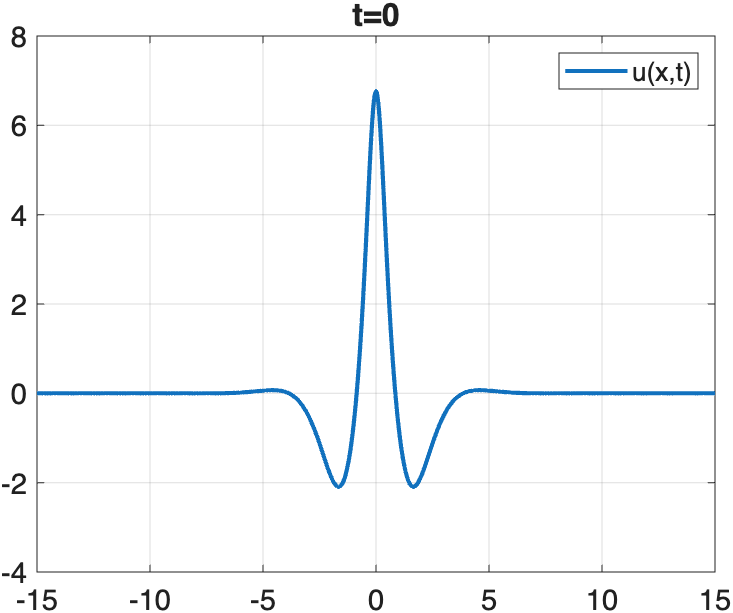} \hspace{12pt} 
\includegraphics[width=0.22\textwidth]{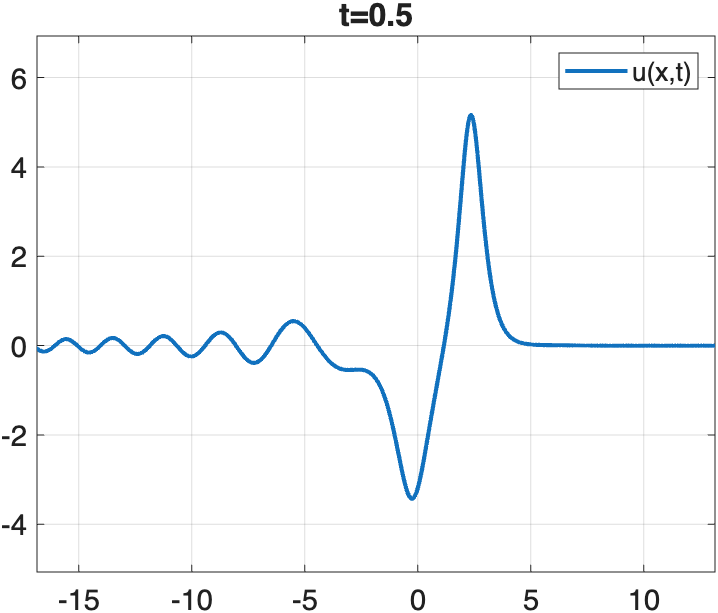} \hspace{12pt} 
\includegraphics[width=0.22\textwidth]{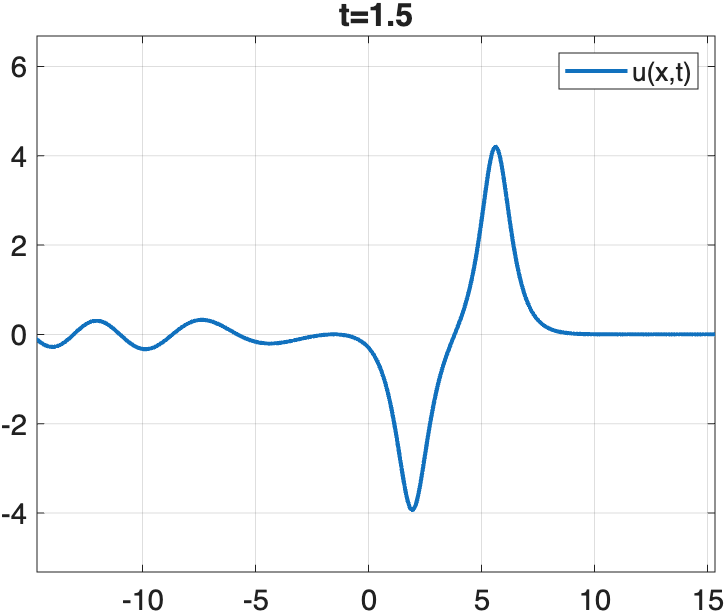} \hspace{12pt} 
\includegraphics[width=0.22\textwidth]{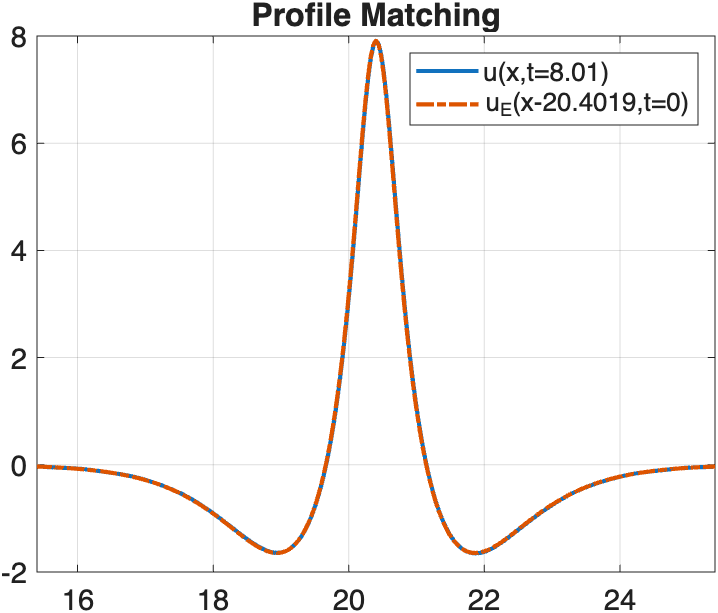} \\ \vspace{10pt}
\includegraphics[width=0.22\textwidth]{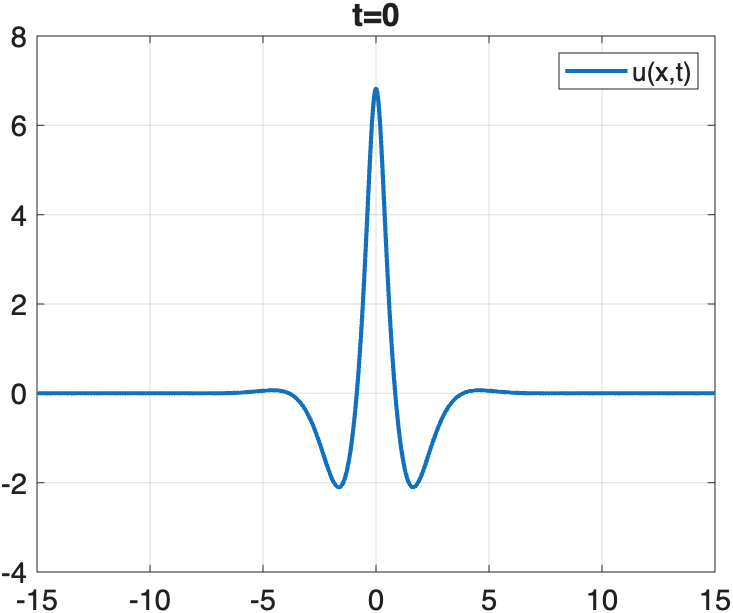} \hspace{12pt} 
\includegraphics[width=0.22\textwidth]{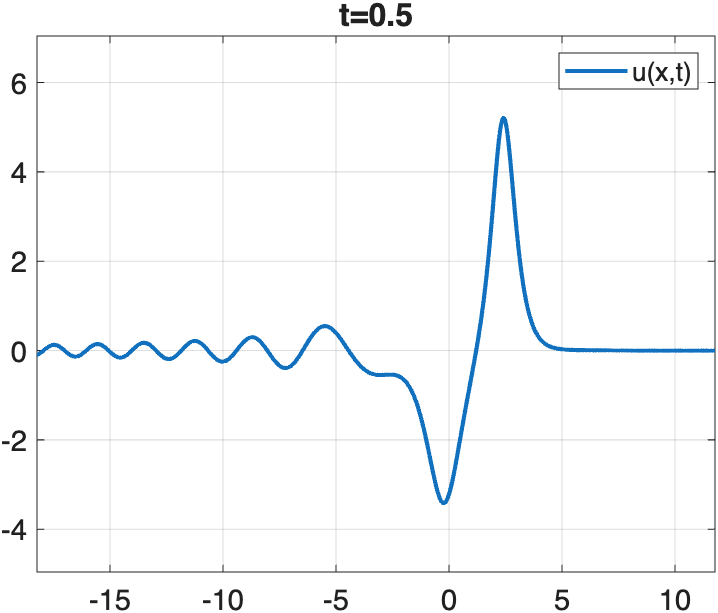} \hspace{12pt} 
\includegraphics[width=0.22\textwidth]{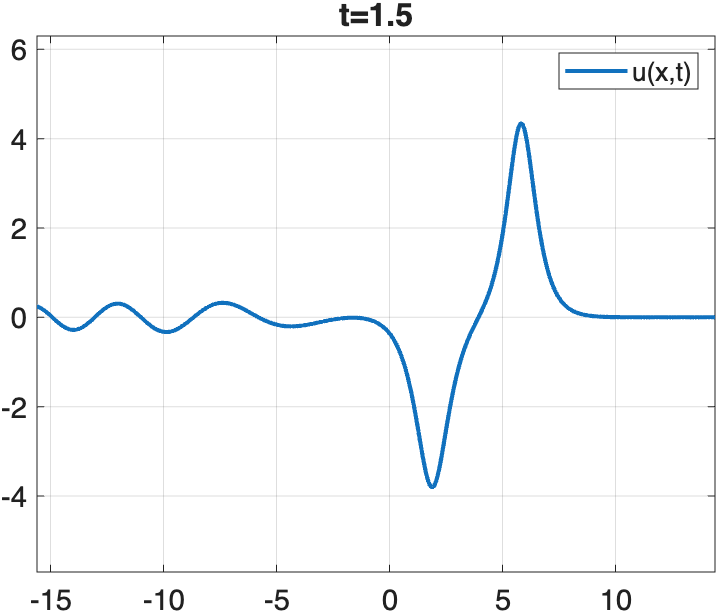} \hspace{12pt} 
\includegraphics[width=0.22\textwidth]{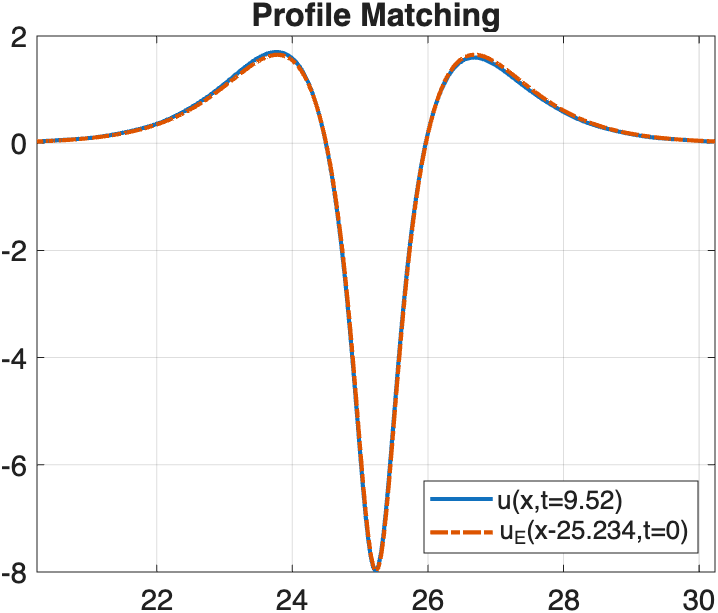} \\ \vspace{10pt}
\includegraphics[width=0.22\textwidth]{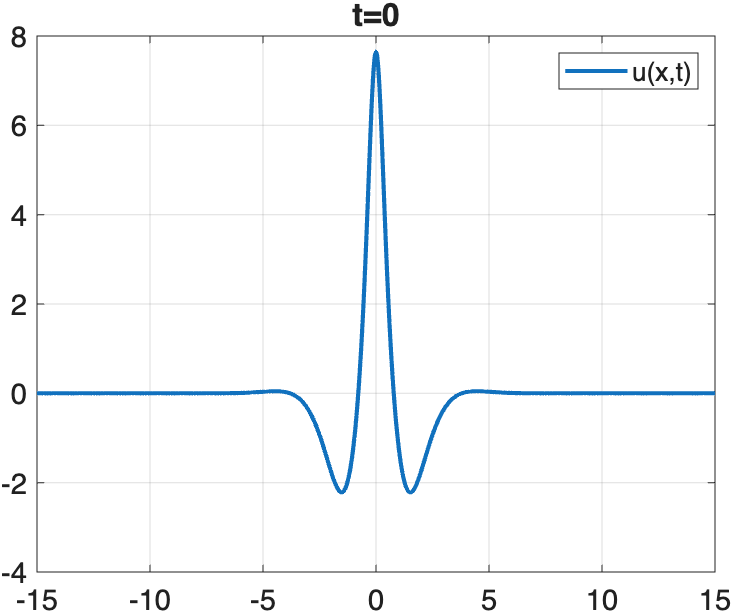} \hspace{12pt} 
\includegraphics[width=0.22\textwidth]{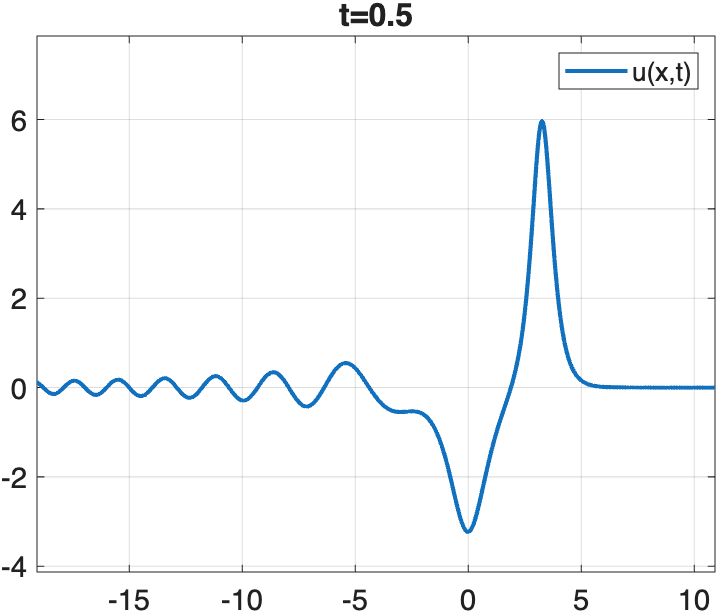} \hspace{12pt} 
\includegraphics[width=0.22\textwidth]{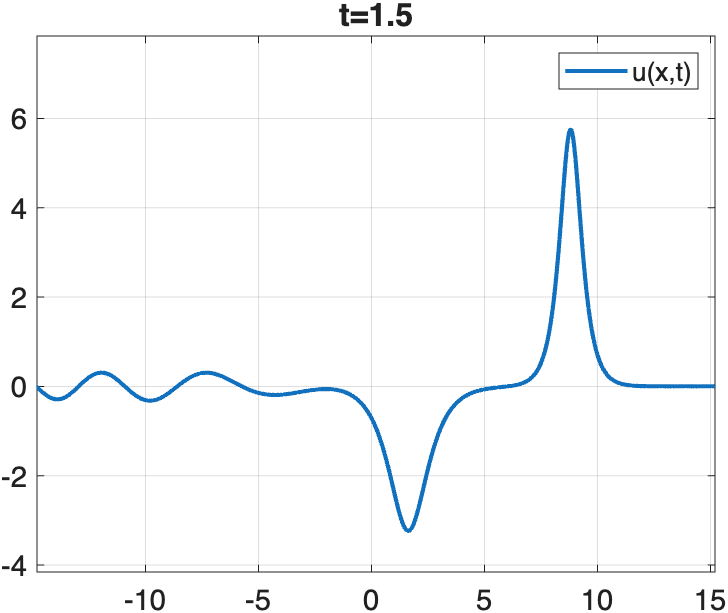} \hspace{12pt} 
\includegraphics[width=0.22\textwidth]{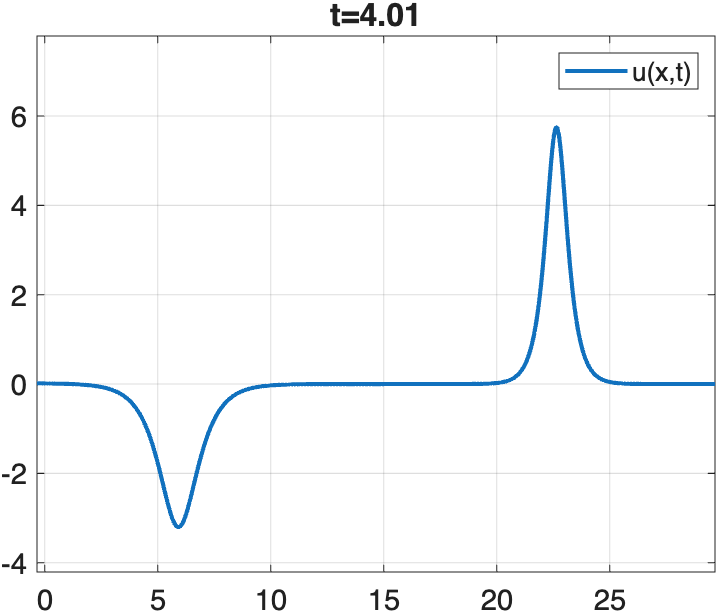} 
\caption{\label{1.2B} {\small Time evolution of $u_0(x) = 1.2B_{1,\beta}(x,0)$: (top row) $\beta = 1.15$ with profile matching to the explicit breather $u_E(x)=B_{0.14978,1.6158}(x-20.4019, 0)$ at $t=8.01$; (center row) $\beta=1.16$ %$u_0(x) = 1.2B_{1,1.16}(x,0)$ 
with profile matching to the explicit breather $u_E(x)=-B_{0.059159,1.6297}(x-25.234, 0)$ at $t=9.52$; (bottom row) $\beta = 1.3$ evolving into one positive and one negative soliton (and radiation). } }
\end{center}
\end{figure}

\begin{figure}[h!]
\begin{centering}
\includegraphics[width=0.26\textwidth]{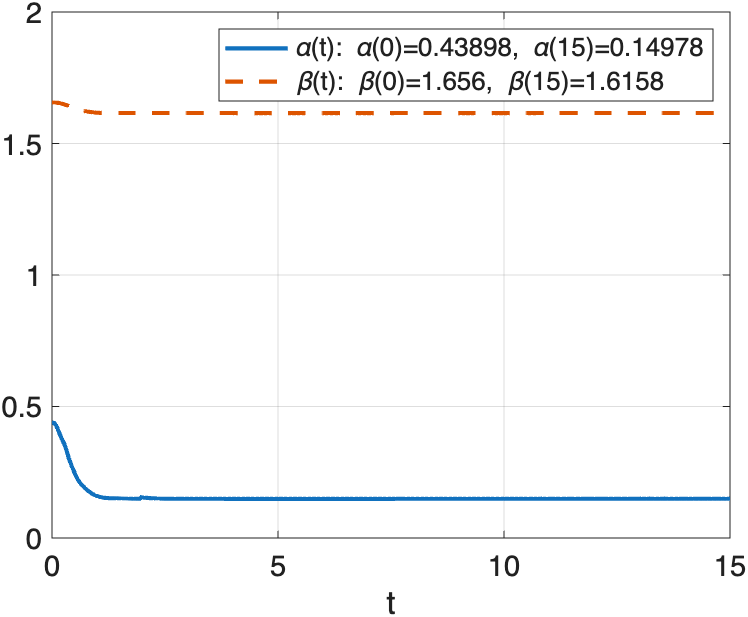} \hspace{20pt}
\includegraphics[width=0.26\textwidth]{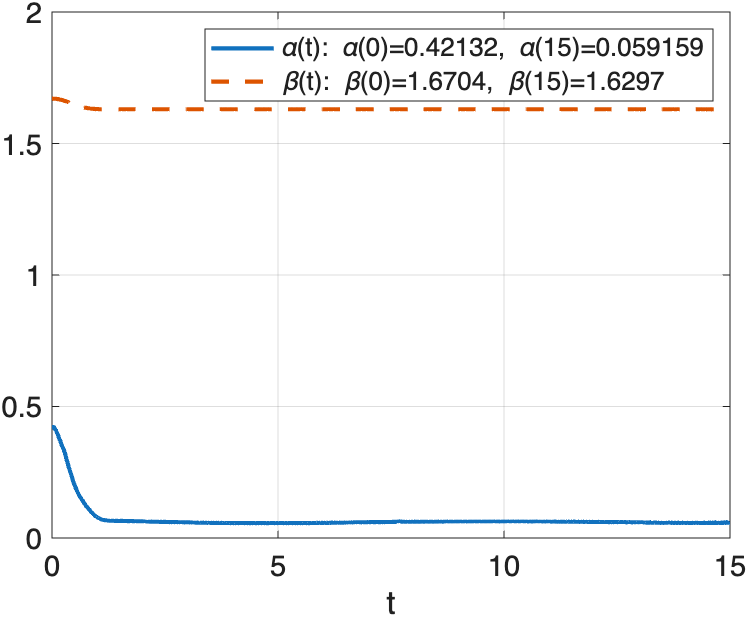} \hspace{20pt}
\includegraphics[width=0.26\textwidth]{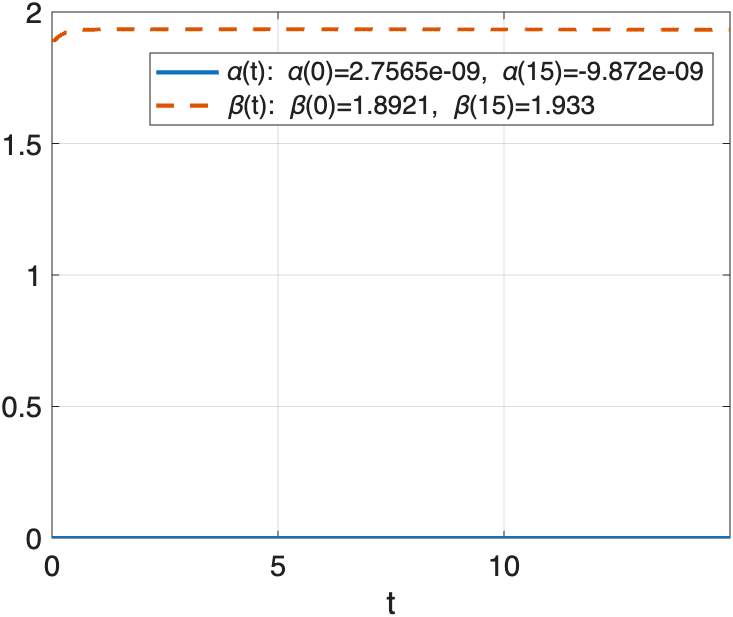}\\ \vspace{5pt} 
\hspace{-1.5pt}
\includegraphics[width=0.255\textwidth]{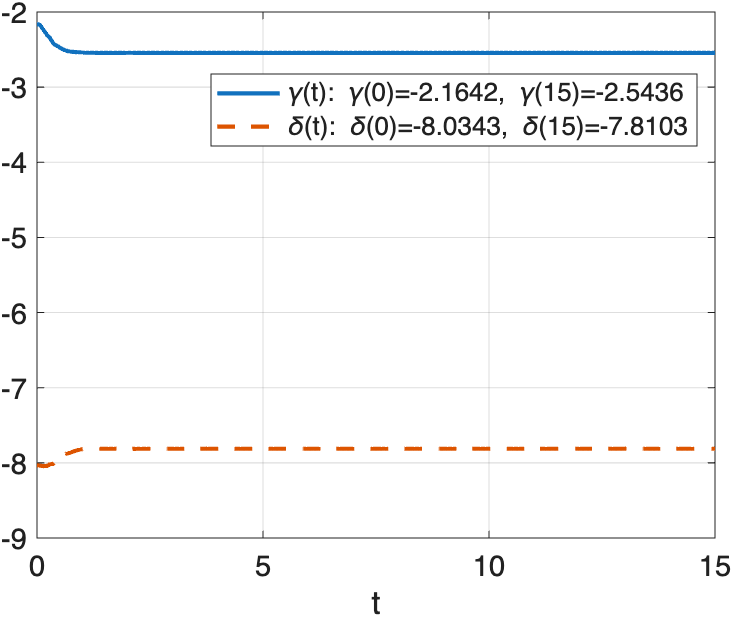} \hspace{23pt}
\includegraphics[width=0.255\textwidth]{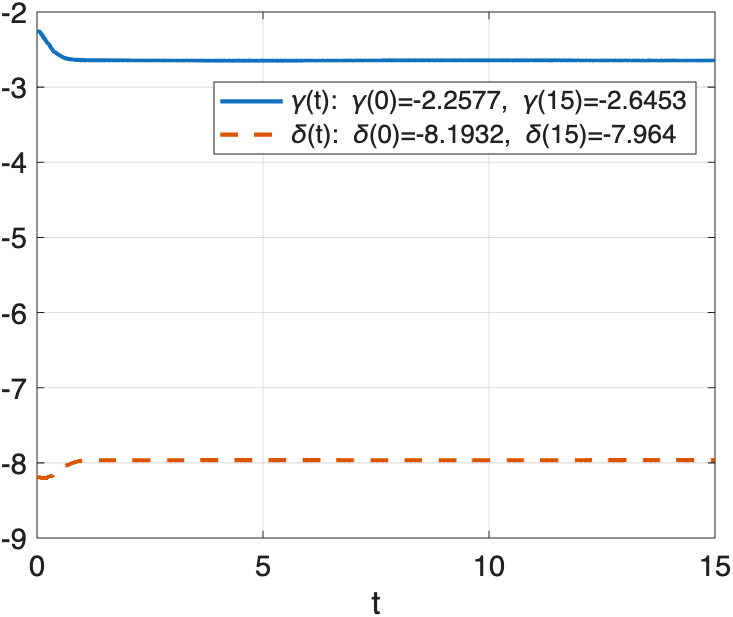} \hspace{20pt}
\includegraphics[width=0.26\textwidth]{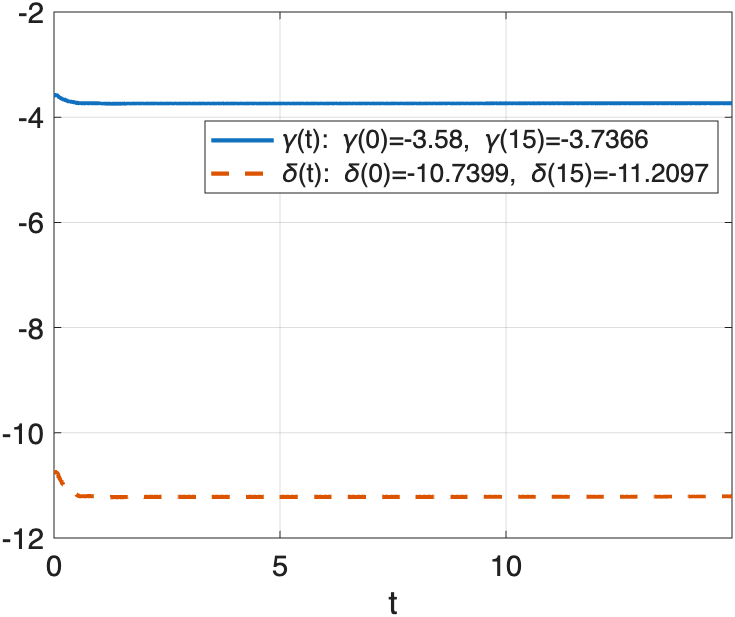} 
\caption{\label{1.2abdg} {\small Tracking of $\alpha(t)$, $\beta(t)$, $\delta(t)$, and $\gamma(t)$ for time evolution of $u_0(x) = 1.2B_{1,\beta}(x,0)$: $\beta=1.15$ (left column), $\beta=1.16$
%$u_0(x) = 1.2B_{1,1.16}(x,0)$ 
(center column), and $\beta =1.3$ %$u_0(x) = 1.2B_{1,1.3}(x,0)$ 
(right column).} }
\end{centering}
\end{figure}

In Figure \ref{1.2B}, the time evolutions of three breathers initialized with three different $\beta$ values are shown in rows. 
In the top row of Figure \ref{1.2B}, $\beta=1.15$ and the evolution starts with radiation shedding to the left while the breather decreases in height. By timestamp $t=1.5$, the profile starts to form a bend around the $x$-axis in between the negative and positive bumps. Since the height of the positive bump is close to the height of the negative bump, both bumps travel together and maintain closeness. This allows for interactions until the positive bump loses mass to the negative bump, hence, activating the breather structure. Following the profile at $t=1.5$, observe that the negative bump grows in amplitude while the positive bump decreases. Simultaneously, a positive bump forms to the left of the negative bump and grows in height. The moment the two positive bumps, one to the left and one to the right of the negative bump, are the same height, the evolving breather reaches its half-period at around $t=4$. Then, the breather completes a full period and is profile matched to an explicit breather at $t=8.01$, see the last subplot of the first row in Figure \ref{1.2B}. The explicit breather is generated with the $\alpha^\prime$ and $\beta^\prime$ values tracked during the evolution, which we provide in the table below as well as its evolution in the first column of Figure \ref{1.2abdg}. 

In the center row of Figure \ref{1.2B}, $\beta=1.16$ we observe similar behavior as in the first row, however, there will be a deeper bend formed in between the positive and negative bumps around $t=1.5$, which affects the breather's periodicity, i.e., how long it takes for the breather to complete its period. With $\beta=1.15$, the breather completes a full period by $t=8.01$, however, with $\beta=1.16$, the breather completes its half-period at $t=9.51$ (thus, the full period is around $19.02$). In the last subplot of the center row in Figure \ref{1.2B}, the breather at its half-period is matched to an explicit breather using the tracked $\alpha^\prime$ and $\beta^\prime$ values from the table below and the middle column of Figure \ref{1.2abdg}. We note that the breather's profile at its half-period is flipped down across the $x$-axis, so for matching, we negate the explicit breather (or take its evolution at half period) to align the profiles. 

In the last row of Figure \ref{1.2B}, $\beta=1.3$ and the evolution of the perturbed breather is significantly different from the previous rows. With $\beta = 1.3$, the evolution starts with radiation shedding to the left, and by $t=0.5$, a bend begins to form near the $x$-axis in between the positive and negative bumps. By $t=1.5$, the profile clearly separates and breaks down into positive and negative solitons. Notice that the positive bump has a higher amplitude than the negative bump, meaning that it will travel faster than the negative bump. This means that the bumps will separate more in time. At $t=4.01$, the solitons both travel to the right at different rates, then the distance between the solitons continues to grow throughout the simulation. 

In Figure \ref{1.2abdg}, we track the evolution of all parameters in time, corresponding to different $\beta$ values in different columns,
noting that as the $\beta$ value increases, $\beta^\prime$ increases and $\alpha^\prime$ decreases (to zero in the last example), and so do $\gamma^\prime$ and $\delta^\prime$.  

Table \ref{T:3} lists all asymptotic values for each initial $\beta$.
\begin{comment}
{\small 
\begin{center}
\begin{tabular}{| c || c | c | c | c |} 
 \hline
 \multicolumn{5}{|c|}{Asymptotic Values (from initial $A=1.2$, $\alpha=1$)} \\
 \hline
 Initial $\beta$ & $\alpha^\prime$ & $\beta^\prime$ & $\gamma^\prime$ & $\delta^\prime$\\ 
 \hline
 1.15 & 0.1497800 & 1.6158 & -2.5436 & -7.8103 \\ 
 \hline
 1.16 & 0.0591590 & 1.6297 & -2.6453 & -7.9640 \\
 \hline
 1.3 & $9.872\times 10^{-9}$ & 1.9330 & -3.7366 & -11.2097 \\
 \hline
\end{tabular}
\end{center} 
}
\end{comment}
{\small
\begin{table}[ht]
\centering
\begin{tabular}{| c || c | c | c | c |} 
 \hline
 Initial $\beta$ & $\alpha^\prime$ & $\beta^\prime$ & $\gamma^\prime$ & $\delta^\prime$\\ 
 \hline
 1.15 & 0.1497800 & 1.6158 & -2.5436 & -7.8103 \\ 
 \hline
 1.16 & 0.0591590 & 1.6297 & -2.6453 & -7.9640 \\
 \hline
 1.3 & $9.872\times 10^{-9}$ & 1.9330 & -3.7366 & -11.2097 \\
 \hline
\end{tabular}
\vspace{-6pt}
\caption{Asymptotic Values from initial $A=1.2$, $\alpha=1$.}
\label{T:3}
\end{table}
}

\begin{figure}[h!]
\begin{center}
\includegraphics[width=0.29\textwidth]{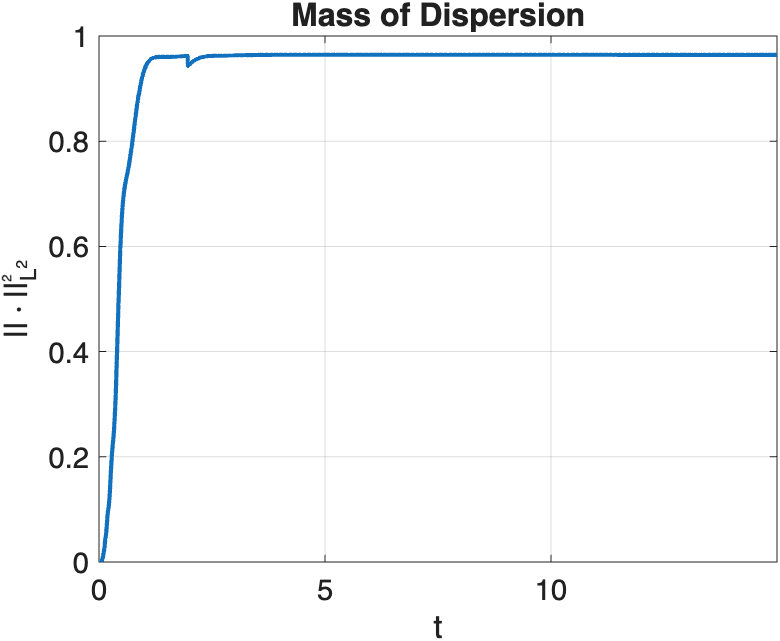} \hspace{15pt}
\includegraphics[width=0.29\textwidth]{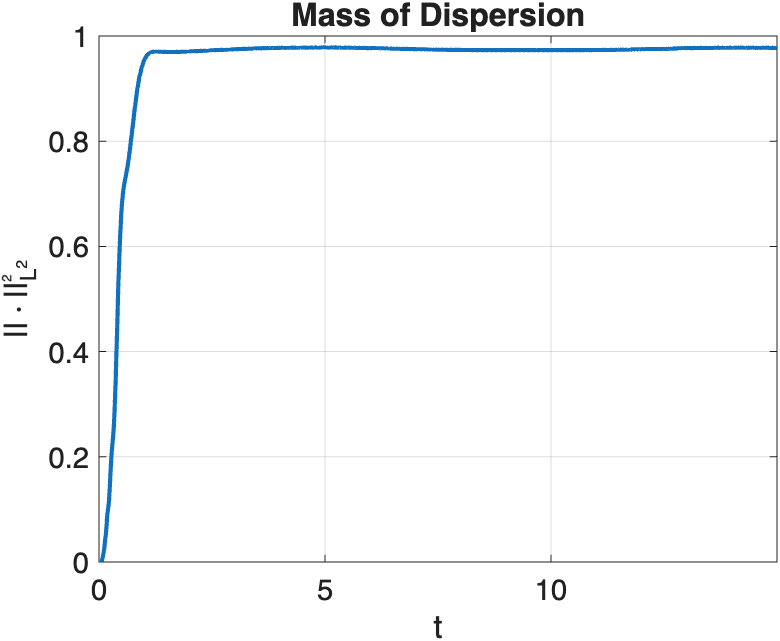} \hspace{15pt}
\includegraphics[width=0.29\textwidth]{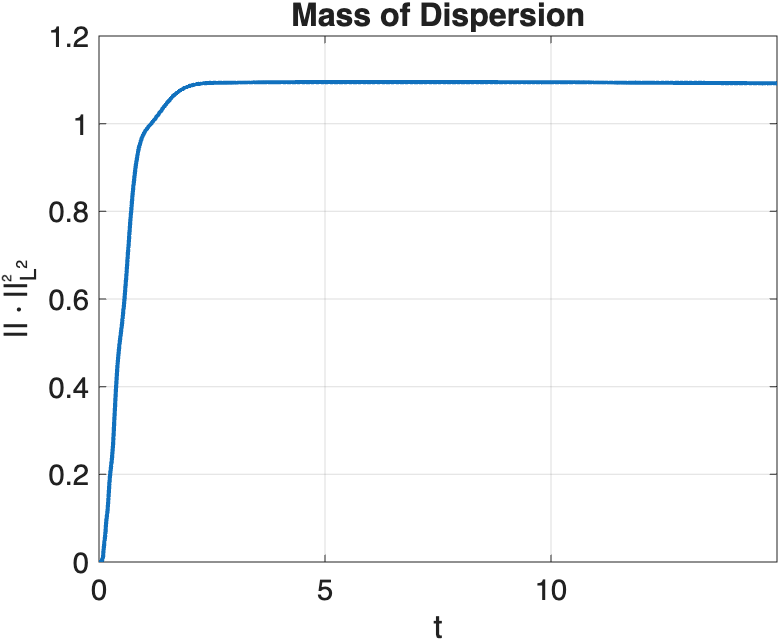}
\\\hspace{15pt} (a) \hspace{145pt} (b) \hspace{145pt} (c)
\caption{\label{1.2rm} Tracking mass of dispersion or radiation from perturbed breather $u_0(x) = 1.2B_{1,\beta}(x,0)$: (a) $\beta=1.15$, %$u_0(x) = 1.2B_{1,1.16}(x,0)$ 
(b) $\beta=1.16$, and %u_0(x) = 1.2B_{1,1.3}(x,0)$ 
(c) $\beta=1.3$.}
\end{center}
\end{figure}

These asymptotic values are used for profile matching in Figure \ref{1.2B}. 
In the first and second columns of Figure \ref{1.2abdg} (corresponding to the top and center rows of Figure \ref{1.2B}), the perturbed breather sheds radiation, which results in a steep decrease in $\alpha(t)$, a slight increase in $\delta(t)$, and a slight decrease in $\beta(t)$ and $\gamma(t)$, then all parameters converge to their respective asymptotic values (listed in the table above), representing the breather approaching its stable state. With $\beta=1.3$, Figure \ref{1.2B} shows that the breather breaks down into two solitons; this is reflected in $\alpha(t)$ converging to 0, $\beta(t)$ increasing slightly, and $\gamma(t)$ and $\delta(t)$ decreasing slightly. Then, the mass of the dispersion is calculated at each time step and plotted as a function of time in Figure \ref{1.2rm}. Table \ref{T:4} lists the initial $\beta$, initial total mass, the calculated local mass (of a breather trying to form), the mass of the dispersion or radiation, and the percentage of mass loss to radiation. 
\begin{comment}
{\small 
\begin{center}
\begin{tabular}{| c || c | c | c | c |} 
 \hline
 \multicolumn{5}{|c|}{Quantifying Radiation (for initial $AB_{1,\beta}(x,0)$ with $A=1.2$) } %, $\alpha=1$)} 
 \\
 \hline
 Initial & Total mass & Local Mass & Mass of Radiation & Percentage\\ 
 %\hline
 $\beta$ & $\|u\|^2_{L^2(\mathbb R)}$ & $\|B_{\alpha^\prime,\beta^\prime}\|^2_{L^2_{loc}}$ & 
 $\|u\|^2_{L^2(\mathbb R)} - \|B_{\alpha^\prime,\beta^\prime}\|^2_{L^2_{loc}}$ & of radiation mass\\ 
 \hline
 1.15 & 39.7440 & 38.7799 & 0.9641 & $2.43\%$ \\ 
 \hline
 1.16 & 40.0896 & 39.1122 & 0.9774& $2.44\%$ \\
 \hline
 1.3 & 44.9280 & 43.8361 & 1.0919 & $2.43\%$ \\
 \hline
\end{tabular} 
\end{center} 
}
\end{comment}
{\small
\begin{table}[ht]
\centering
\begin{tabular}{| c || c | c | c | c |} 
 \hline
 Initial & Total mass & Local Mass & Mass of Radiation & Percentage\\ 
 $\beta$ & $\|u\|^2_{L^2(\mathbb R)}$ & $\|B_{\alpha^\prime,\beta^\prime}\|^2_{L^2_{loc}}$ & 
 $\|u\|^2_{L^2(\mathbb R)} - \|B_{\alpha^\prime,\beta^\prime}\|^2_{L^2_{loc}}$ & of radiation mass\\ 
 \hline
 1.15 & 39.7440 & 38.7799 & 0.9641 & $2.43\%$ \\ 
 \hline
 1.16 & 40.0896 & 39.1122 & 0.9774& $2.44\%$ \\
 \hline
 1.3 & 44.9280 & 43.8361 & 1.0919 & $2.43\%$ \\
 \hline
\end{tabular}
\vspace{-6pt}
\caption{Quantifying Radiation for initial $AB_{1,\beta}(x,0)$ with $A=1.2$.}
\label{T:4}
\end{table}
}

Notice the curves in Figure \ref{1.2rm} are much smoother than in Figure \ref{1.1rm}. This is due to the negative $\gamma$ values, larger perturbation of $A=1.2$, and longer simulation runtime. Since $\gamma$ is negative, the perturbed breather travels right, in the opposite direction of the radiation.

\begin{figure}[h!]
\begin{center}
\includegraphics[width=0.27\textwidth]{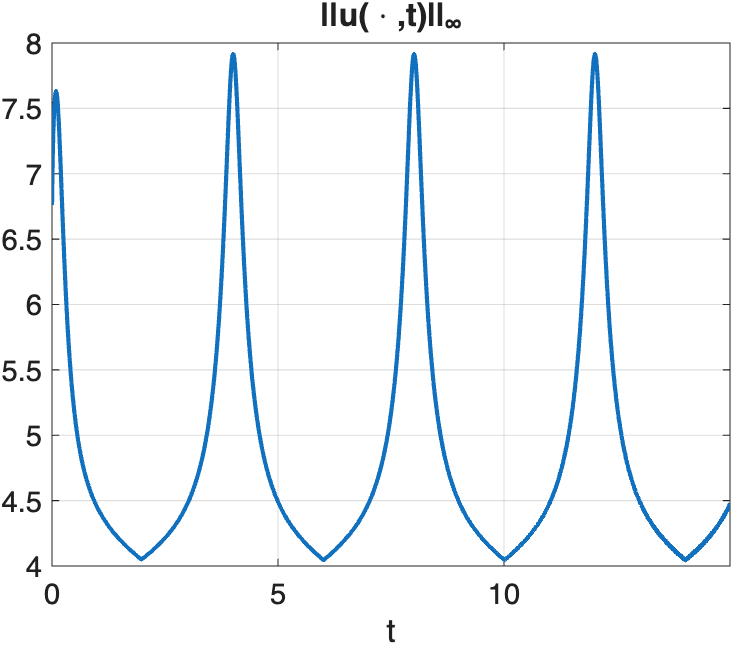} \hspace{20pt}
\includegraphics[width=0.27\textwidth]{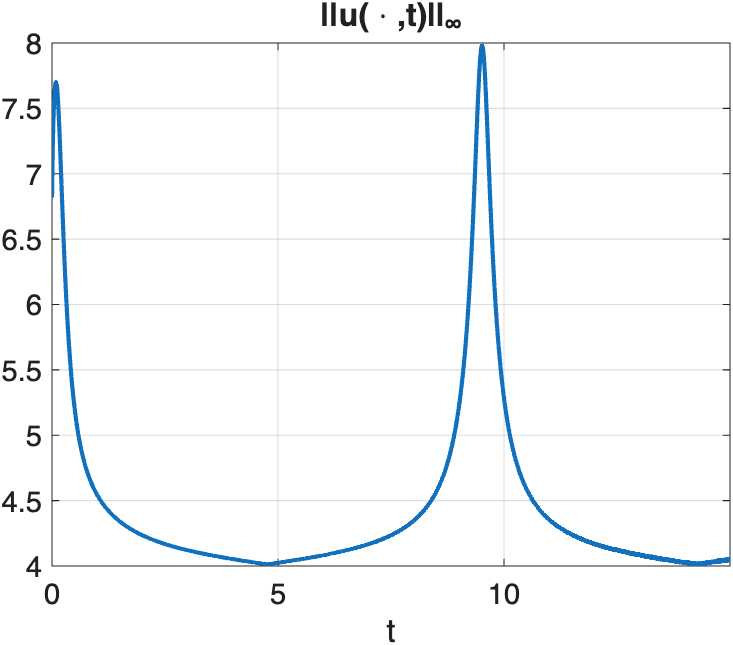} \hspace{20pt}
\includegraphics[width=0.27\textwidth]{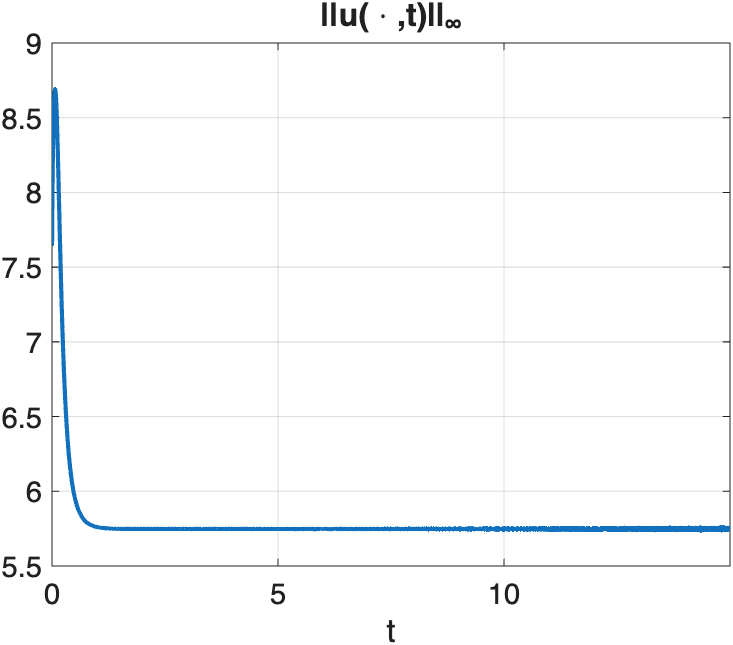}
\\\hspace{4pt} (a) \hspace{142pt} (b) \hspace{142pt} (c)
\caption{\label{1.2L} $L^{\infty}$ norms up to the time $t=15$ for $u_0(x) = 1.2B_{1, \beta}(x,0)$: (a) $\beta=1.15$, %$u_0(x) = 1.2B_{1,1.16}(x,0)$ 
(b) $\beta=1.16$, and 
%$u_0(x) = 1.2B_{1,1.3}(x,0)$ 
(c) $\beta=1.3$.}
\end{center}
\end{figure}

In Figure \ref{1.2L}, observe the effects of changing the $\beta$ values on the $L^\infty$ norms. In subplots (a) and (b), the $L^{\infty}$ norms closely align to the usual structure of breathers, i.e., periodic in time. The breather with the fastest period corresponds to subplot (a), then the breather with a slower period corresponds to subplot (b). If initial $\beta$ is significantly larger, the period slows down significantly and eventually periodicity disappears; this corresponds to subplot (c), where the breather structure breaks down and fails to breathe, splitting into two solitons (one positive and one negative). The $L^\infty$ norm in subplot (c) tracks the height of the tallest bump, i.e., the positive soliton. 

The results in this part are generated by perturbing the mKdV breather with a coefficient of $A=1.2$ and varying the parameter $\beta$: this allows us to investigate how $\beta$ controls the structure of a breather and its effect on the eventual breather stability and dynamics. For $A=1.2$, we can find the critical $\beta^*$ such that for $\beta<\beta^*$, the perturbed breather will shed radiation left and converge to an asymptotic final state (see Figure \ref{1.2B} with $\beta=1.15 \; \text{and} \; 1.16$). For $\beta>\beta^*$, the perturbed breather sheds radiation left, then a kink-type of bend forms between the positive and negative bumps, and the breather structure is lost, splitting into a positive soliton and a negative soliton (see Figure \ref{1.2B} with $\beta=1.3$). Note that the breakdown into two solitons resembles another solution to the mKdV equation, known as \textit{double pole solutions}. At the numerically observed threshold $\beta=\beta^*$, the perturbed breather is expected to converge to a \textit{double pole solution}. Reasoning note: this threshold behavior is supported numerically here, but a rigorous proof of convergence at the exact critical value is not provided. It would be interesting to investigate the connection between perturbed breathers and double pole solutions, since it is known that as $\alpha$ converges to $0$, the breather solution converges to a double pole solution \cite{am2013}.  
\smallskip

$\blacklozenge$ \underline{$A=1.3$:}~
Next, we investigate the effects of varying $\alpha$ with even higher amplitude perturbations. We consider the same initial data as in the previous example, but take $A=1.3$ and fix $\beta=1$. The computational parameters here are $L=200\pi$, $N=2^{16}$, and $dt=0.0025$. 
\begin{figure}[h!]
\begin{center}
\vspace{8pt}
\includegraphics[width=0.22\textwidth]{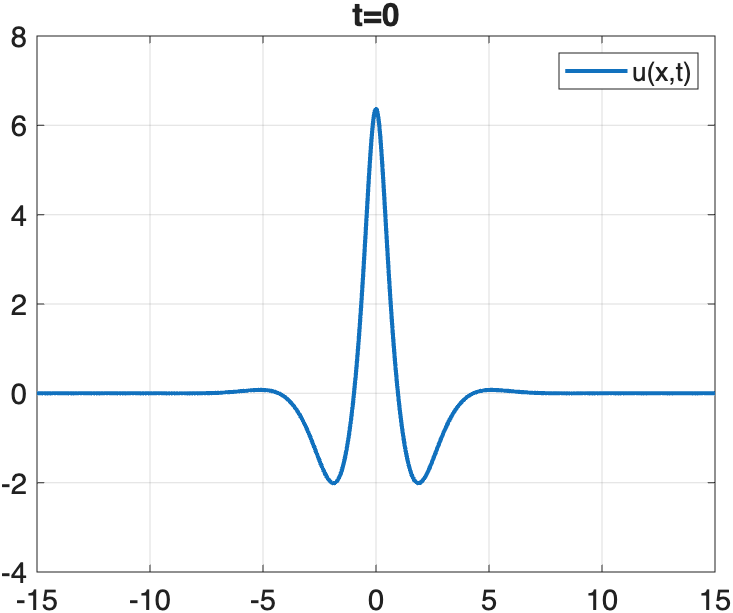} \hspace{12pt} 
\includegraphics[width=0.22\textwidth]{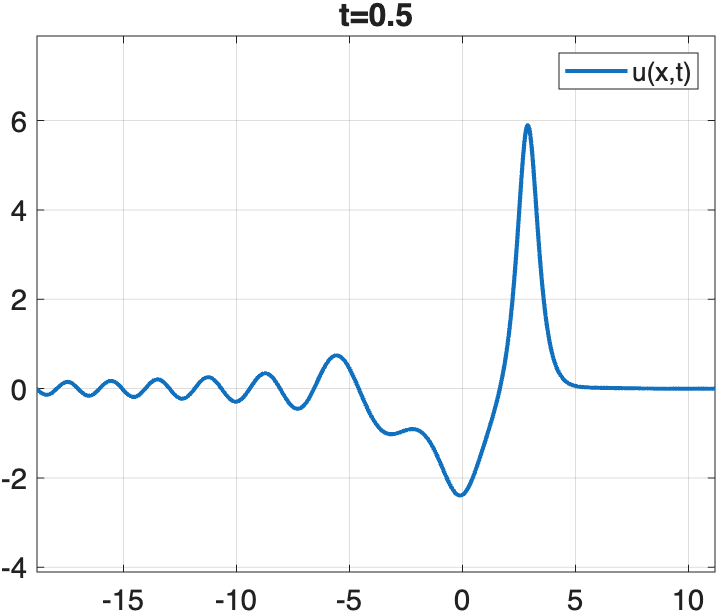} \hspace{12pt} 
\includegraphics[width=0.22\textwidth]{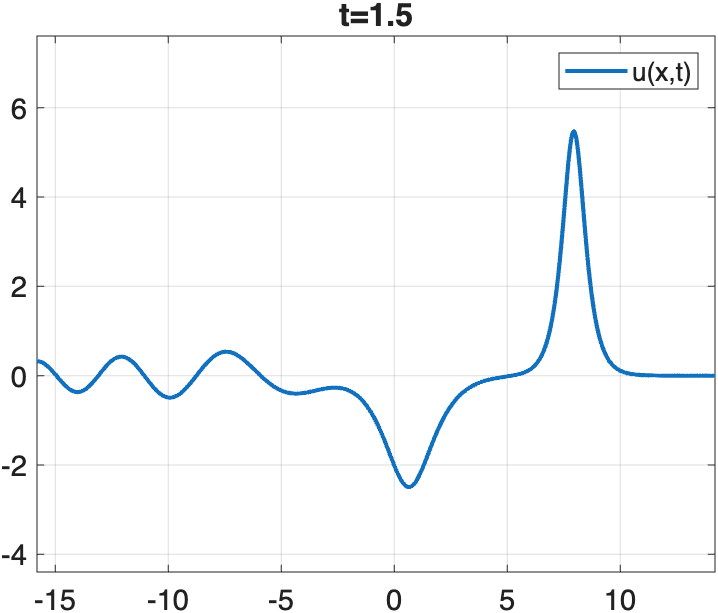} \hspace{12pt} 
\includegraphics[width=0.22\textwidth]{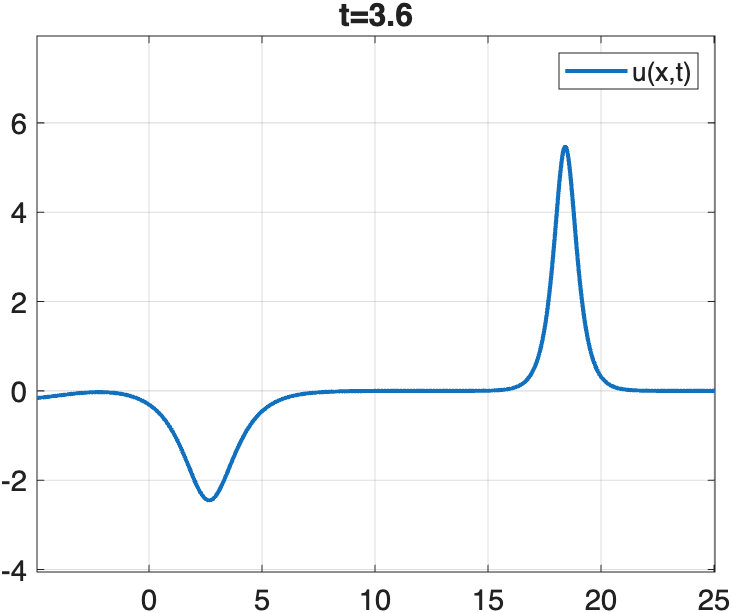} \\ \vspace{10pt}
\includegraphics[width=0.22\textwidth]{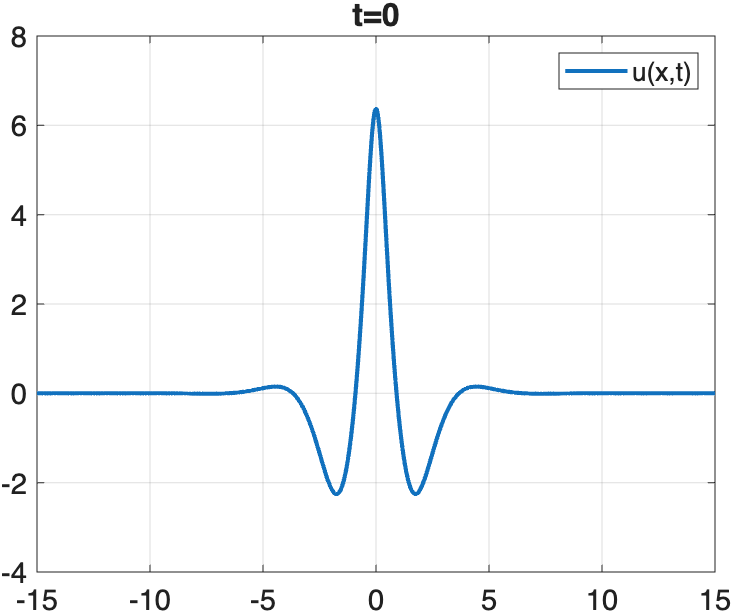} \hspace{12pt} 
\includegraphics[width=0.22\textwidth]{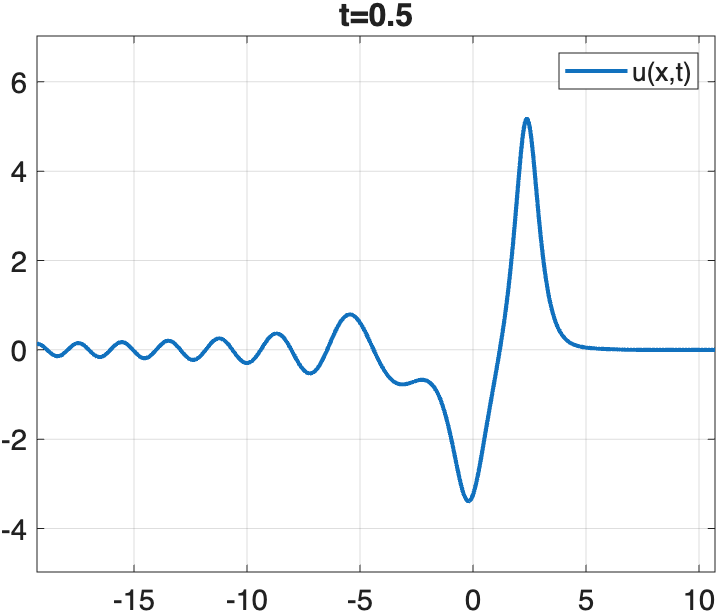} \hspace{12pt} 
\includegraphics[width=0.22\textwidth]{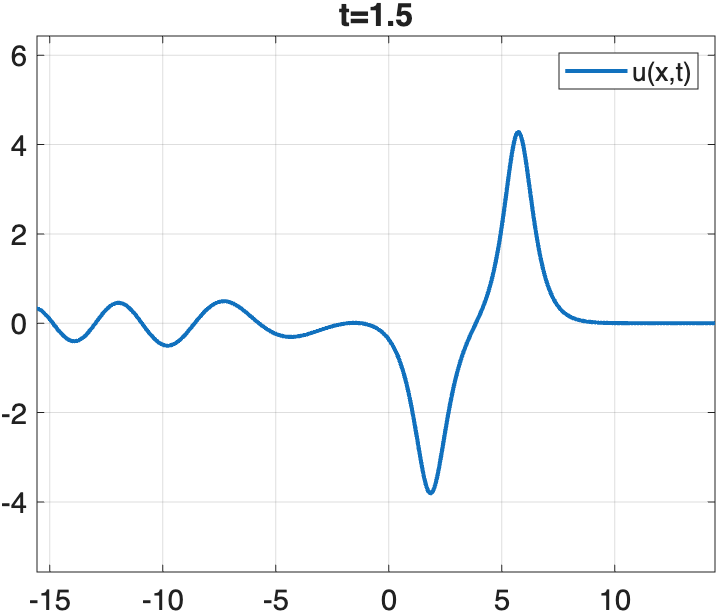} \hspace{12pt} 
\includegraphics[width=0.22\textwidth]{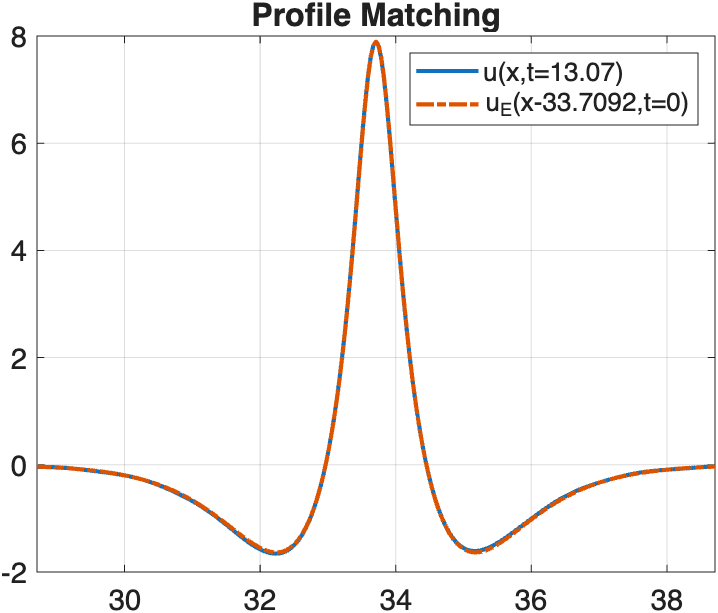} \\ \vspace{10pt}
\includegraphics[width=0.22\textwidth]{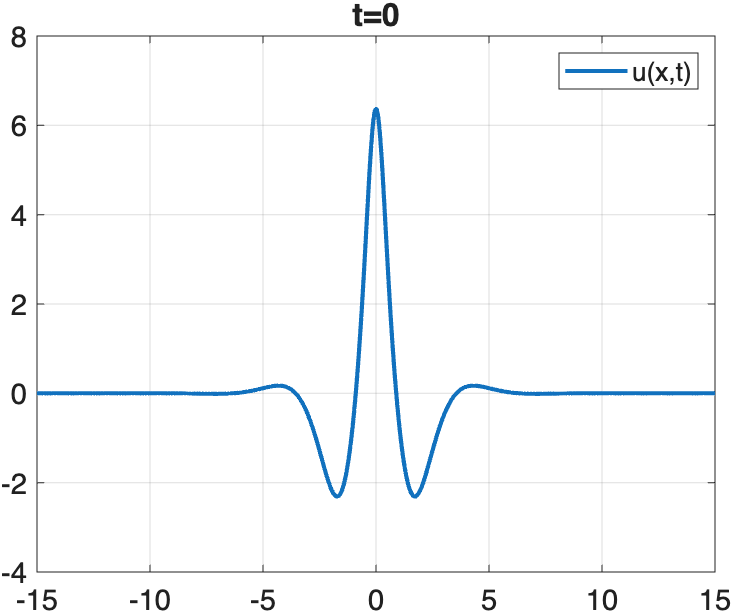} \hspace{12pt} 
\includegraphics[width=0.22\textwidth]{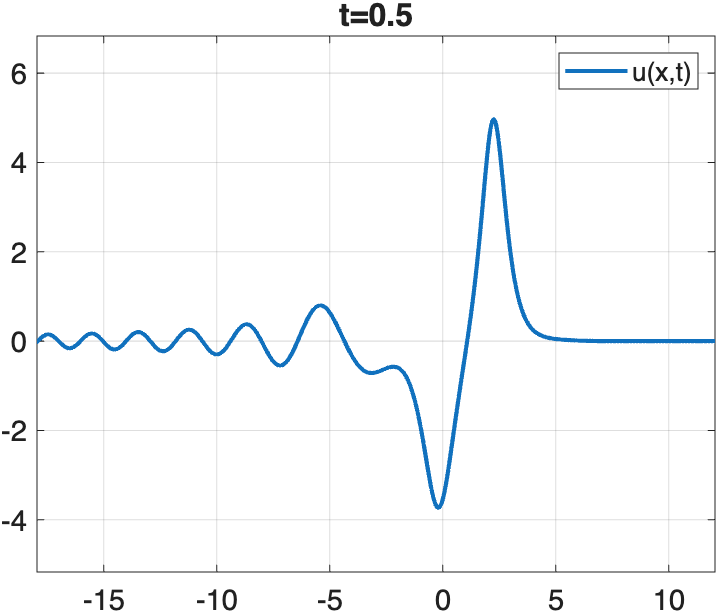} \hspace{12pt} 
\includegraphics[width=0.22\textwidth]{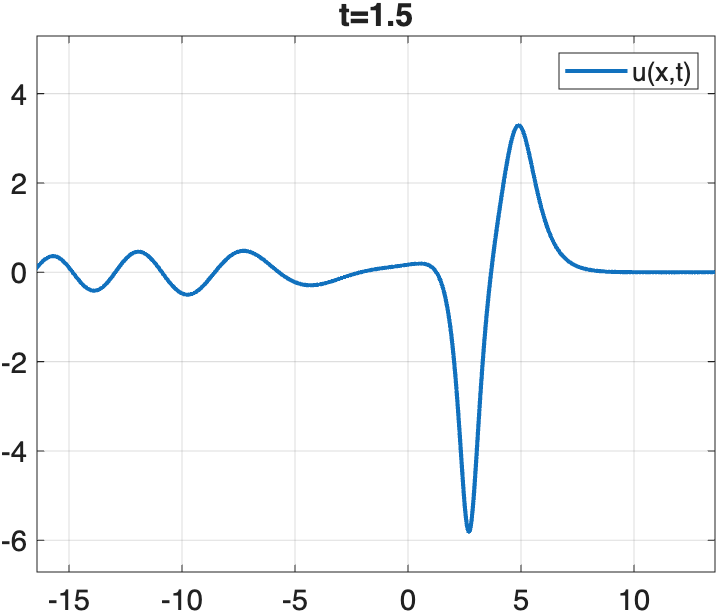} \hspace{12pt} 
\includegraphics[width=0.22\textwidth]{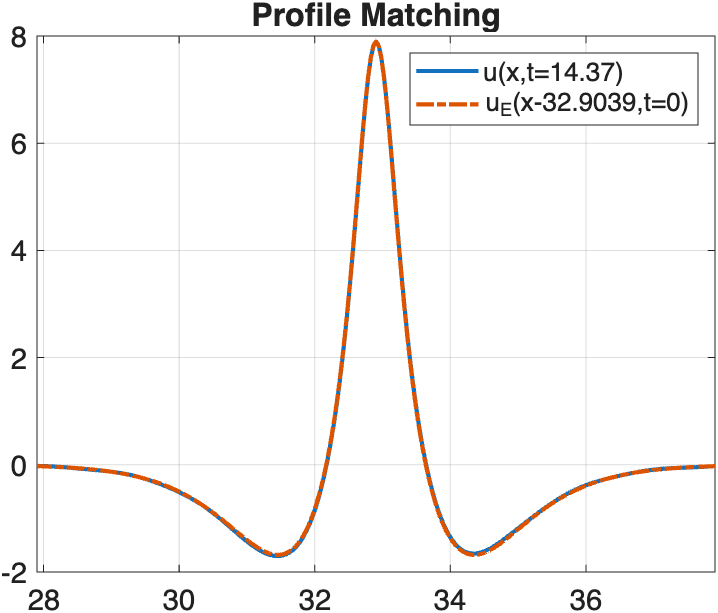} \\
\caption{\label{1.3B} {\small Time evolution of $u_0(x)=1.3B_{0.9,1}(x,0)$ (top row), $u_0(x)=1.3B_{1.08,1}(x,0)$ (center row) and profile matching with the explicit breather $u_E(x)=B_{0.093433, 1.6122}(x-33.7092,0)$ (last subplot), and $u_0(x)=1.3B_{1.12,1}(x,0)$ (bottom row) and profile matching with the explicit breather $u_E(x)=B_{0.32415,1.6123}(x-32.9039,0)$ (last subplot).} }
\end{center}
\end{figure}
In Figure \ref{1.3B}, the time evolution of the breathers each with different values $\alpha=0.9$, $1.08$, and $1.12$, respectively, are shown. Recall that the parameter $\alpha$ controls the number of internal oscillations of the breather profile: larger $\alpha$ values produce more internal oscillations in the profile. 

In the first row of Figure \ref{1.3B}, $\alpha=0.9$ and one can observe a similar evolution to the example in Figure \ref{1.2B} (bottom row) with $\beta=1.3$. The breather starts by shedding radiation to the left, a kink-type bend forms between the positive and negative bumps at $t=0.5$, then by $t=1.5$, the breather structure breaks down, separating into one positive and one negative soliton. At $t=3.6$, the solitons separate more in time due to their difference in height, and hence, difference in speed. 

\begin{figure}[h!]
\begin{centering}
\includegraphics[width=0.26\textwidth]{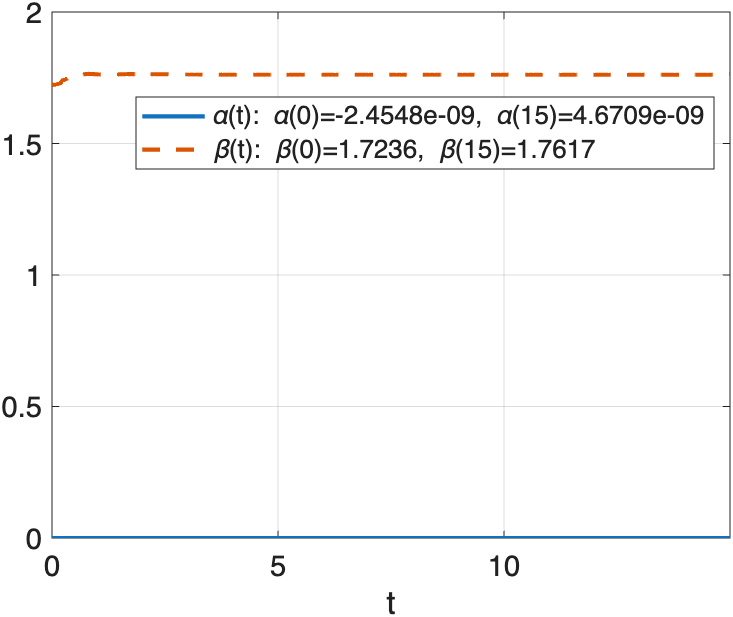} \hspace{20pt}
\includegraphics[width=0.26\textwidth]{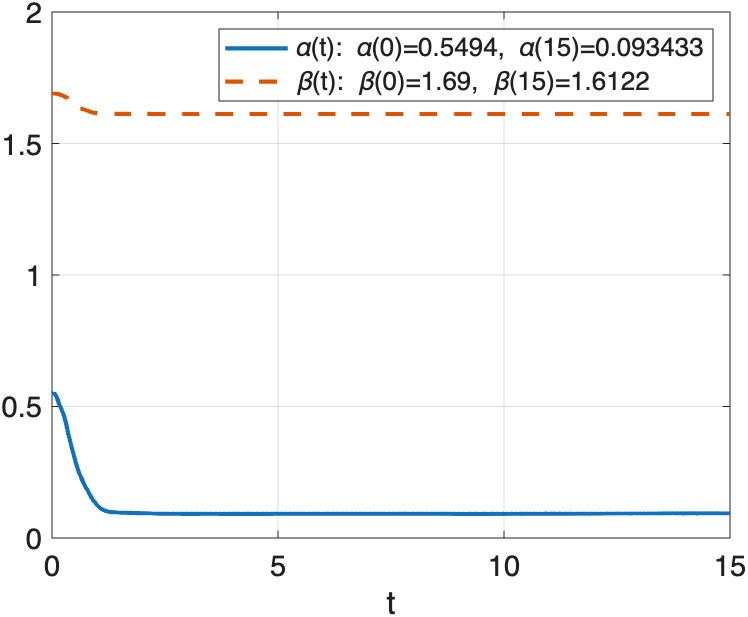} \hspace{20pt}
\includegraphics[width=0.26\textwidth]{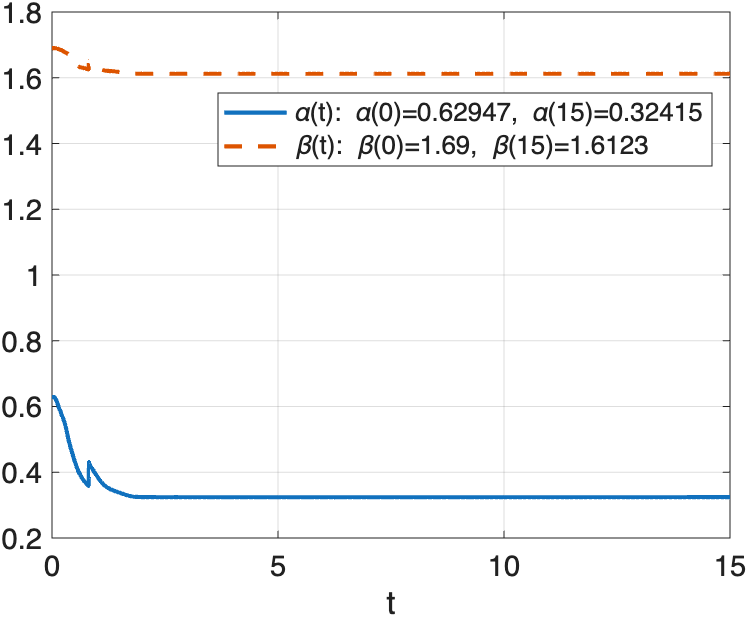}\\ \vspace{5pt} 
\includegraphics[width=0.26\textwidth]{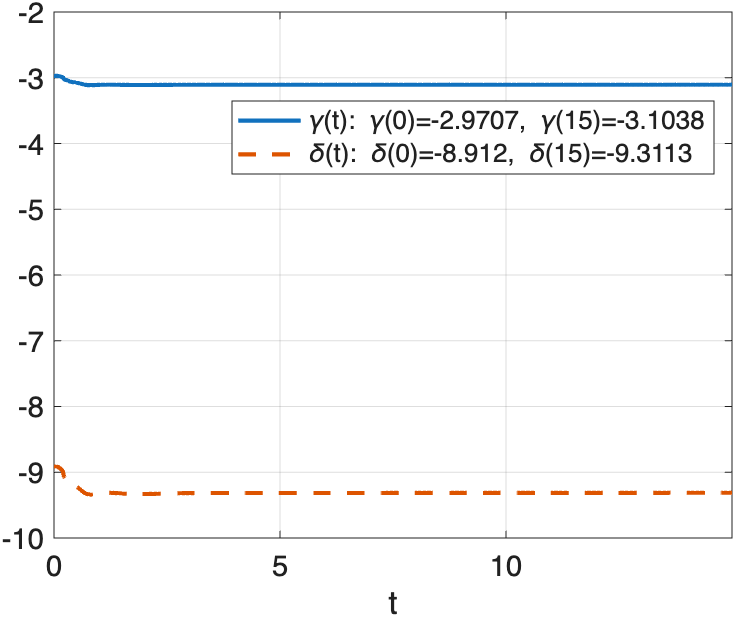} \hspace{23pt}
\includegraphics[width=0.255\textwidth]{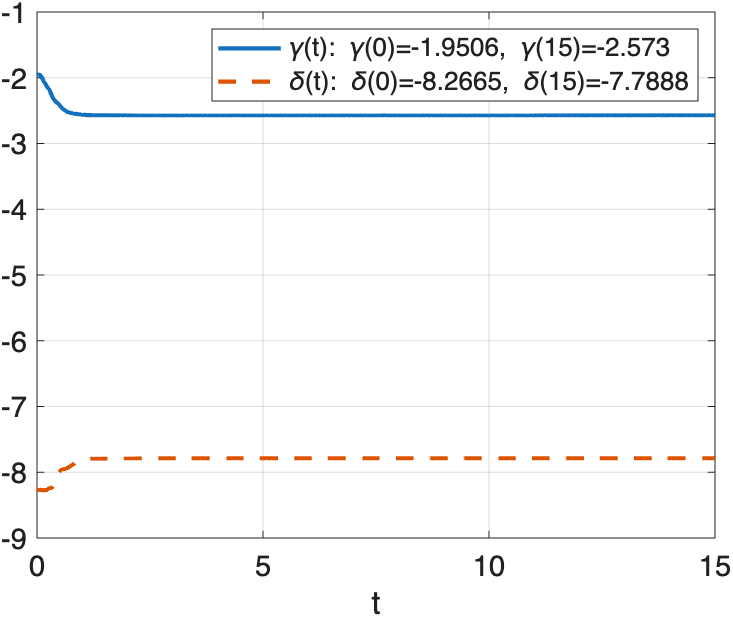} \hspace{23pt}
\includegraphics[width=0.255\textwidth]{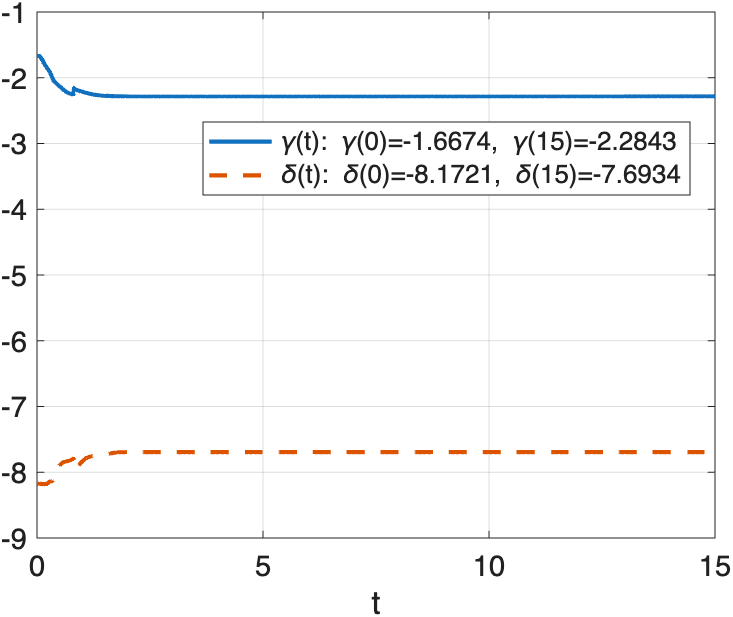} 
\caption{\label{1.3abdg} {\small Tracking $\alpha(t)$, $\beta(t)$, $\delta(t)$, and $\gamma(t)$ for time evolution of $u_0(x)=1.3 B_{\alpha,1}(x,0)$: $\alpha = 0.9$ (left column), %$u_0(x)=1.3B_{1.08,1}(x,0)$ 
$\alpha=1.08$ (center column), and 
$\alpha=1.12$ 
%$u_0(x)=1.3B_{1.12,1}(x,0)$ 
(right column).} }
\end{centering}
\end{figure}

In the second row of Figure \ref{1.3B}, $\alpha$ is increased to $1.08$, and again, the breather sheds radiation to its left. At $t=0.5$, the amplitude of the negative bump is greater than it was for $\alpha=0.9$ at $t=0.5$. For this reason, there is less of a bend between the positive and negative bumps. At $t=1.5$, the bend slightly deepens, but the positive and negative bumps remain close, which means the breather is maintaining its structure and will successfully breathe in its time evolution. At $t=1.5$, we see that the positive and negative bumps have nearly equal heights; this visually suggests that the breather structure will stay intact, but it is near the limit between `breathing' and breaking down into two separate solitons. Then, at $t=13.07$, the breather completes its first full period; we do the profile matching at that time to the explicit breather, 
$$
u_E(x)=B_{0.093433, 1.6122}(x-33.7092,0),
$$ 
where $\alpha^\prime$ and $\beta^\prime$ are chosen based on the tracking, see center column of Figure \ref{1.3abdg}.

\begin{figure}[h!]
\begin{center}
\includegraphics[width=0.27\textwidth]{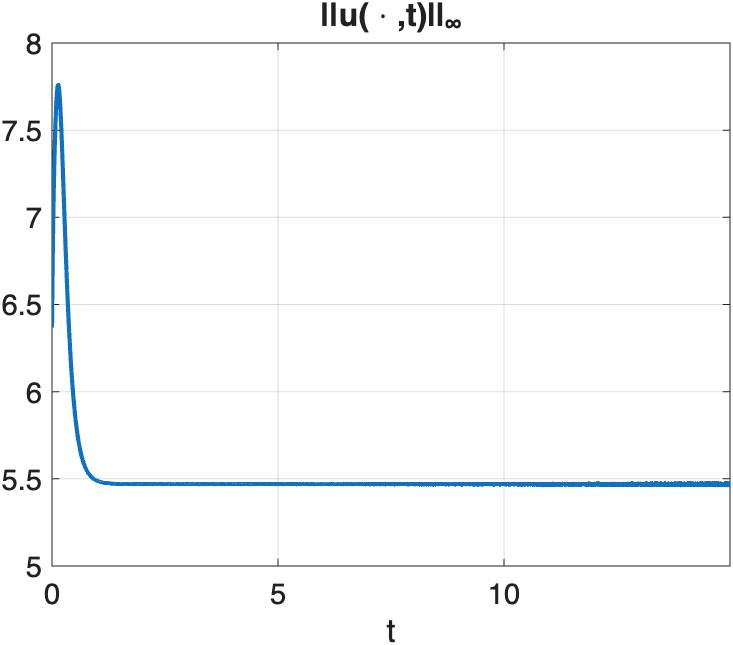} \hspace{20pt}
\includegraphics[width=0.27\textwidth]{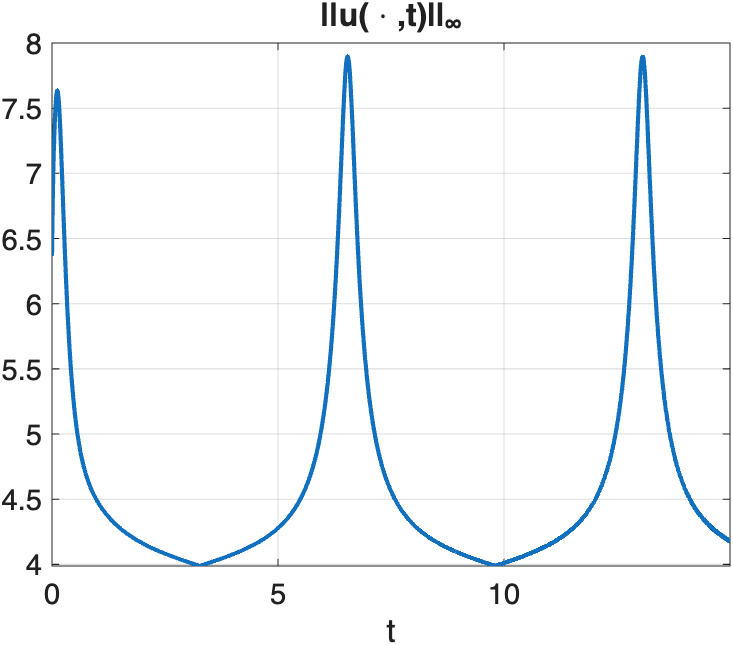} \hspace{20pt}
\includegraphics[width=0.27\textwidth]{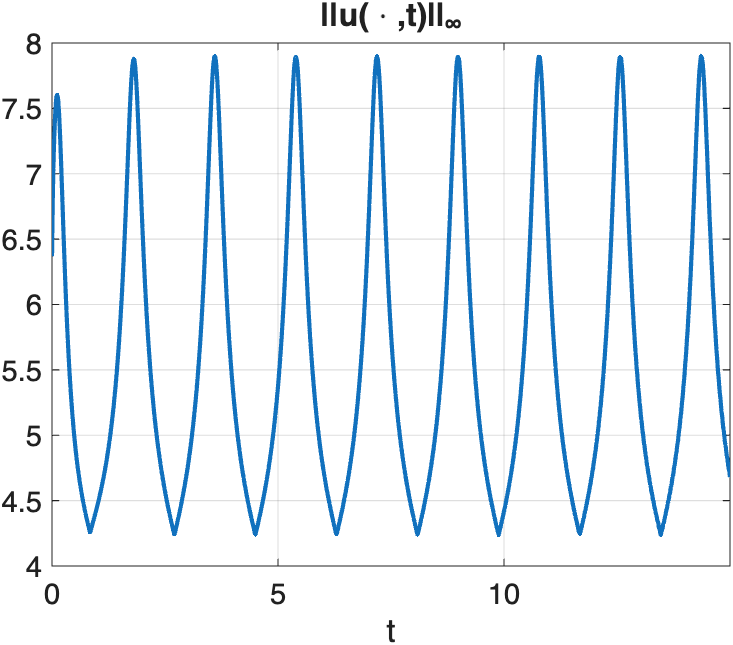}
\\\hspace{5pt} (a) \hspace{141pt} (b) \hspace{142pt} (c)
\caption{\label{1.3L}{\small $L^{\infty}$ norms up to the time $t=15$ for $u_0(x)=1.3B_{\alpha,1}(x,0)$: 
(a) $\alpha=0.9$, 
(b) $\alpha=1.08$, %u_0(x)=1.3B_{1.08,1}(x,0)$ (b), 
and (c) $\alpha = 1.12$. %$u_0(x)=1.3B_{1.12,1}(x,0)$.
}}
\end{center}
\end{figure}

In the last row of Figure \ref{1.3B}, $\alpha=1.12$ and the evolution of this perturbed breather is much quicker in shedding its radiation, then trying to stabilize and travel to the right. This breather completes 4 full periods by $t=14.37$, see the $L^{\infty}$ norm in subplot (c) of Figure \ref{1.3L}. At that time $t=14.37$ the evolved breather is matched to an explicit breather generated by the tracked $\alpha^\prime$ and $\beta^\prime$ values. Observe that at $t=0.5$, there is no bend between the positive and negative bumps as it happened in the above cases. Furthermore, the bumps are closer together than in the previous rows with $\alpha=0.9\;\text{ and} \; 1.08$, i.e., the `breathing' happens faster. Comparing the cases in Figure \ref{1.3B}, we find that varying $\alpha$ has similar results as varying $\beta$ with perturbation $A>1$ in Figure \ref{1.2B}. Numerically, we find that the critical $\alpha^*$ is approximately $1.08$, i.e., for initial $\alpha<\alpha^*$, the perturbed breather breaks down into one positive and one negative soliton, and for $\alpha>\alpha^*$, the perturbed breather converges to an asymptotically stable breather that can be matched to an explicit breather.

In Figure \ref{1.3abdg}, we plotted time evolution of parameters $\alpha(t)$, $\beta(t)$, $\delta(t)$, and $\gamma(t)$, which correspond to the 3 cases from Figure \ref{1.3B}. Table \ref{T:5} lists the asymptotic values for each case of initial $\alpha$. 
\begin{comment}
{\small 
\begin{center}
\begin{tabular}{| c || c | c | c | c |} 
\hline
\multicolumn{5}{|c|}{Asymptotic Values (from initial $A=1.3$, $\beta=1$)} \\
\hline
 Initial $\alpha$ & $\alpha^\prime$ & $\beta^\prime$ & $\gamma^\prime$ & $\delta^\prime$\\ 
 \hline
 0.90 & 4.6709e-09 & 1.7617 & -3.1038 & -9.3113 \\ 
 \hline
 1.08 & 0.093433 & 1.6122 & -2.5730 & -7.7888 \\
 \hline
 1.12 & 0.32415 & 1.6123 & -2.2843 & -7.6934 \\
 \hline
\end{tabular}
\end{center} 
}
\end{comment}
{\small
\begin{table}[ht]
\centering
\begin{tabular}{| c || c | c | c | c |} 
 \hline
 Initial $\alpha$ & $\alpha^\prime$ & $\beta^\prime$ & $\gamma^\prime$ & $\delta^\prime$\\ 
 \hline
 0.90 & 4.6709e-09 & 1.7617 & -3.1038 & -9.3113 \\ 
 \hline
 1.08 & 0.093433 & 1.6122 & -2.5730 & -7.7888 \\
 \hline
 1.12 & 0.32415 & 1.6123 & -2.2843 & -7.6934 \\
 \hline
\end{tabular}
\vspace{-6pt}
\caption{Asymptotic Values from initial $A=1.3$, $\beta=1$.}
\label{T:5}
\end{table}
}

Recall in Figure \ref{1.2B} and \ref{1.2abdg}, as the initial $\beta$ is increased, it makes $\beta^\prime$ increase and $\alpha^\prime$ decrease. Here, for initial $\alpha<\alpha^*$, $\beta(t)$ increases and $\gamma^\prime$, $\delta^\prime$ decrease. For $\alpha>\alpha^*$, $\alpha(t)$, $\beta(t)$, $\gamma(t)$ decrease and $\delta(t)$ increases. Also, notice that for initial $\alpha>\alpha^*$, $\alpha(0)$ increases while $\beta(0)=1.69$, see the top row of the center and right columns of Figure \ref{1.3abdg}.

\begin{figure}[h!]
\begin{center}
\includegraphics[width=0.29\textwidth]{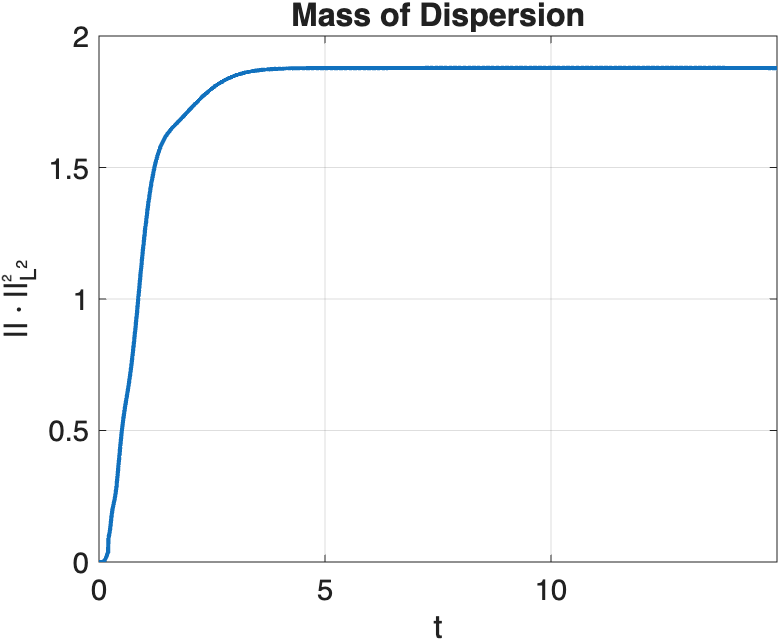} \hspace{10pt}
\includegraphics[width=0.29\textwidth]{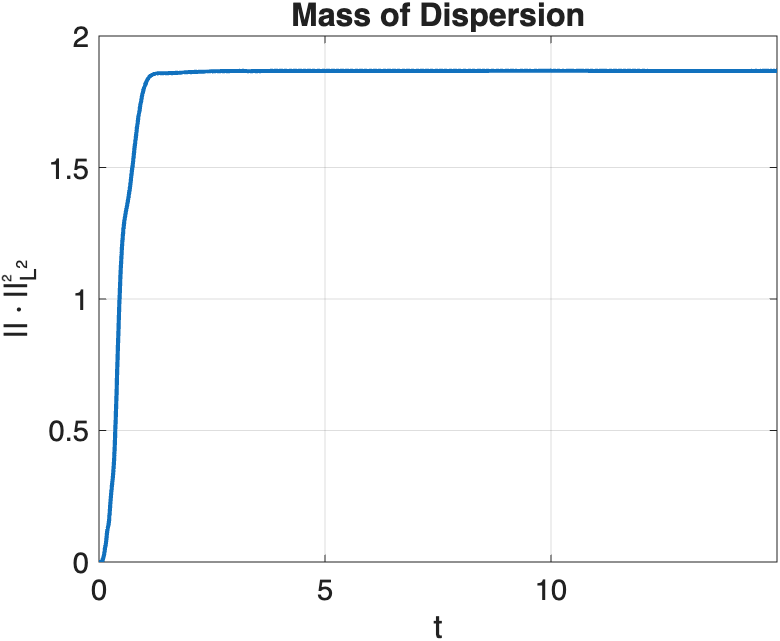} \hspace{10pt}
\includegraphics[width=0.29\textwidth]{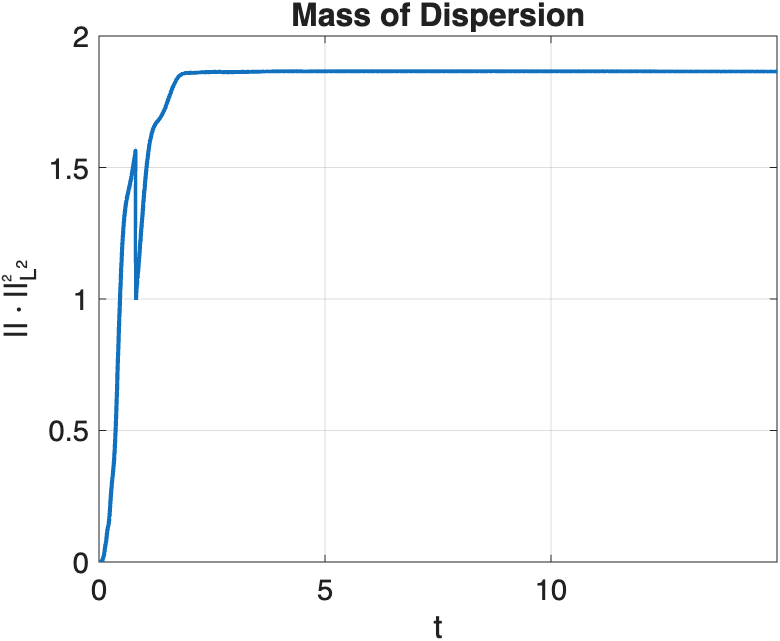}
\\\hspace{15pt} (a) \hspace{140pt} (b) \hspace{140pt} (c)
\caption{\label{1.3rm} {\small Tracking mass of dispersion or radiation from the perturbed initial data $u_0(x)=1.3B_{\alpha,1}(x,0)$: 
(a) $\alpha=0.9$, %$u_0(x)=1.3B_{1.08,1}(x,0)$ 
(b) $\alpha=1.08$, and 
(c) $\alpha=1.12$. %$u_0(x)=1.3B_{1.12,1}(x,0)$.
} }
\end{center}
\end{figure}

In Figure \ref{1.3rm}, the mass of the radiation or dispersion is calculated at each time step and plotted as a function of time. Table \ref{T:6} lists the values for the initial mass, local mass, radiation mass, and the percentage of loss due to radiation.
\begin{comment}
{\small
\begin{center}
\begin{tabular}{| c || c | c | c | c |} 
 \hline
 \multicolumn{5}{|c|}{Quantifying Radiation (for initial $AB_{\alpha,1}(x,0)$ with $A=1.3$) } \\
 \hline
 Initial & Total mass & Local Mass & Mass of Radiation & Percentage\\ 
 %\hline
 $\alpha$ & $\|u\|^2_{L^2(\mathbb R)}$ & $\|B_{\alpha^\prime,\beta^\prime}\|^2_{L^2_{loc}}$ & 
 $\|u\|^2_{L^2(\mathbb R)} - \|B_{\alpha^\prime,\beta^\prime}\|^2_{L^2_{loc}}$ & of radiation mass\\ 
 \hline
 0.90 & 40.5600 & 38.6819 & 1.8781 & $4.63\%$ \\ 
 \hline
 1.08 & 40.5600 & 38.6928 & 1.8672 & $4.60\%$ \\
 \hline
 1.12 & 40.5600 & 38.6950 & 1.8650 & $4.60\%$ \\
 \hline
\end{tabular} 
\end{center} 
}
\end{comment}
{\small
\begin{table}[ht]
\centering
\begin{tabular}{| c || c | c | c | c |} 
 \hline
 Initial & Total mass & Local Mass & Mass of Radiation & Percentage\\ 
 $\alpha$ & $\|u\|^2_{L^2(\mathbb R)}$ & $\|B_{\alpha^\prime,\beta^\prime}\|^2_{L^2_{loc}}$ & 
 $\|u\|^2_{L^2(\mathbb R)} - \|B_{\alpha^\prime,\beta^\prime}\|^2_{L^2_{loc}}$ & of radiation mass\\ 
 \hline
 0.90 & 40.5600 & 38.6819 & 1.8781 & $4.63\%$ \\ 
 \hline
 1.08 & 40.5600 & 38.6928 & 1.8672 & $4.60\%$ \\
 \hline
 1.12 & 40.5600 & 38.6950 & 1.8650 & $4.60\%$ \\
 \hline
\end{tabular}
\vspace{-6pt}
\caption{Quantifying Radiation for initial condition $AB_{\alpha,1}(x,0)$ with $A=1.3$.}
\label{T:6}
\end{table}
}
Recall in Figure \ref{1.2rm}, as initial $\beta$ increases, the initial mass, local mass, and dispersion/radiation mass increase along with the initial $\beta$. Here, as initial $\alpha$ increases, the initial mass is fixed at $\|u_0\|^2_{L^2}=40.56$, while the local mass and mass of dispersion both decrease. 

The $L^{\infty}$ norms are included in Figure \ref{1.3L} to show the periodicity in each of the cases. For $\alpha=0.9$ (less than critical $\alpha^*$), the evolution splits into two solitons and loses its breather structure, then the $L^\infty$ norm continues to track the tallest soliton (in this case, the positive soliton). In the case of $\alpha=1.08$ (close to critical $\alpha^*$), the perturbed breather converges to an asymptotically stable breather with a slow period. For our last example of $\alpha=1.12$ (greater than $\alpha^*$), the perturbed breather converges to a stable breather with faster periodicity.
\smallskip

In this part, we investigated the effects of perturbing the amplitude of an exact breather and varying its parameters $\alpha$ and $\beta$. Table \ref{T:7} lists the critical $\alpha^*$ and $\beta^*$ values, corresponding to each amplitude perturbation value $A$.
\begin{comment}
{\footnotesize
\begin{center}
\begin{tabular}{| c || c | c |} 
 \hline
 \multicolumn{3}{|c|}{Approximate Critical Values} \\
 \hline
 Perturbation A: & Critical $\alpha^*$ ($\beta=1$): & Critical $\beta^*$ ($\alpha=1$): \\ 
 \hline
 1.1 & 0.60 & 1.68 \\ 
 \hline
 1.2 & 0.87 & 1.16 \\
 \hline
 1.3 & 1.08 & 0.92 \\
 \hline
 1.4 & 1.27 & 0.78 \\
 \hline
 1.5 & 1.45 & 0.69 \\
 \hline
\end{tabular} 
\end{center} 
%\vspace{10pt}
}
\end{comment}
{\footnotesize
\begin{table}[ht]
\centering
\begin{tabular}{| c || c | c |} 
 \hline
 Perturbation A: & Critical $\alpha^*$ ($\beta=1$): & Critical $\beta^*$ ($\alpha=1$): \\ 
 \hline
 1.1 & 0.60 & 1.68 \\ 
 \hline
 1.2 & 0.87 & 1.16 \\
 \hline
 1.3 & 1.08 & 0.92 \\
 \hline
 1.4 & 1.27 & 0.78 \\
 \hline
 1.5 & 1.45 & 0.69 \\
 \hline
\end{tabular}
\vspace{-6pt}
\caption{Approximate Critical Values}
\label{T:7}
\end{table}
}

For perturbations $A>1$, critical values of $\alpha^*$ and $\beta^*$ indicate the values, at which the perturbed breathers will either break down into two solitons (one negative and one positive) or converge to an asymptotically stable mKdV breather (or converge to the mKdV {\it double pole} solution). For $\alpha^*$, the initially perturbed breather will stabilize if initial $\alpha>\alpha^*$ and break down if initial $\alpha<\alpha^*$. For $\beta^*$, the perturbed breather will stabilize if initial $\beta<\beta^*$ and will break down if the initial $\beta>\beta^*$. From the table, one can note that for higher perturbations of the initial data, the critical $\alpha^*$ increases and $\beta^*$ decreases. For larger perturbations, the profile of the solution is stretched, increasing the height. Then, to avoid disappearing and encourage stable solutions, a smaller initial $\beta$ is needed because smaller $\beta$ combats the increase in height, or a greater initial $\alpha$ is needed to introduce more oscillations in the profile, which encourages stability. Note that these critical values are approximate (to order of $\mathcal{O}(h^2)$). With higher numerical accuracy, these critical-value estimates can be refined. At the critical threshold, the numerics suggest that the perturbed breather sheds radiation to the left and then converges to a \textit{double pole} solution; this remains a numerical observation rather than a proof. This behavior of a breather solution converging to a double pole solution is known to happen when $\alpha$ converges to $0$, \cite{am2013}; we have shown in this section and Figure \ref{1.2B} and \ref{1.3B} that $\alpha^\prime$ converges to $0$ for a range of perturbations and initial $\alpha$ and $\beta$ values. 

%As an outlook, 
Looking ahead, it would be interesting to investigate exact critical values of $\alpha$ and $\beta$ to understand the connection between breather solutions and double pole solutions in the mKdV equation \eqref{mKdV}.

\section{Perturbations of the nonlinear term}\label{combined}

\subsection{The mKdV with combined nonlinearities}
While the mKdV \eqref{mKdV} is well-known for having coherent structures such as solitons and breathers, we study the effects of perturbations of the nonlinear term on the existence and stability of breather-type solutions. Given a perturbation strength $\epsilon$, we consider two types of nonlinear term perturbations:
$$
\mathcal{N}_1(u)=u^2u_x+\epsilon\, u^{p-1}u_x,
$$ 
and 
$$
\mathcal{N}_2(u)=u^2u_x+\epsilon\, |u|^{p-1}u_x.
$$ 
In the first case, the perturbed mKdV equation \eqref{mKdV-n} becomes
\begin{equation}\label{E:Per1}
u_t + \big(u_{xx}+\tfrac13u^3+\epsilon \tfrac1{p} u^p\big)_x = 0,
\end{equation}
and in the second case, since $u$ is real-valued, it can be written as 
\begin{equation}\label{E:Per2}
u_t + \big(u_{xx}+\tfrac13u^3+\epsilon \tfrac1{p}|u|^{p-1}u\big)_x = 0.
\end{equation}
The mass in both equations is conserved and is given by the same quantity as in \eqref{mass}.
Although the energies for the solutions of the above equations differ from the mKdV energy \eqref{energy}, they are still conserved. For equation \eqref{E:Per1}, the energy is given by
\begin{equation}\label{combinedenergy1}
E_1[u](t):=\frac{1}{2} \int_{\mathbb{R}}^{} u^2_x(t,x) \,dx - \frac{1}{12} \int_{\mathbb{R}}^{} u^4(t,x) \,dx - \frac{\epsilon}{p(p+1)} \int_{\mathbb{R}}^{} u^{(p+1)}(t,x) \,dx= E_1[u](0).
\end{equation}
For equation \eqref{E:Per2}, the corresponding energy is
\begin{equation}\label{combinedenergy2}
E_2[u](t):=\frac{1}{2} \int_{\mathbb{R}}^{} u^2_x(t,x) \,dx - \frac{1}{12} \int_{\mathbb{R}}^{} u^4(t,x) \,dx - \frac{\epsilon}{p(p+1)} \int_{\mathbb{R}}^{} |u(t,x)|^{p+1} \,dx= E_2[u](0).
\end{equation}

For this paper, we only consider $p=2$ and study breather behavior in both cases \eqref{E:Per1} and \eqref{E:Per2}. 
The first equation, when $p=2$, is a variant of the Gardner equation, which is integrable and, similar to the mKdV, known to have breathers \cite{a2018}. The second equation is non-integrable, though the mass and the energy are conserved.

\subsubsection{The mKdV with subcritical perturbation $p=2$ (a la Gardner equation).}
We take $p=2$ and $\epsilon=\frac{1}{10}$ in \eqref{E:Per1}, that is, 
$$
\mathcal{N}(u)=u^2u_x+\frac{1}{10}uu_x,
$$ 
and consider the mKdV breather \eqref{mKdV-B} as the initial condition 
$$
u_0(x)=B_{1,1}(x,0).
$$ 
The computational parameters are $L=300\pi$, $N=2^{16}$, and $dt=0.0025$. (While it is known that the Gardner equation has explicit breathers, for the purpose of this work and comparison, we start with the mKdV breathers, and do not rely on the explicit expression for a Gardner breather, so that the same diagnostic can also be applied to a non-integrable case.)

\begin{figure}[h!]
\begin{center}
\includegraphics[width=0.23\textwidth]{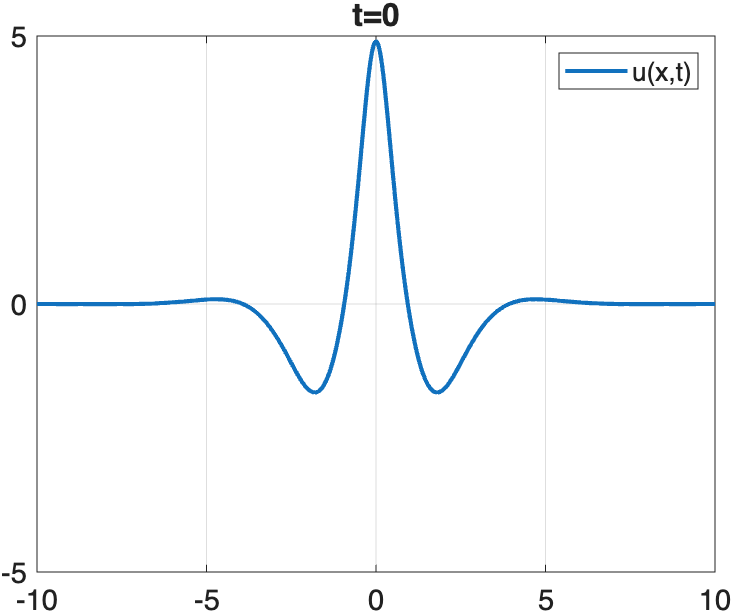} \hspace{2pt} 
\includegraphics[width=0.23\textwidth]{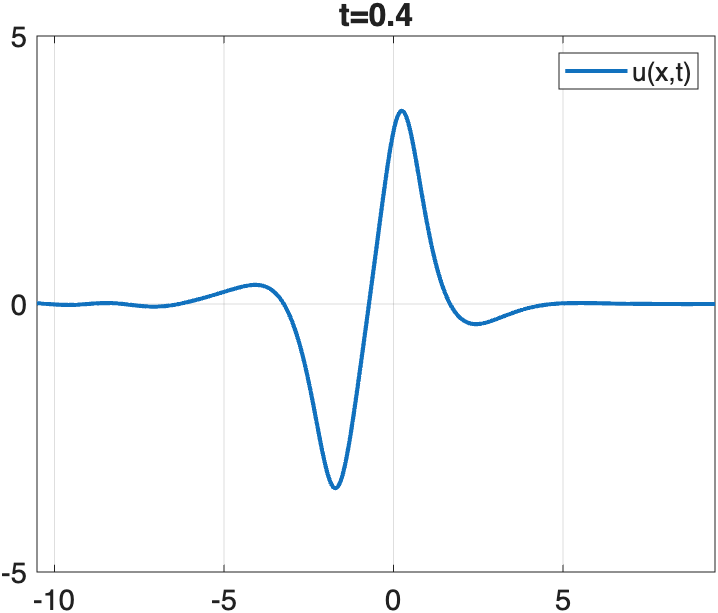} \hspace{2pt} 
\includegraphics[width=0.23\textwidth]{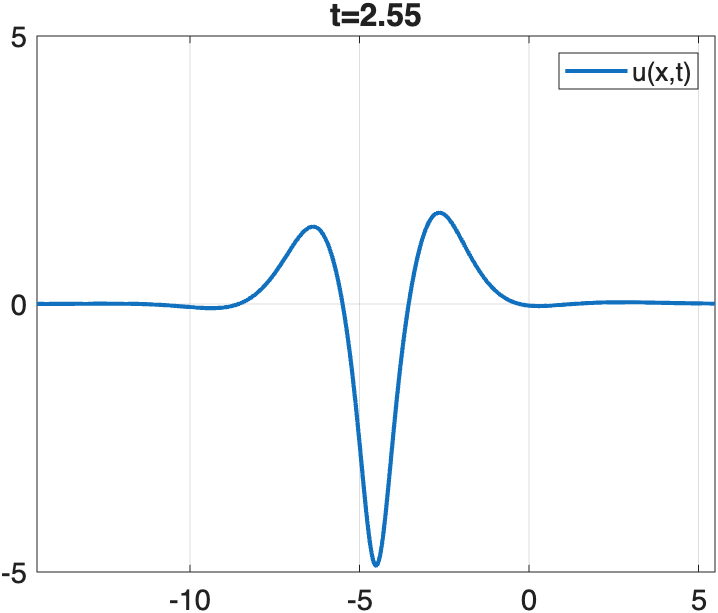} \hspace{2pt} 
\includegraphics[width=0.23\textwidth]{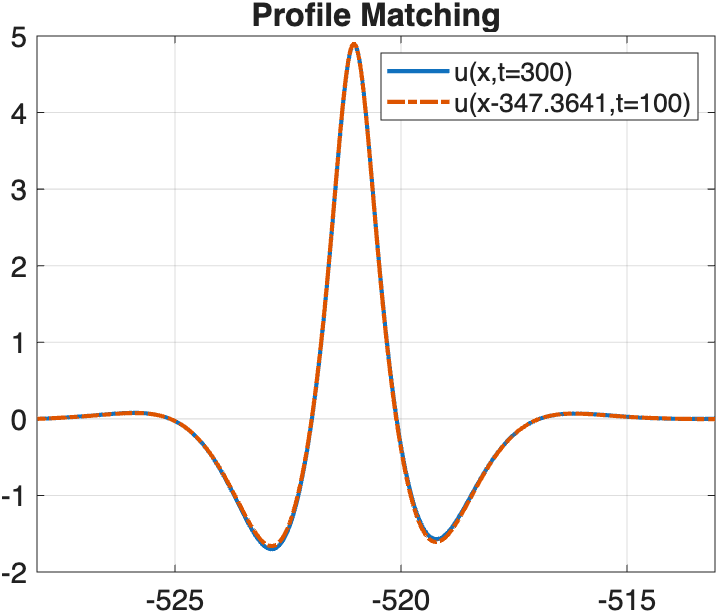} \\ 
\caption{\label{p2e0.1} {\small Time evolution of mKdV breather \eqref{mKdV-B} $u_0(x)=B_{1,1}(x,0)$ in the perturbed mKdV with $\mathcal{N}(u)=u^2u_x+\frac{1}{10} uu_x$ with profile matching of $u(x-347.3641,t=100)$ to $u(x,t=300)$ (last subplot).} }
\end{center}
\end{figure}

In Figure \ref{p2e0.1}, the profile of the initial data at $t=0$ is the same as the one seen in Figure \ref{B} at $t=0$.
At $t=0.4$, very small-scale radiation disperses to the left while the breather-like structure forms and starts to `breathe' and travel left. (The radiation here appears since we do not consider an explicit Gardner's breather, so some mass is shed to radiation). By $t=2.55$, the solution has approached the \textit{negative} breather profile. To confirm stability, the solution is tracked for sufficiently long times, up to $t=300$. 
Here, we do not track the governing breather parameters (e.g., $\alpha$, $\beta$, $\gamma$, $\delta$), since for that we would need to introduce Gardner's breather; instead we treat it as if it were a non-integrable case and perform profile matching between different (far apart) periods. 
Thus, in the last subplot of Figure \ref{p2e0.1}, we match two profiles from two sufficiently spaced time steps ($t=100$ and $t=300$) to justify and show the periodicity and stability of the breather. 

\begin{figure}[h!]
\begin{center}
\includegraphics[width=0.4\textwidth,height=.35\textwidth]{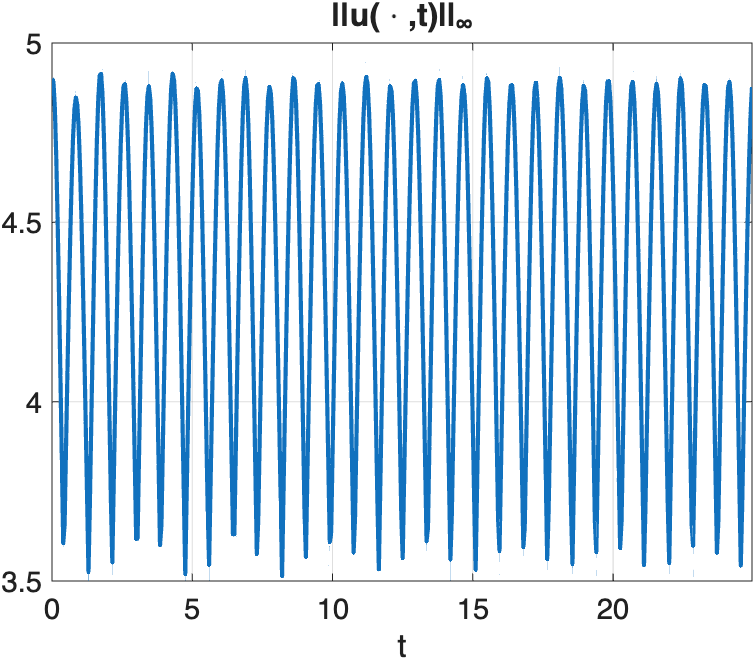} \hspace{10pt}
\includegraphics[width=0.44\textwidth,height=.36\textwidth]{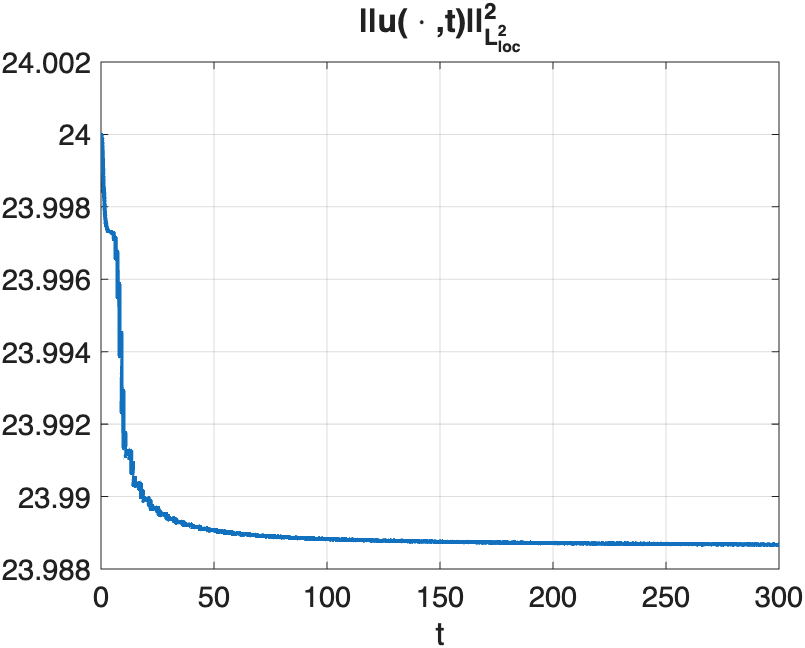} \\
%\hspace{10pt} \\
\hspace{-5pt}
\includegraphics[width=0.415\textwidth,height=.34\textwidth]{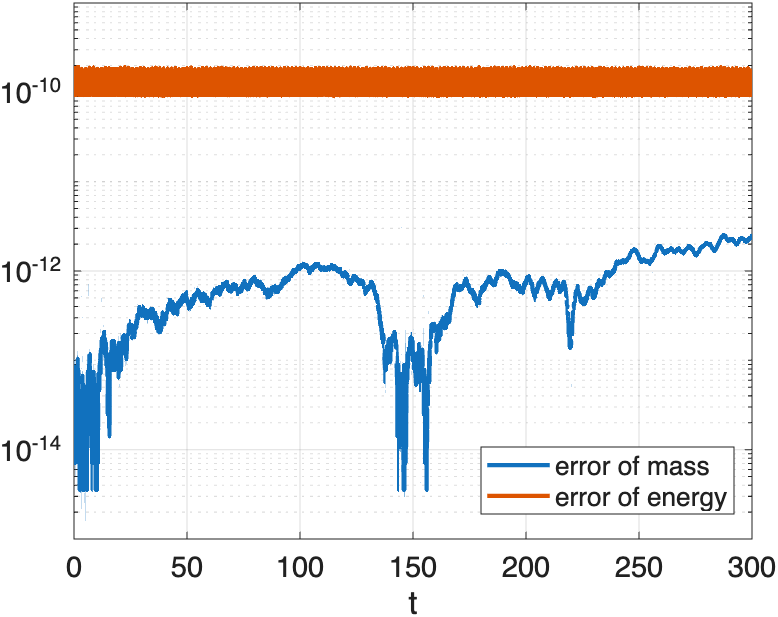} \hspace{5pt}
\includegraphics[width=0.43\textwidth,height=.36\textwidth]{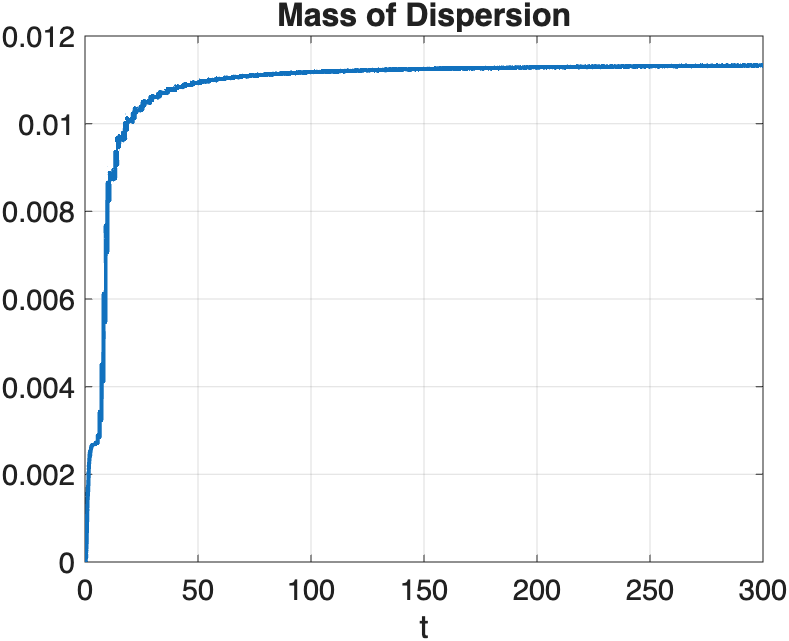}
\caption{\label{p2e0.1L} {\small Time evolution of $u_0(x)=B_{1,1}(x,0)$ in the perturbed mKdV with $\mathcal{N}(u)=u^2u_x+\frac{1}{10} uu_x$: $L^\infty$ norm up to $t=25$ (top left), tracking of local mass to $t=300$ (top right), mass and energy errors (bottom left), tracking of mass of dispersed radiation to $t=300$ (bottom right).} }
\end{center}
\end{figure}

In the top left of Figure \ref{p2e0.1L} we show the $L^{\infty}$ norm evolving up to time $t=25$; 
for visual purposes, a snippet of the $L^{\infty}$ norm is shown in order to distinguish the oscillatory pattern (other plots in the same figure are given up to $t=300$). In the top right subplot of Figure \ref{p2e0.1L}, the local mass of the evolving breather is plotted as a function of time, which seems to stabilize around $t=200$. Since we observed small-scale radiation that dispersed left in Figure \ref{p2e0.1}, one can notice that before $t=50$, the local mass drops down (the computation gives a drop of about $0.011$), then it starts to stabilize  around $t=100$. We track the time evolution further to make sure it levels out, which it does (around $t=200$). Then, to quantify the perturbation, we compare the initial mass and the final (computationally) local mass 
$$
\|u_0(x)\|^2_{L^2(\mathbb{R})}=24 \quad \text{and} \quad \|u(x,t=300)\|^2_{L^2_{loc}}=23.9887.
$$ 
Knowing the local mass at each time step, we can track the growing mass of the radiation as a function of time throughout the simulation and observe its asymptotic leveling off. The mass of the radiation around $t=100$ is about $0.0112$, see further plateauing on the bottom right of Figure \ref{p2e0.1L}, thus,  
supporting the presence of a stable breather-like structure, periodic in time and localized in space over the simulated time interval. 

To show the accuracy of our computations, 
we show the errors of mass and energy as functions of time in the bottom left subplot of Figure \ref{p2e0.1L}. The error of energy remains consistent around $10^{-10}$, and the error of mass is even smaller, of machine precision, between $10^{-14}$ and $10^{-12}$, capturing dynamics of the solution throughout the time evolution. Both errors of mass and energy remain small throughout the simulation, thus, supporting the accuracy, stability, and reliability of our numerical results in Figures \ref{p2e0.1} and \ref{p2e0.1L}.

\subsubsection{The mKdV with subcritical perturbation $p=2$ with absolute value.}

We consider an absolute value in the nonlinear term \eqref{E:Per2} with $p=2$ and assess whether stable breather-like solutions can exist in a non-integrable model. With absolute values, the potential term $\mathcal{N}(u)$ is symmetric. Following the work from \cite{frrsy2022}, we conjecture the existence of breather solutions for equations with symmetric potentials. This example provides positive numerical evidence for this conjecture. To compare with the previous integrable example, we keep $p=2$ and $\epsilon=\frac{1}{10}$ in \eqref{E:Per2} and initialize the time evolution with the exact mKdV breather \eqref{mKdV-B} 
$$
u_0(x)=B_{1.3,1}(x,0).
$$ 
The computational parameters in this simulation are $L=500\pi$, $N=2^{17}$, and $dt=0.0025$. We note that with the addition of absolute value and increase of $\alpha$ from $1$ to $1.3$, the breather-like solution travels faster to the left in the time evolution. Therefore, we increase the size of the computational domain and number of points to ensure simulations can run for sufficiently long times and high accuracy. With these computational parameters, we are able to simulate and identify a stable breather-like solution for a non-integrable system. 
\begin{figure}[h!]
\begin{center}
\includegraphics[width=0.23\textwidth]{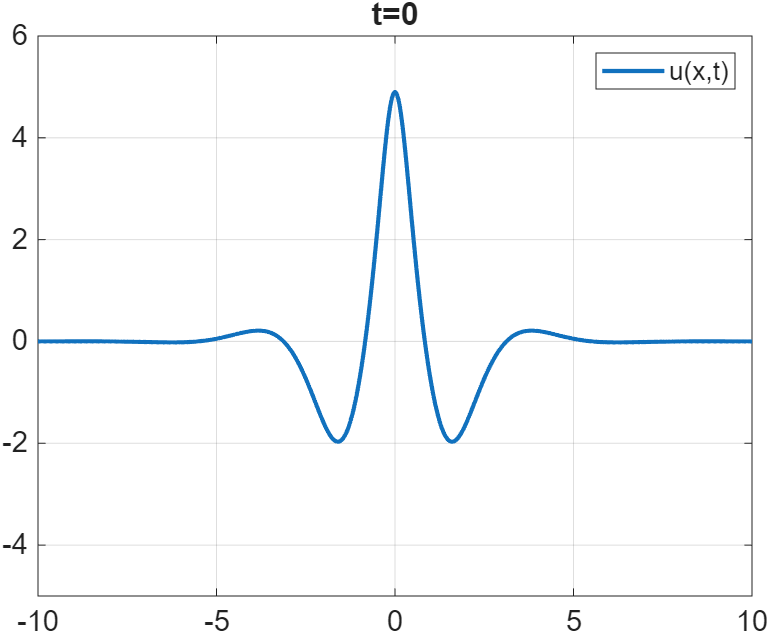} \hspace{2pt} 
\includegraphics[width=0.23\textwidth]{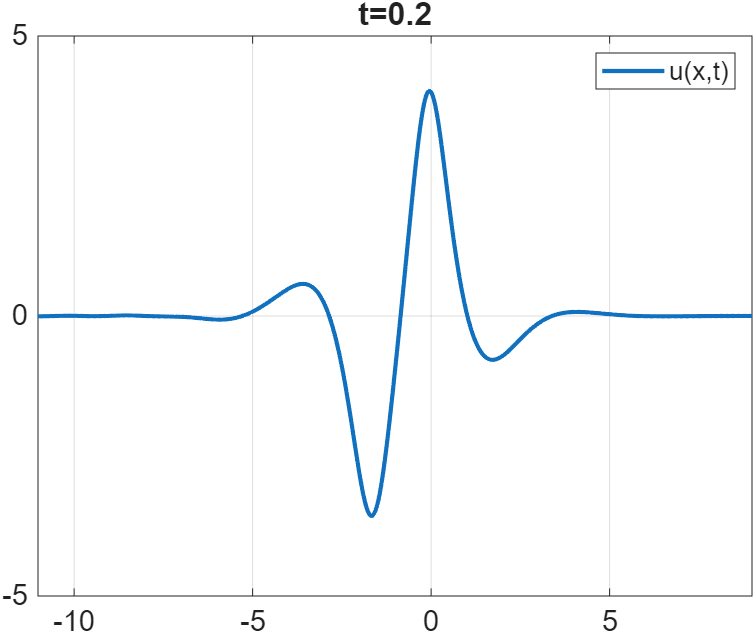} \hspace{2pt} 
\includegraphics[width=0.23\textwidth]{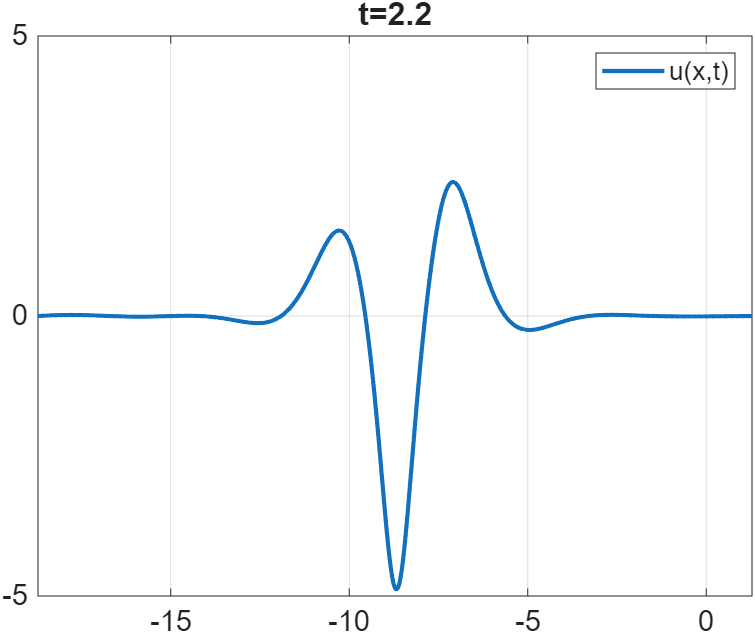} \hspace{2pt} 
\includegraphics[width=0.23\textwidth]{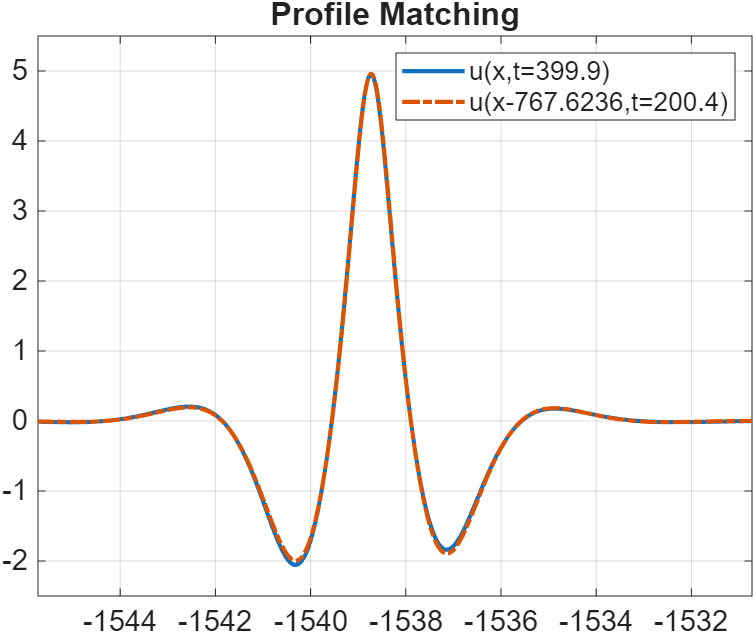} \\ 
\caption{\label{p2e0.1abs} {\small Time evolution of mKdV breather \eqref{mKdV-B} $u_0(x)=B_{1.3,1}(x,0)$ in the perturbed mKdV with $\mathcal{N}(u)=u^2u_x+\frac{1}{10} |u|u_x$ with profile matching of $u(x-767.6236,t=200.4)$ to $u(x,t=399.9)$ (last subplot).} }
\end{center}
\end{figure}

\begin{figure}[h!]
\begin{center}
\includegraphics[width=0.4\textwidth,height=0.33\hsize]{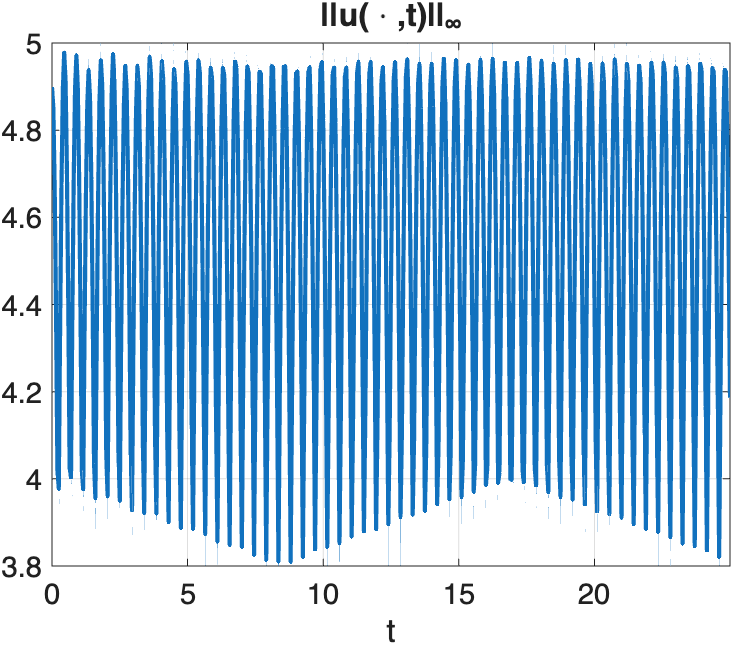} \hspace{10pt}
\includegraphics[width=0.4\textwidth,height=0.33\hsize]{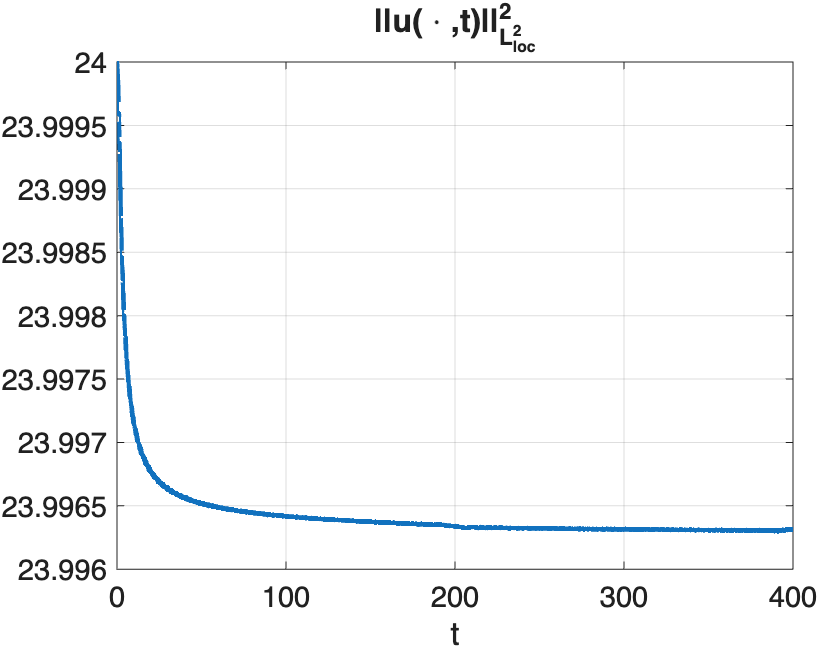} \\ %\hspace{10pt}
\includegraphics[width=0.4\textwidth,height=0.33\hsize]{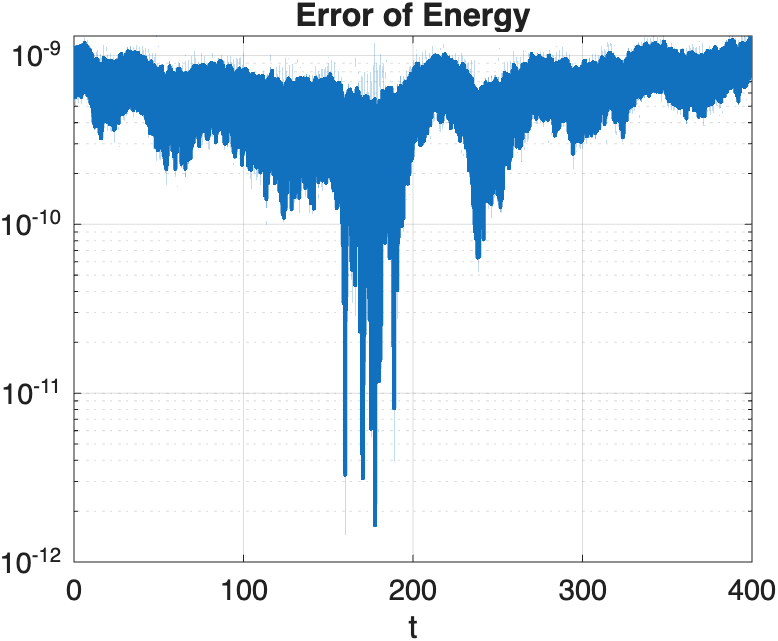} \hspace{10pt}
\includegraphics[width=0.4\textwidth,height=0.33\hsize]{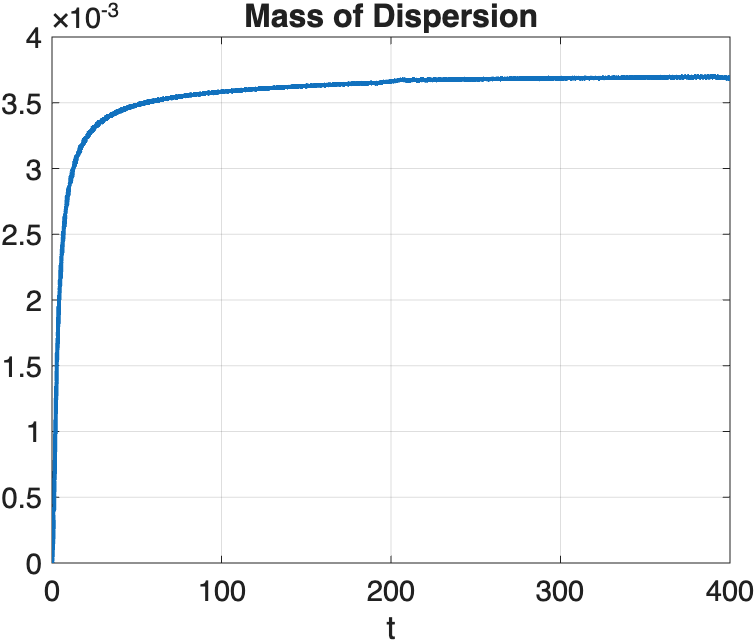}
\caption{\label{p2e0.1absL} {\small Time evolution of $u_0(x)=B_{1.3,1}(x,0)$ in the perturbed mKdV with $\mathcal{N}(u)=u^2u_x+\frac{1}{10} |u|u_x$: $L^\infty$ norm up to $t=25$ (top left), tracking of local mass to $t=400$ (top right), mass and energy errors (bottom left), tracking of mass of dispersed radiation to $t=400$ (bottom right).} }
\end{center}
\end{figure}

In Figure \ref{p2e0.1abs}, the solution profile at $t=0$ is slightly different from that in Figure \ref{p2e0.1} due to the different initial value of $\alpha$, however, this change in the initial $\alpha$ does not change the initial mass. In Figure \ref{p2e0.1abs}, the solution begins by shedding small-scale radiation to its left, then we observe the solution travel to the left. By $t=2.2$, the breather-like solution achieves the \textit{negative} breather profile. The solution in Figure \ref{p2e0.1abs} travels much faster to the left reaching maximum absolute height near $x=-8.5$ at $t=2.2$, while in Figure \ref{p2e0.1}, the solution reached its maximum absolute height near $x=-4.5$ at $t=2.55$. In the last subplot of Figure \ref{p2e0.1abs}, the solution's profile at an earlier time $t=200.4$ (shifted by $-767.6236$) is matched to the profile at a much later time $t=399.9$. These timestamps are after the breather-like solution stabilizes (confirmed by the local mass plot from Figure \ref{p2e0.1absL}), so they were chosen to show the convergence, periodicity, and stability of the breather-like solution. 

In Figure \ref{p2e0.1absL}, the $L^{\infty}$ norm, local mass, mass of the radiation, and error of energy are shown to justify the stability of the breather-like solution seen in Figure \ref{p2e0.1abs}. Since the solution is highly oscillatory, the $L^{\infty}$ norm is a snippet, plotted to $t=25$, to show the oscillations more clearly. % visible. 
The $L^{\infty}$ norm is consistent for the entire simulation, however, it is not sufficient at capturing the small changes that come from the perturbation of $\mathcal{N}(u)$. To quantify this perturbation, we computed the total and local mass:
$$
\|u_0\|^2_{L^2(\mathbb{R})}=24 \quad \text{and} \quad \|u(x,t=400) \|^2_{L^2_{loc}}=23.9963,
$$ 
making the mass of the radiation about $0.0037$, see bottom right subplot of Figure \ref{p2e0.1absL}. % also %Then, in the middle subplot of Figure \ref{p2e0.1abs}, the 
Tracking the local mass captures the initial changes the solution undergoes in order to converge to a stable, final state. The initial local mass starts at $24$, then steadily decreases. By $t=200$, the local mass stabilizes. Meanwhile, the mass of the dispersion first steadily increases (because the solution is shedding radiation), then plateaus by around $t=200$.  
To show the accuracy of our simulation, we plot the error of energy as a function of time in the bottom left subplot of Figure \ref{p2e0.1absL}. The error of energy captures the dynamics of the solution and remains mostly between $10^{-10}$ and $10^{-9}$. To make the energy-error estimate more precise, by the end of our simulation we had 
$$
%dt=0.0025 \quad \text{and} \quad 
Err(E)=1.3417\times 10^{-9}.
$$
The error of mass has similar accuracy, so we omit it in the figure. Since the domain size is $L=500\pi$ in Figures \ref{p2e0.1abs} and \ref{p2e0.1absL}, the error of energy and error of mass improve by increasing the number of points $N$ in the spatial domain, though this does significantly increase computational time. 

The subplots of Figure \ref{p2e0.1absL} provide numerical evidence for the existence and stability of breather-like solutions in this non-integrable system.

\section{Conclusions}

In this paper, we present a systematic numerical study of breather solutions to the modified Korteweg-de Vries (mKdV) equation, focusing on their structural properties, stability under perturbations, and interaction dynamics. We investigate perturbed breathers and characterize their long-time behavior, including scenarios in which the %final
emergent coherent structures propagate to the left, to the right, or remain stationary. To describe these asymptotic states (or final computational states), we develop practical procedures for extracting effective asymptotic parameters from numerical simulations. 

A central component of our approach is tracking localized structures and extracting their local mass. If computed globally in space, this mass would be conserved; computed locally, it shows asymptotic convergence to a final state and allows us to diagnose the persistence and stability of a breather-like coherent structure. 
These quantities also enable a quantitative assessment of radiative shedding: we measure the outgoing (left-propagating) radiation, and estimate its mass, which we then report as a fraction of the initial breather mass transferred to radiation under perturbations (and also as a percentage of the initial mass, which can be useful for future studies). We furthermore study multi-breather and breather-soliton interactions, emphasizing the robustness of the observed structures and their stability during collisions. 

Finally, we examine the effect of perturbing the nonlinearity and its implications for the persistence of breather-like dynamics. In particular, we provide numerical evidence that breather behavior can persist in {non-integrable} mKdV-type models, and we identify and quantify both the coherent breather component and the accompanying radiation. For example, it would be interesting to investigate and adapt diagnostics from this work to related equations in the sine-Gordon/mKdV hierarchy. The numerical evidence presented in this paper motivates future analytical work to rigorously justify the observed stability and long-time asymptotics as well as other numerical explorations of coherent structures.

\bigskip

\noindent \textbf{Acknowledgments.} The research of this project started during the Summer 2022 REU program ``AMRPU @ FIU'' that took place at the Department of Mathematics and Statistics, Florida International University, and was supported by the NSA grant H982302210016 and NSF (REU Site) grant DMS-2050971 (PI: S. Roudenko). In particular, support of C.H. and D.S. came from these grants. 
S.R. and K.Y. were partially supported by the NSF grants DMS-1927258 and DMS-2055130. In addition, the research of C.H., D.S., and S.R. on this project was also partially supported by the NSF grant DMS-2452782. 
\bigskip

\noindent{\bf Conflict of Interest:} The authors declare that they have no conflicts of interest.
\bigskip

\bigskip

\noindent{\bf ORCID}
\smallskip

\noindent{\it Chandler Haight}
\qquad {https://orcid.org/0009-0006-0035-9146}

\noindent{\it Svetlana Roudenko}
\quad {https://orcid.org/0000-0002-7407-7639}

\noindent{\it Diana Son}
\qquad {https://orcid.org/0009-0009-2164-1265}

\noindent{\it Kai Yang} 
\qquad {https://orcid.org/0000-0002-2289-9403}

%\bigskip

%\newpage
%\addcontentsline{toc}{section}{References}

\end{document}